\documentclass[%
aip,
amsmath,amssymb,
reprint,%
nofootinbib,
]{revtex4-1}
\usepackage{graphicx}% Include figure files
\usepackage{dcolumn}% Align table columns on decimal point
\usepackage{bm}% bold math
\usepackage[utf8]{inputenc}
\usepackage[T1]{fontenc}
\usepackage{mathptmx}
\usepackage{etoolbox}
\usepackage{hyperref}
\usepackage{cleveref}
\usepackage{xcolor}
\usepackage{booktabs}
\usepackage{placeins}

\newcommand{\md}{\mathrm{d}}
\newcommand{\joint}{q}
\newcommand{\marginal}{\hat{p}}
\newcommand{\cond}{m}
\newcommand{\ele}{x}
\newcommand{\elec}{x_t}
\newcommand{\elem}{x_s}
\newcommand{\ind}{i}

\newcommand{\grad}{\nabla}

\makeatletter
\def\@email#1#2{%
 \endgroup
 \patchcmd{\titleblock@produce}
  {\frontmatter@RRAPformat}
  {\frontmatter@RRAPformat{\produce@RRAP{*#1\href{mailto:#2}{#2}}}\frontmatter@RRAPformat}
  {}{}
}%
\makeatother

\preprint{AIP/123-QED}
\draft % marks overfull lines with a black rule on the right

\begin{document}

% Use the \preprint command to place your local institutional report number 
% on the title page in preprint mode.
% Multiple \preprint commands are allowed.
%\preprint{}

\title{Operator Forces For Coarse-Grained Molecular Dynamics
%\protect\footnotemark[5]
} %Title of paper

% repeat the \author .. \affiliation  etc. as needed
% \email, \thanks, \homepage, \altaffiliation all apply to the current author.
% Explanatory text should go in the []'s, 
% actual e-mail address or url should go in the {}'s for \email and \homepage.
% Please use the appropriate macro for the type of information

% \affiliation command applies to all authors since the last \affiliation command. 
% The \affiliation command should follow the other information.

\author{Leon Klein}
\thanks{equal contribution}
%\thanks{Corresponding author; Electronic mail: leon.klein@fu-berlin.de}
\author{Atharva Kelkar}
\thanks{equal contribution}
\author{Aleksander Durumeric}
\thanks{Corresponding author; Electronic mail: alekepd@gmail.com}
\author{Yaoyi Chen}
\affiliation{Department of Mathematics and Computer Science, Freie Universität, Berlin,
Germany}
\author{Frank Noé}
\thanks{Corresponding author; Electronic mail: franknoe@microsoft.com}
\affiliation{Microsoft Research AI for Science, Berlin, Germany}
\affiliation{Department of Mathematics and Computer Science, Freie Universität, Berlin,
Germany}
\affiliation{Department of Physics, Rice University, Houston, Texas 77005, USA}

% Collaboration name, if desired (requires use of superscriptaddress option in \documentclass). 
% \noaffiliation is required (may also be used with the \author command).
%\collaboration{}
%\noaffiliation
\date{\today}

\begin{abstract}
\emph{{The following article has been submitted to The Journal of Chemical Physics.}}\\\\
Coarse‐grained (CG) molecular dynamics simulations extend the length and time scale of atomistic simulations by replacing groups of correlated atoms with CG beads. Machine‐learned coarse‐graining (MLCG) has recently emerged as a promising approach to construct highly accurate force fields for CG molecular dynamics. However, the calibration of MLCG force fields typically hinges on force matching, which demands extensive reference atomistic trajectories with corresponding force labels. In practice, atomistic forces are often not recorded, making traditional force matching infeasible on pre-existing datasets. Recently, noise‐based kernels have been introduced to adapt force matching to the low-data regime, including situations in which reference atomistic forces are not present. While this approach produces force fields which recapitulate slow collective motion, it introduces significant local distortions due to the corrupting effects of the noise-based kernel. In this work, we introduce more general kernels based on normalizing flows that substantially reduce these local distortions while preserving global conformational accuracy. We demonstrate our method on small proteins, showing that flow‐based kernels can generate high-quality CG forces solely from configurational samples.
\end{abstract}

% insert suggested PACS numbers in braces on next line

\maketitle %\maketitle must follow title, authors, abstract and \pacs
%\footnotetext[5]{The following article has been submitted to The Journal of Chemical Physics.}
% Body of paper goes here. Use proper sectioning commands. 
% References should be done using the \cite, \ref, and \label commands

% If in two-column mode, this environment will change to single-column format so that long equations can be displayed. 
% Use only when necessary.
%\begin{widetext}
%$$\mbox{put long equation here}$$
%\end{widetext}

% Figures should be put into the text as floats. 
% Use the graphics or graphicx packages (distributed with LaTeX2e).
% See the LaTeX Graphics Companion by Michel Goosens, Sebastian Rahtz, and Frank Mittelbach for examples. 
%
% Here is an example of the general form of a figure:
% Fill in the caption in the braces of the \caption{} command. 
% Put the label that you will use with \ref{} command in the braces of the \label{} command.
%
% \begin{figure}
% \includegraphics{}%
% \caption{\label{}}%
% \end{figure}

% Tables may be be put in the text as floats.
% Here is an example of the general form of a table:
% Fill in the caption in the braces of the \caption{} command. Put the label
% that you will use with \ref{} command in the braces of the \label{} command.
% Insert the column specifiers (l, r, c, d, etc.) in the empty braces of the
% \begin{tabular}{} command.
%
% \begin{table}
% \caption{\label{} }
% \begin{tabular}{}
% \end{tabular}
% \end{table}

%%%%%%%%%%%%%%%%%%%%%%%%%%%%%%%%%%%%%%%%%%%%%%%%%%%%%%%%%%%%%%%%%%%%%%%
\section{Introduction}
Molecular Dynamics (MD) simulations provide microscopic insight into biomolecular systems, playing a crucial role in drug discovery~\citep{adcock2006molecular,hospital2015molecular,hollingsworth2018molecular}. Atomistic MD numerically integrates the equations of motion for every atom using time steps on the order of femtoseconds, 
%requiring all‑atom positions and forces at each step, 
an approach that quickly becomes computationally challenging for large systems \citep{voelz2010molecular,lindorff2011fast,plattner2017complete,robustelli2022molecular}.
% \Yaoyi{MD with Langevin dynamics uses stochastic equations, but others might not.}
Coarse‑graining (CG) offers a remedy by representing groups of atoms as individual beads (e.g. replacing each amino acid using only the corresponding C$\alpha$ atom), thereby reducing the number of variables present in the equations of motion and permitting larger time steps \citep{clementi2008coarse,noid2013perspective,jin2022bottom,noid2023perspective,borges2023pragmatic,marrink2023two}. However, obtaining the CG forces required to define the equations of motion is not trivial, as standard force fields are designed for all‑atom models.

Depending on the systems and phenomena under investigation, CG force fields can be obtained by tuning parameters to fit either macroscopic observables (``top-down'') \citep{marrink2023two,borges2023pragmatic} or the behavior of a finer resolution, usually atomistic, reference system (``bottom-up'') \citep{jin2022bottom,noid2023perspective}. A principled way to obtain CG forces in bottom-up CG is through variational force matching (also known as multiscale CG): one minimizes the mean‐squared error between instantaneous all‐atom forces projected onto the CG coordinates and those given by a CG force field \citep{izvekov2005multiscale,izvekov2005multiscale2,noid2008multiscale, lu2011multiscale}. In practice, this requires having all‐atom force labels for a large number of reference configurations.
%, often on the order of millions of frames.
However, archiving MD trajectories can require a substantial amount of storage capacity, motivating discarding force information when trajectories are saved to reduce their file size. Furthermore, these storage constraints can also lead to pruning the recorded positions to retain only a subset of the original system (e.g., the removal of all solvent atoms) to further reduce overhead, preventing reevaluation of forces on saved configurations.
%These difficulties motivate training approaches that avoid reference to atomistic forces. 

Recently, bottom-up machine‐learning (ML) approaches for learning CG force‐fields (MLCG) have emerged
%, creating surrogate models to predict CG forces when trained on all‐atom data
\citep{lemke2017neural,wang2019machine,husic2020coarse,chen2021implicit,wang2021multi,majewski2022machine,arts2023two,durumeric2023machine,charron2023navigating, wellawatte2023neural, kohler2023flow, kramer2023statistically, chennakesavalu2023ensuring, mirarchi2024amaro}. MLCG force fields trained using force matching have achieved state‐of‐the‐art performance~\cite{charron2023navigating}; however, their need for force-equipped atomistic configurations represents an heightened barrier to application due to an increased need for training data~\citep{durumeric2023machine}. For example, modeling individual miniproteins requires approximately one million reference configurations~\cite{kramer2023statistically}.
%which are both costly to generate and rarely archived in existing simulation repositories \citep{durumeric2023machine}. 
Consequently, there is growing interest in learning MLCG force fields without atomistic forces~\citep{shell2008relative, carmichael2012new, kohler2023flow, arts2023two, lemke2017neural, thaler2022deep, thaler2021learning, ding2022contrastive, plainer2025consistentsamplingsimulationmolecular}, including building on existing ML training approaches well-suited for no-force training.~\citep{hinton2002training, hyvarinen2005estimation, gutmann2010noise, vincent2011connection, song2020sliced}.
\citet{durumeric2024learning} proposed combining CG force matching and denoising score matching~\citep{vincent2011connection, song2021, ho2020denoising}, a standard ML technique for learning force fields. In practice, the approach uses a noise kernel $\kappa_N(R'|R)$  with $R' \sim\mathcal{N}(R, \sigma^2)$ to noise configurations $R$; this defines a ``noise-force''  via $\nabla_{R'}\log\kappa_N(R'|R)$ which may be combined with atomistic forces to create a dataset for downstream force field training.
%Moreover, these noise forces can be combined with CG forces derived from the underlying all atom configurations. 
However, in order to define a useful learning objective, models must be trained on both noise-induced force information and the corresponding noise-corrupted configurations. Training on such distorted configurations produces CG MD trajectories that inherit the configurational artifacts induced by the noising kernel,
%which are particularly in detailed metrics like bond‑length distributions,
which can be problematic. 
For example, directly mapping generated CG structures to low-energy all-atom conformations \citep{badaczewska2020computational,shmilovich2022temporally,wang2022generative,jones2023diamondback,yang2023chemically,pang2024simple, sahrmann2025emergence} can be infeasible, as significant local distortions prevent low-energy reconstructions. 
%For example, training transferable MLCG models \citep{charron2023navigating} typically requires forces from mostly undistorted configurations \citep{durumeric2023machine} \Alek{The transf. MLCG paper did use non-noise forces, but it's complicated to talk about here. Briefly, the finer resolution combined with noise-based distortion made the prior calculation unstable. Maybe you can omit it and just talk about the point in the following sentence. It is a real reason to use the methods here, but I don't know how easy it will be to say.}

In the numerical results of \citet{durumeric2024learning}, only kernels based on random Gaussian noise are considered; however, the described theory extends to any kernel where the corresponding derivative can be computed. Here, we explore kernels based on generative models, which have the capacity to introduce far fewer local distortions. Generative models have achieved remarkable success across domains, ranging from image synthesis \citep{goodfellow2014generative, ho2020denoising} and language generation with Large Language Models \citep{vaswani17_atten_is_all_you_need, devlin2019bert,brown2020language, achiam2023gpt} to physical‑science applications such as protein folding \citep{jumper2021highly, abramson2024accurate} and accelerated sampling of Boltzmann distributions \citep{noe2019boltzmann}. We focus on exact likelihood generative models, specifically normalizing flows \citep{RezendeEtAl_NormalizingFlows, papamakarios2021normalizing}, which facilitate computing differentiable sample probabilities and thus allow the corresponding forces to be calculated.
%-->Kernel forces

In this work, we demonstrate the effectiveness of generative model-based kernels in coarse-grained force matching using two different construction techniques, both implemented via normalizing flows.
On alanine dipeptide and two small proteins, we show that generative model-based kernels preserve local structure, whereas the previously proposed random noise kernels introduce significant distortions in local behavior (e.g., bond lengths) despite preserving slow motion as described via \emph{time-lagged independent component analysis} (TICA) features \citep{perez2013identification}. 
Furthermore, we show that the distortions induced by simple noising prevent backmapping to low energy all atom conformations, and that this difficulty is substantially reduced when using our generative approach.
Finally, we demonstrate that an existing transferable generative model, which was originally trained on a different all atom task, can be used in our framework to generate CG forces.
% We make the following main contributions: \Leon{Maybe this is not a common thing to do for journals}
% \begin{itemize}
%     \item We show that arbitrary differentiable kernels can be used to compute coarse-grained forces, and we illustrate this with two examples built based on normalizing flows.
%     \item On alanine dipeptide and two proteins, we show that random noise kernels, while preserving TICA features, distort local coordinates (e.g. bond lengths) significantly, whereas generative model based kernels reduce these distortions. Furthermore, we show that these distortions prevent backmapping to low energy all atom conformations. 
%     \item We further demonstrate that an existing transferable generative model, which was originally trained on a different all atom task, can be used in our framework to generate CG forces. 
%     %\item We show in the absence of atomistic forces but also combine atomistic and synthetic forces, which is especially beneficial in low data regimes. 
% \end{itemize}

%\Leon{where to put this now??}
%Notably, \citet{kohler2023flow} also employ a flow model to generate forces without atomistic labels. Unlike our approach, where forces are obtained from the conditional flow distribution, they derive forces directly from the unconditional push‐forward distribution, since their flow is trained to sample the Boltzmann distribution. However, they were unable to employ these forces for CGSchNet training. Instead, they applied them only within the CGNet framework \citep{wang2019machine}.

\section{Theory}\label{sec:theory}
\begin{figure}[ht]
\centering
\includegraphics[width=0.9\columnwidth]{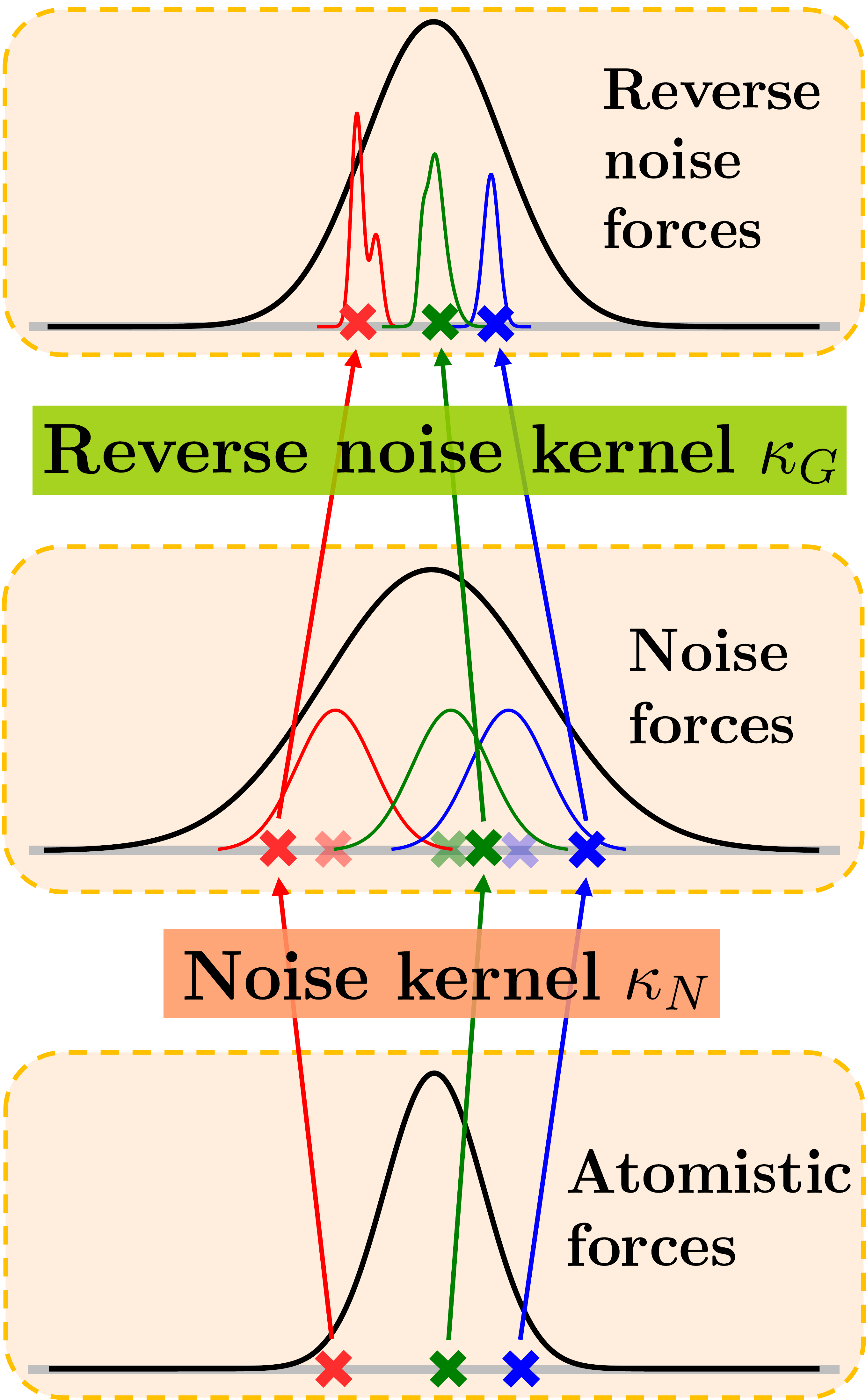}
\caption{Illustration of the three force‐generation strategies for a single unimodal distribution (e.g., a bond length): \textbf{Bottom panel:} Three points sampled from the data distribution. When atomistic forces are available, they can be used directly in subsequent applications. \textbf{Middle panel:} Each of the three samples is perturbed by Gaussian noise (noise kernel), broadening the distribution. The original sample points, which now serve as the centers of the Gaussian kernels, are displayed semi-transparent. Forces computed from this noising process correspond to the distorted distribution and thus inherit its local inaccuracies.
\textbf{Top panel:} We apply a learned reverse noise kernel (via a conditional flow) to correct the perturbed samples, yielding a distribution closer to the original. Forces derived from this reverse kernel result in fewer local distortion in subsequent applications.
}
\label{fig:intro}
\end{figure}
In this section, we summarize the theoretical foundations underlying our approach. We begin by reviewing the traditional coarse-grained force matching framework, continue by describing how noise‐based kernels can be used to obtain coarse‐grained forces without atomistic labels, and finally introduce more general kernel constructions, which we build upon in \cref{sec:method}.

\subsection{Force matching}
Force matching is a widely used method for constructing CG or surrogate potentials by minimizing the discrepancy between forces predicted by a model and reference forces obtained from simulations,  experiments, or other methods \citep{ercolessi1994interatomic,izvekov2005multiscale,noid2008multiscale}. In the CG scenario, each CG configuration $R\in \mathbb{R}^{3N}$ is obtained from the all-atom coordinates $r\in\mathbb{R}^{3n}$ via a (linear) map $R=Mr$. Similarly, atomistic forces $\mathfrak{f}(r)$ can be projected to CG forces via a force-mapping operator $M_{\mathfrak{f}}$. A parametrized CG force field $\hat{F}_{\theta}(R)$ can then be learned by minimizing the force matching objective 
\begin{align}\label{eq:fore-matching}
    \mathcal{L}(\theta) = \left\langle \left\| M_{\mathfrak{f}} {\mathfrak{f}}(r) - \hat{F}_{\theta}(Mr) \right\|_2^2 \right\rangle_r
\end{align}
where $\langle \cdot \rangle_r$ denotes the ensemble average over the all atom ensemble  \citep{izvekov2005multiscale,noid2008multiscale}. 
%The most common approach projects the atomistic forces $\mathfrak{f}(r)$ via a mapping $M_{\mathfrak{f}}$, such that $F(R) = M_{\mathfrak{f}} \mathfrak{f}(r)$ .
The construction of a valid $M_{\mathfrak{f}}$ for a chosen CG coordinate mapping $M$ has been discussed in details in \citet{ciccotti2005blue,noid2008multiscale, kramer2023statistically}. In this work, we explore alternative strategies for generating CG forces when atomistic forces are not available. These methods can then be used to generate training data for downstream use in the force matching framework. 

\subsection{Forces from noise kernels}
Recently, \citet{durumeric2024learning} proposed a modified force matching approach for coarse-grained systems distorted via noise kernels. We adopt their notation here. The core idea parallels popular denoising diffusion models \citep{hyvarinen2005estimation,vincent2011connection,ho2020denoising,song2021}: one perturbs each molecular configuration and uses knowledge about the applied distortion to produce force information. In the case of sampling distorted configurations through the addition of Gaussian noise with covariance $\sigma^2 I$,
% $R' \sim \mathcal{N}\bigl(R,\sigma^2 I\bigr)$
% with variance $\sigma^2$, 
this corruption process is described by the noise kernel
\begin{align}
\kappa_N(R'\mid R)
= \frac{1}{(2\pi\sigma^2)^{\frac{3N}{2}}}
\exp\Bigl(-\tfrac{|R'-R|^2}{2\sigma^2}\Bigr).
\end{align}
Perturbing the data distribution $p(R)$ with the kernel gives rise to a distorted data distribution 
\begin{align}\label{eq:nodistortion}
\hat{p}(R')=\int \kappa_N(R'\mid R) p(R) dR.
\end{align}
The work of \citet{vincent2011connection, song2021, durumeric2024learning} show that if one applies force matching to corrupted configurations $R'$ using the score of the distortion kernel,
\begin{equation}\label{eq:gaussforce}
F_N(R', R) = \nabla_{R'}\log\kappa_N(R'\mid R),
    %= -\frac{R'-R}{\sigma^2}
\end{equation}
then the resulting force matching procedure targets the \emph{distorted} data distribution $\hat p$. In this setup, the force matching objective becomes
\begin{align}\label{eq:noise-force-matching}
    \mathcal{L}_N(\theta) = \left\langle \mathbb{E}_{\kappa_N} \left\| F_N(R', Mr) - \hat{F}_{\theta}(R') \right\|_2^2 \right\rangle_r,
\end{align}
where $\mathbb{E}_{\kappa_N}$ denotes the expectation over all distorted configurations $R'$ sampled from the noise kernel $\kappa_N(R'\mid Mr)$.
An illustration is provided in \cref{fig:intro}, and a detailed derivation can be found in \cref{app:derivation}. Although they only employ noise kernels in their work, the theoretical results in \citet{durumeric2024learning} hold for more general kernels representing conditional probability densities with tractable log derivatives. 

\subsection{Forces from general kernels}\label{sec:general-kernels}
Despite the successful creation of CG force fields through Gaussian noising (\citet{durumeric2024learning}), training using forces produced through \cref{eq:gaussforce} is inherently distorted by the addition of said Gaussian noise. Although lowering the magnitude of $\sigma$ reduces this distortion, doing so increases the amount of data needed for force field training. However, it is possible to describe kernels that result in no distribution-level distortion; that is, they result in a $\hat{p}$ that is equivalent to $p$. 
As described in this section, while it is straightforward to obtain samples from these kernels, it is not trivial to directly evaluate their corresponding log derivatives.

In this work we approximate optimal non-distorting kernels by training conditional generative models to reproduce datasets describing the behavior of the targeted kernel.
There are numerous ways to construct generative models for this purpose. Since we require tractable derivatives, we employ exact‐likelihood generative models, in our case normalizing flows. We discuss the technical details of these generative models in  \cref{sec:normalizing-flows}, and continue here by describing the non-distorting processes we aim to emulate.

\subsubsection{Reverse noise kernels} 
The first approach builds directly on corrupting data using Gaussian noise. We begin by perturbing each configuration $R$ with Gaussian noise to obtain $R'\sim\kappa_N(R'\mid R)$, and then reverse the noise sample $R'$ by pushing it towards the data distribution via a generative kernel given by
% The resulting composite kernel is
% \begin{align}
%     \kappa_G(\hat{R}\mid R)=\int\kappa_F(\hat{R}\mid R')\kappa_N(R'\mid R) dR'.
% \end{align}
%\begin{align}
%\hat{p}(\hat{R})=\int\int\kappa_F(\hat{R}\mid R')\kappa_N(R'\mid R) p(R) dR'dR.
%\end{align}
\begin{align}
    \kappa_G(R\mid R')= \frac{\kappa_N(R'\mid R) p(R)}{\hat{p}(R')}.
\end{align}
Consequently, applying the exact reverse kernel would recover the original (unperturbed) data distribution. In practice, however, this reverse kernel is intractable, so we approximate it with a normalizing‐flow model. The resulting distribution $\hat p(\hat R)$ will differ from the true data distribution $p(R)$ but should more closely resemble it than the corrupted distribution. Forces from the reverse noise kernel can be obtained in the same way as before, namely via 
\begin{align}\label{eq:reverse-force}
    F_G(R,R')=\nabla_{R}\log\kappa_G(R\mid R'),
\end{align}
with the corresponding force matching objective
\begin{align}\label{eq:reversenoise-force-matching}
    \mathcal{L}_G(\theta) = \left\langle \mathbb{E}_{\kappa_G} \left\| F_G(R, R') - \hat{F}_{\theta}(R) \right\|_2^2 \right\rangle_{R'},
\end{align}
where $\langle \cdot \rangle_{R'}$ denotes the ensemble average over the distorted configurations $R'$ sampled from the noise kernel (\cref{eq:nodistortion}). An illustration is provided in \cref{fig:intro}, and a detailed derivation can be found in \cref{app:derivation}. 

\subsubsection{Transition kernels}
Instead of creating data pairs through the synthetic addition of noise, one may instead consider deriving forces from lag-time dependent transition kernels. For a lag time $\tau$ and configuration $R_t$ at time $t$, a transition kernel can defined as
$\kappa_T\bigl(R_{t+\tau}\mid R_t\bigr)$. As molecular dynamics is well-characterized as a Markov process with a stationary density at sufficient lag times \citep{prinz2011markov,bowman2013introduction}, the corresponding transition density fulfills 
\begin{align}
p(R_{t+\tau})=\int \kappa_T\bigl(R_{t+\tau}\mid R_t\bigr) p(R_{t}) dR_{t}
\end{align}
when $p$ is the equilibrium distribution. 
The corresponding force on the later configuration is then similar to before
$F(R_{t+\tau}, R_t) \;=\; \nabla_{R_{t+\tau}} \log \kappa_T\bigl(R_{t+\tau}\mid R_t\bigr)$.
This approach requires transition kernels with tractable derivatives and access to trajectory data rather than independent samples. A major benefit is that pretrained models with exact likelihoods already exist for this task \citep{klein2023timewarp}. These models can be used directly to compute CG forces on trajectories they were trained on or to which they transfer.

\section{Method}\label{sec:method}
In this section, we detail how we construct noise‐ and flow‐based kernels and how we employ the resulting force estimates to train a force matching CG model. Moreover, we outline our evaluation protocol for the experiments. 

\subsection{CGSchNet}\label{sec:cgschnet}
One particularly successful MLCG force matching model is \emph{CGSchNet} \citep{husic2020coarse}, which learns an energy function $\hat{U}_{\theta}(R)$ and obtains forces via $\hat{F}_{\theta}(R) = -\nabla_R \hat{U}_{\theta}(R)$. Its architecture builds on the message‐passing network \emph{SchNet} \citep{schutt2018schnet}, encoding interparticle geometric relationships. In addition to the learned SchNet energy, CGSchNet incorporates a fixed “prior” energy term, $\hat{U}_{\text{prior}}(R)$, typically composed of bonded interactions (bonds, angles, dihedrals) and short‐range repulsions \citep{wang2019machine,husic2020coarse}. Usually the prior potentials adopt rather simple functional forms (e.g., harmonic potentials) and the parameters are fitted against the statistics of corresponding features of the training set. Previous studies have demonstrated that these priors are essential for stable and accurate CGSchNet performance \citep{wang2019machine,husic2020coarse,charron2023navigating}, since they compensate the lack of data samples available for learning high-free-energy barriers around physical states. Consequently, the total model energy is
$
    \hat{U}_{\theta}(R) = \hat{U}_{\text{SchNet}}(R)\;+\;\hat{U}_{\text{prior}}(R),
$
where $\hat{U}_{\text{prior}}(R)$ remains fixed throughout training.

\subsection{Normalizing flows}\label{sec:normalizing-flows}
Normalizing flows \cite{RezendeEtAl_NormalizingFlows, papamakarios2021normalizing} are a type of deep generative model used to estimate complex probability densities $p(x)$ by transforming a base distribution $q(x)$ through an invertible transformation $f_{\theta}:\mathbb{R}^n \to \mathbb{R}^n$, resulting in the push-forward distribution $\hat{p}_{\theta}(x)$.
Importantly $\hat{p}_{\theta}(x)$ can be evaluated via the change of variable equation
\begin{equation}
\hat{p}_{\theta}(x)=q\left(f_{\theta}^{-1}(x)\right)\left|\det J_{f_{\theta}^{-1}}(x)\right|,
\label{eq:change_of_variable}
\end{equation}
where $f_{\theta}^{-1}(\cdot)$ denotes the inverse transformation and $J_{f_{\theta}^{-1}}(\cdot)$ the Jacobian of $f_{\theta}^{-1}(\cdot)$. 
Hence, also the gradient of $\hat{p}_{\theta}(x)$ is tractable, which we use to derive forces. 

We can also employ a flow as a conditional model by conditioning the transformation $f_{\theta}(\cdot \mid x')$, which yields a conditional push-forward distribution $\hat{p}_{\theta}(x \mid x')$, which can be used to approximate our proposed kernels. There are two primary ways to parametrize the flow transformation $f_{\theta}$: coupling flows and continuous normalizing flows (CNFs), both of which we investigate in this work.

\subsubsection{Coupling flows}
Coupling flows \citep{dinh14_nice, dinh16_densit_estim_using_real_nvp} employ a sequence of coupling layers $l_{\theta}^i$ to build the invertible transformation $f_{\theta}$. At each layer the input $x = (x_1, x_2)$ is partitioned into two blocks, and one block is transformed conditioned on the other:
\begin{equation}
   y_1 = x_1, 
\quad
y_2 = l_{\theta}^i(x_2| x_1). 
\end{equation}
The full flow is then the composition of $n$ such layers where the conditioning is alternated,
\begin{equation}
f_{\theta}(x) \;=\; (l_{\theta}^n\circ l_{\theta}^{n-1}\circ\cdots\circ l_{\theta}^1)(x).
\end{equation}
Because each layer’s Jacobian matrix is block-triangular, its determinant is simply the product of diagonal terms, making the change-of-variables formula (\cref{eq:change_of_variable}) efficient to evaluate.  
Training can be performed via maximum‐likelihood training:
\begin{align}\label{eq:likelihood}
    \mathcal{L}_{\mathrm{lik}}(\theta) =  \mathbb{E}_{x\sim p(x)}\left[ \log \hat{p}_{\theta}(x)\right].
\end{align}

We adopt the coupling flow architecture proposed by \citet{klein2023timewarp}, the \emph{Timewarp} model, which builds on the transformer framework of \citet{vaswani17_atten_is_all_you_need} and is tailored specifically for enhancing molecular simulations. Originally designed to accelerate sampling by predicting large time‐step transitions, this conditional normalizing coupling flow learns the distribution
$\hat{p}_{\theta}(R_{t+\tau}\mid R_t)$
for large lag times $\tau$, and thus directly furnishes the transition kernel from which forces can be derived.

\subsubsection{Continuous normalizing flows}
Continuous normalizing flows (CNFs) \citep{chen2018neural, grathwohl2018ffjord} can be viewed as the continuous‐time analogue of coupling flows. The invertible mapping $f_{\theta}(x)$ is defined by integrating a time‐dependent vector field over the unit interval:
\begin{equation}
    f_{\theta}(x) \;=\; x_0 \;+\; \int_{0}^{1} v_{\theta}\bigl(x_t, t\bigr)\,\mathrm{d}t,  
\end{equation}
where $v_{\theta}\colon \mathbb{R}^{n}\times[0,1]\to\mathbb{R}^{n}$ is a smooth, time‐dependent vector field and $x_0\sim q(x)$ is sampled from the base distribution. This continuous formulation allows flexible transformation of probability densities by solving an ordinary differential equation driven by $v_{\theta}$.

We adopt the CNF architecture of \citet{klein2023equivariant} and \citet{klein2024tbg}, which builds on equivariant graph neural networks (EGNN) \citep{satorras2021graph, satorras2021n} and is specifically tailored for molecular systems. Because this flow model is rotation‐equivariant, training requires no data augmentation yet produces a rotationally invariant push‐forward distribution \citep{kohler2020equivariant}. However, these models are trained to generate independent Boltzmann‐distributed samples and do not support conditioning on an input configuration. 
To remedy this, we propose a simple modification that enables modeling the conditional distribution $\hat{p}_{\theta}(x|x')$ via a conditional flow model $f_{\theta}(x|x')$.
Concretely, we treat the conditioning configuration $x'$ as an additional input to the time‐dependent vector field $v_{\theta}(x, t| x')$ by adding the conditioning configuration to the input and subtracting it again from the output:
\begin{align}
    v_{\theta}(x, t| x') &= \phi_{\text{EGNN}}(x+x', t) - x'
\end{align}
The corresponding conditional flow transformation becomes
\begin{align}
       \hat{x}= f_{\theta}(x_0|x') = x' + x_0 + \int_{0}^{1}v_{\theta}(x_t, t| x')\,\mathrm{d}t, 
\end{align}
where $x_0$ is drawn from the base distribution (usually a standard normal). The flow learns to map random noise into an update relative to the conditional configuration $x'$.
%For full architectural details, see \cref{app:CNF}.

Continuous normalizing flows can also be trained via the maximum‐likelihood objective (\cref{eq:likelihood}). However, a substantially more effective approach is \emph{flow matching} \citep{lipman2022flow, albergo2023stochastic, liu2022flow}, which we describe in detail in \cref{app:CNF}. 

For further implementation details on both flow architectures, see \cref{app:implementatoin}. 
Despite their differing network designs, once trained, both flow models can generate CG forces following the same procedures.

\subsection{Approximating general kernels with flow models}
Below, we demonstrate how the conditional flow models can be employed to approximate our proposed kernels.

\subsubsection{Obtaining forces with flow models via reverse noise kernels}\label{sec:forces-via-denoising-kernels}
We use both the coupling flow as well as the CNF model to approximate the reverse noise kernel $\kappa_G(R\mid R')$ via the conditional push‐forward distribution $\hat{p}_{\theta}(\hat{R}\mid R')$. 
For the coupling flow (Timewarp) model, we retain the original architecture but repurpose it for the reverse noise kernel. Instead of predicting future configurations, we train the flow to stochastically reverse the noise corruption step. Given a noised configuration $R'$, the model learns the conditional push‐forward distribution $\hat{p}_{\theta}(\hat{R}\mid R')$.
%which serves as our desired reverse noise kernel for force matching.
%Because the model is not inherently rotation‐invariant, we apply random rotational data augmentation during training.
%Importantly, by structuring the EGNN input as a molecular graph and applying this “add‐and‐subtract” conditioning, we obtain a simple yet powerful conditional CNF architecture that, to our knowledge, has not been proposed before. 
Our conditional CNF model is inherently designed for this task, as the noised configuration $R'$ serves directly as the conditioning input $x'$, which results in the desired conditional push‐forward distribution $\hat{p}_{\theta}(\hat{R}\mid R')$.

Once a flow model has been trained, we generate CG forces in three steps, using only samples from the CG data distribution for conditioning:

\begin{enumerate}
\item Corrupt a reference configuration by adding Gaussian noise:
$R' = R + \varepsilon,\quad \varepsilon \sim \mathcal{N}(0,\sigma^2).$
\item Reverse the process with the trained flow model:\\
$   \hat{R} = f_{\theta}(x_0 \mid R'),\quad x_0 \sim \mathcal{N}(0,\sigma^2).$
\item Compute the CG force via the score of the conditional push‐forward distribution:
$ \hat{F}_{\theta}(\hat{R}, R') \;=\; \nabla_{\hat{R}}\log \hat{p}_{\theta}(\hat{R}\mid R').$
\end{enumerate}

These model‐derived forces can then be used to train any MLCG force matching framework, such as CGSchNet, entirely without requiring atomistic force labels.

\subsubsection{Obtaining forces with flow models via transition kernels}\label{sec:forces-via-transition-kernels}
If we have instead access to a flow model trained to predict large time‐step transitions, we can similarly derive CG forces as follows. We draw configuration pairs $(R_{t+\tau},R_t)$ from the CG data distribution, ensuring they are separated by the same lag time $\tau$ used during training. For each pair, the force is given by
\begin{equation}
F_{\theta}(R_{t+\tau},R_t)=
    \nabla_{R_{t+\tau}}\log \hat{p}_{\theta}\bigl(R_{t+\tau}\mid R_t\bigr), 
\end{equation}
and can be used to train MLCG models as before. 
We did not train any new flow models for this task. Instead, we directly use the pretrained Timewarp model from \citet{klein2023timewarp}.

Although we did not explore this variant here, one could generate novel configurations $\tilde R$ from the trained flow model and compute forces directly via
$
    F_{\theta}(\tilde R, R_t)
    = \nabla_{\tilde R}\,\log\hat p_{\theta}(\tilde R \mid R_t),
$
thereby eliminating the need for pre-existing sample pairs.

\subsection{Evaluation metrics}
We assess model performance using two complementary metrics. 

First, the \textit{PMF RMS error} measures the root‑mean‑square deviation between two dimensional free energy surfaces, namely Ramachandran plots \citep{ramachandran1963stereochemistry} or TICA \citep{perez2013identification} projections, of the generated versus reference ensembles. This provides a quantitative way to compare these distributions. We note that there are many implementation choices when estimating these divergences on finite samples, such as the number of histogram bins or the treatment of empty bins, which can substantially affect the numerical values obtained. We adopt the choices made in \citet{durumeric2024learning}, which we find closely mirror the visual differences in our observed distributions. However, this approach down-weights outliers, inadvertently favoring models with larger distortions (e.g., based on noise kernels). 
%For full details, see \cref{app:evaluation-metrics}.

Second, the \textit{Wasserstein distance} between generated and reference bond‑length distributions quantifies distortions in local structural features. Moreover, we also access the level of local distortion by trying to backmap generated CG conformations to all atom ones. 

Depending on the requirements of a given task, one may accept greater global fidelity at the expense of local distortion, or vice versa, prioritize exact local features even if global accuracy suffers. The Pareto front \citep{deb2011multi} marks those models for which no improvement in one metric is possible without compromising the other.

\subsection{Backmapping}\label{sec:backmapping}
As a test of local CG geometry for the investigated small protein systems, we perform backmapping \citep{badaczewska2020computational,shmilovich2022temporally,wang2022generative,jones2023diamondback,yang2023chemically,pang2024simple, sahrmann2025emergence}, reconstructing the backbone from CG configurations without relying on specialized libraries. Adding atomistic detail to CG configurations makes them practically useful, but this process becomes challenging when the CG structures suffer from distortions. At the C$\alpha$ resolution, we restore only the backbone carbon and nitrogen atoms, then energy‐minimize over bond, angle, and dihedral terms while keeping the original C$\alpha$ positions fixed. The force field terms are taken from \texttt{CHARMM22*} \citep{piana2011robust}, the force field for the atomistic simulations of the investigated Chignolin and Trp-Cage systems. We then compare these backbone energies to those of reference structures processed identically, i.e., mapping the all‐atom coordinates to the CG representation, reintroducing the missing backbone atoms, and performing energy minimization. Focusing solely on the backbone decouples our analysis from side‐chain reconstruction, which typically depends on tailored backmapping protocols, yet still reveals whether local distortions in the CG model induce systematic energy penalties even before side‐chain placement.

For the more fine grained alanine dipeptide system we reconstruct all atoms and only omit explicit solvent, as the system is significantly smaller. We similarly minimize the full force field and compare final conformational energies to the all‐atom reference ensemble.

\begin{figure*}[t]
\centering
\includegraphics[width=0.8\textwidth]{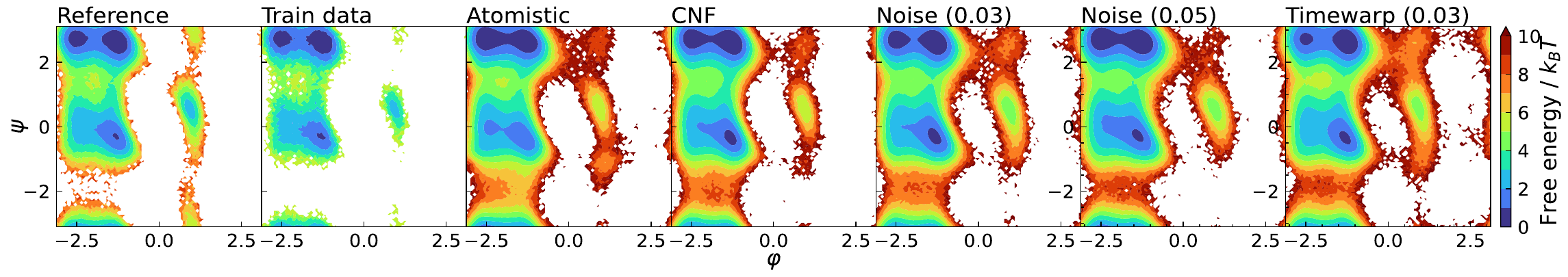}
\caption{Ramachandran plots for alanine dipeptide for the different models trained on $10\%$ of the training set.
}
\label{fig:ala2_rama_small}
\end{figure*}

\section{Results}\label{sec:results}
In this section we compare simulations performed with CGSchNet models based on different training forces.
We label the different CGSchNet models according to the source of their training forces: 
\begin{itemize}
    \item \textbf{Atomistic:} Forces are directly derived from the force of the  all atom configurations.
    \item \textbf{Noise:} Forces are generated with noise kernels. We denote the noise variance $\sigma^2$ in parentheses.
    \item \textbf{Timewarp:} Forces are generated with the reverse noise kernel based on the Timewarp coupling flow model, again specifying the variance for the corresponding noise kernel in parentheses.
    \item \textbf{CNF:} Forces derived from the reverse noise kernel based on the conditional continuous normalizing flow. Since we consistently use $\sigma^2=0.05$ across datasets, we omit it.
\end{itemize}
%We compare training forces generated with noise kernels (Noise) and our proposed general kernels based on the two normalizing flow models (Timewarp and the conditional CNF). As a additional reference, we also use forces derived form all atom configuration (Atomistic).
For all systems, we assess the impact of dataset size by training on the complete dataset as well as on reduced subsets containing $10\%$ and $2\%$ of the available training examples, following \citet{durumeric2024learning}. 
We evaluate our reverse‐noise kernel, implemented with both the Timewarp and CNF models, on alanine dipeptide as well as the mini-proteins Chignolin and Trp‐cage.
Additionally, we apply our transition kernel, modeled with the pretrained Timewarp model from \citet{klein2023timewarp}, to dipeptides, which we discuss at the end of this section.

%When applicable, we indicate the noise variance in parentheses immediately following the model name or the method used to generate the CG forces.

%We briefly explain each investigated system below.

\begin{figure*}[t]
\centering
\includegraphics[width=0.9\textwidth]{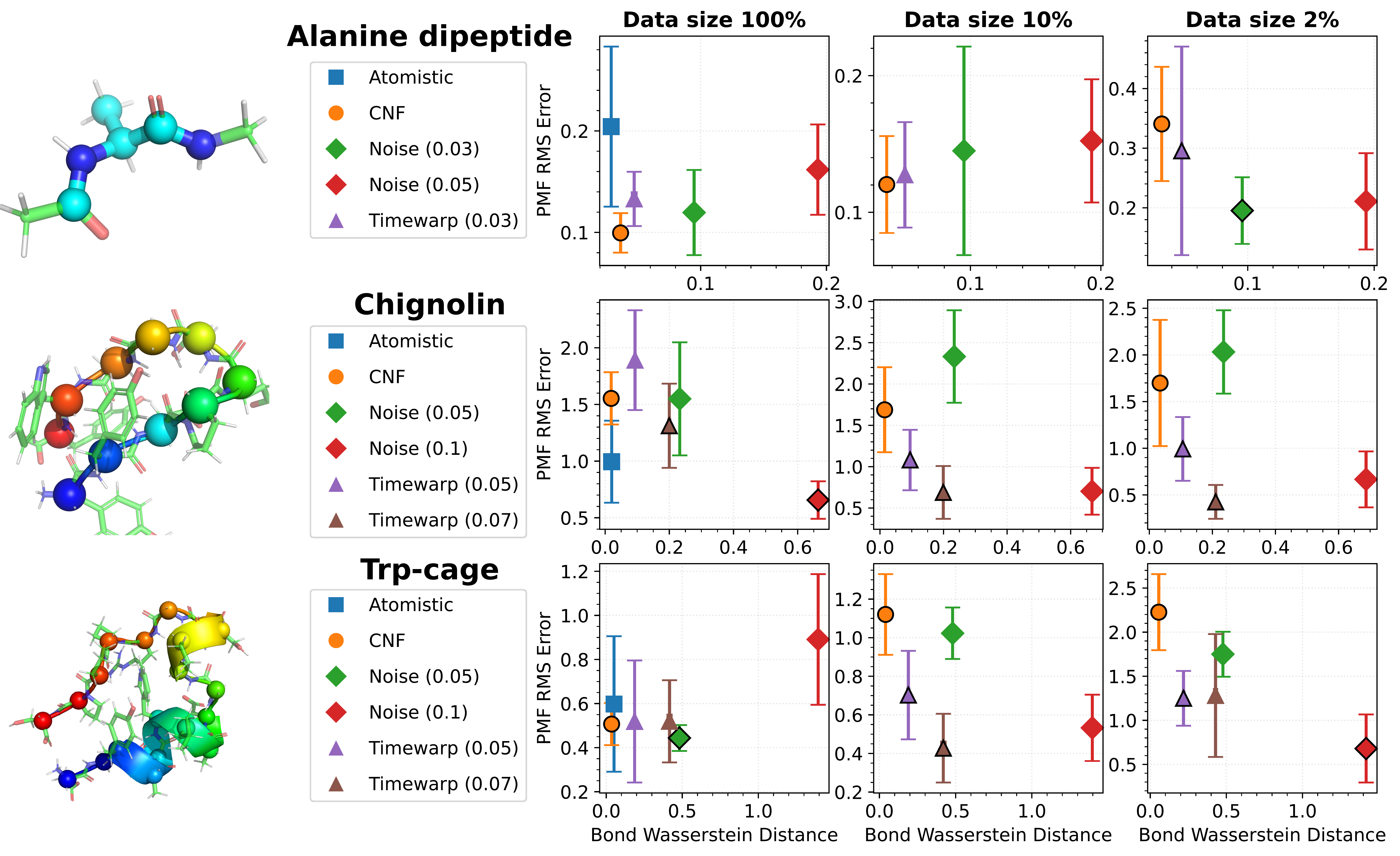}
\caption{Comparison of global‐feature accuracy (dihedral‐angle distributions or TICA projections) versus local‐feature accuracy (bond‐length distributions).
\textbf{Left column:} The three benchmark systems and their CG representations: Alanine dipeptide, Chignolin, and Trp‐cage. Each row corresponds to one system, with panels showing results for the different training set sizes. Pareto‐optimal models are indicated by bold markers, with membership determined solely based on the mean performance. The Atomistic model, having been trained with atomistic forces, is excluded from the Pareto sets and is presented solely as a reference. Furthermore, we report the atomistic results only on the full training set, since its performance on the reduced datasets is inferior to the other models. 
Error bars represent the standard deviation across three CGSchNet models; for the kernel‐based methods, each CGSchNet model was trained on a different set of forces generated with the respective kernels. Note that the error bars for the bond Wasserstein distances are too small to be visible.
}
\label{fig:compare}
\end{figure*}

\subsection{Alanine dipeptide}
In our first experiment, we investigate alanine dipeptide, which is commonly used to benchmark ML methods for MD simulations \citep{husic2020coarse,dibak2021temperature,kohler2021smooth,klein2023equivariant,invernizzi2022skipping}. We here use a trajectory generated with MD for $1\mu$s in explicit solvent \citep{wang2019machine} consisting of $10^{6}$ configurations, which are used as the training set and the reference distribution. Following \citet{husic2020coarse}, we coarse‑grain to six heavy atoms (\cref{fig:compare}), which still capture the essential $\varphi$ and $\psi$ dihedral angles governing the system’s dynamics. For this system, the CGSchNet model includes prior terms only for bonds and angles, omitting priors for dihedral and repulsion interactions. 

\begin{figure*}[t]
\centering
\includegraphics[width=0.9\textwidth]{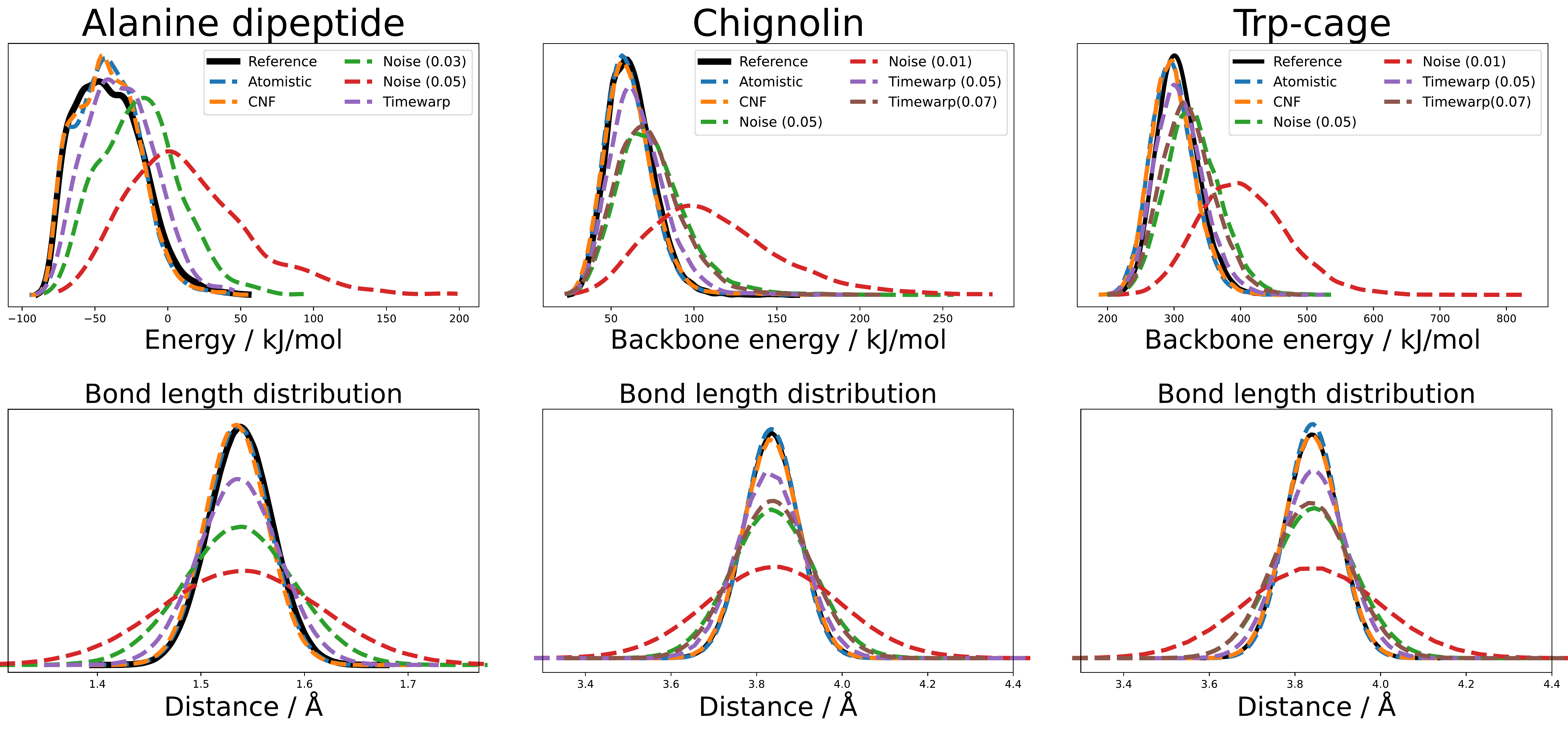}
\caption{\textbf{Top row:} Distribution of minimized energies for backmapped CG configurations across the three benchmark systems, comparing all models.
 \textbf{Bottom row:} Corresponding example bond distributions.
}
\label{fig:energies}
\end{figure*}

The Timewarp and the CNF model produce far lower local distortions than the Noise models for all training set sizes (\cref{fig:compare} and \cref{fig:energies}). In particular, the CNF model reproduces bond‑length distributions very close to the reference, similar to ones generated with the Atomistic model. Although reducing the noise variance in the corruption kernels can further diminish local distortions, it generally impairs global-feature fidelity and sometimes even destabilizes the simulation (see \cref{app:smaller-noise-levels}). 
Furthermore, Timewarp and CNF produce Ramachandran distributions similar to those obtained with the Noise models across all training‐set sizes (see \cref{fig:ala2_rama_small} and for all training set sizes \cref{fig:ala-ramachandran} in \cref{app:additional-plots}). Although the flow‐based kernels underperform on the PMF‐RMS metric for the smallest training split, they match or outperform the Noise models on the larger splits (\cref{fig:compare}). Nevertheless, the sizeable error bars for this system preclude a definitive ranking. Importantly, as the training set size shrinks, models based on atomistic forces degrade far more severely than those on kernel‐derived forces. This observation aligns with \citet{durumeric2024learning}, who similarly report that reliable CG simulations under limited data are achievable only when using kernel‐based forces. 
%Although, the Atomistic models are outperformed even when using the full training set on the PMF‐RMS metric, they exhibit the smallest local distortions.

\subsection{Chignolin}
\begin{figure*}[t]
\centering
\includegraphics[width=0.9\textwidth]{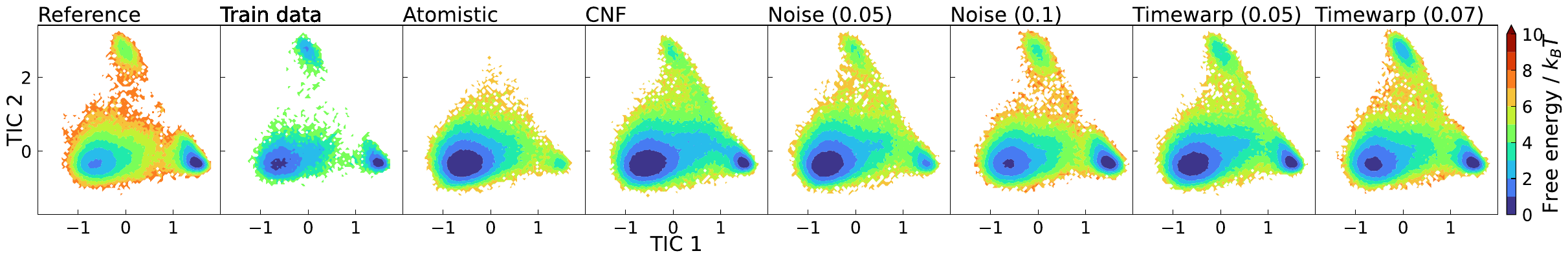}
\caption{TICA projections for Chignolin for the different models trained on $10\%$ of the training set.
}
\label{fig:cln_tica_small}
\end{figure*}

In next experiment, we transition to the C$\alpha$ CG representation and examine the miniprotein  Chignolin \citep{honda2008crystal} (\cref{fig:compare}). The dataset was created with all‐atom explicit solvent simulations comprising numerous short trajectories \citep{wang2019machine}, after which a Markov state model was constructed to reweight these conformations and generates a reference ensemble representing the Boltzmann distribution as in \citet{husic2020coarse}. Consistent with \citet{durumeric2024learning}, we employ the unweighted simulation data for training and benchmark our results against the weighted reference. The full training set consists of about $2\cdot 10^6$ samples.
%Notably, benchmarking against the training data itself yields comparable outcomes.
We observe that the quality of the TICA projections varies across models and training‐set sizes (\cref{fig:cln_tica_small} and for all training set sizes \cref{fig:cln_tica} in \cref{app:additional-plots}). Models trained with kernel‐derived forces consistently sample all metastable states for every training‐set size, whereas the Atomistic models only recover all states when trained on the full dataset. Although global‐feature accuracy deteriorates as the training‐set size decreases for all approaches, local‐feature accuracy remains essentially unchanged (\cref{fig:compare}).

For both Noise and Timewarp models, larger noise variances yield improved performance across all data splits. Specifically, increasing the variance reduces the PMF‐RMS error, as models based on noise‐kernels with the highest variance (Noise (0.1)) outperform all other models on the full dataset, while for smaller training sets, the Timewarp (0.07) attains the lowest PMF‐RMS error. We note that the PMF‐RMS metric down‐weights outliers, which become more frequent as the noise level increases. Moreover, higher‐variance kernels introduce substantially larger local distortions in the subsequent models.

These observations suggest a tradeoff between global and local fidelity among the models considered. Although the Timewarp models achieve lower PMF‐RMS errors compared to the CNF models, they incur greater local distortions. We mark Pareto‐optimal models in the accuracy landscape (\cref{fig:compare}); both flow‐based and noise‐kernel‐based methods contribute models to the Pareto front, highlighting their complementary strengths.

\subsection{Trp-cage}
Trp‐cage is a $20$‐residue miniprotein renowned for its rapid folding kinetics (\cref{fig:compare}). Training data were obtained from all‐atom explicit‐solvent MD simulations kindly shared by the authors of \citet{majewski2022machine}. These data consist of 3,940 independent short trajectories, each seeded from diverse starting conformations, with an aggregate simulation time of 195 $\mu$s. Following \citet{durumeric2024learning}, we built a Markov state model to reweight the trajectory ensemble by transition probabilities, thereby recovering the Boltzmann distribution as our reference.

While the local distortions across models mirror those observed for Chignolin, the global‐feature fidelity exhibits a slightly distinct pattern (\cref{fig:trp_tica}). On the full training set, the Noise (0.1) model, the one with the highest variance, performs worst overall, yet it ranks among the top performers for the reduced training sets (\cref{fig:compare}). The remaining models behave more consistently with the trends previously reported for Chignolin.

%For the smallest training set size, we observe more pronounced differences in global conformational features, characterized by the TICA projections. In particular, for smaller training splits, models trained with noise‐kernel‐based forces achieve slightly improved global accuracy, although they still exhibit significant local distortions, as quantified by the Wasserstein distance of bond‐length distributions. 

\subsection{Backmapping generated CG configurations}
As detailed in \cref{sec:backmapping}, we assess backmapped conformations by comparing their energies to all‐atom references: for alanine dipeptide, we compare total energy distributions, and for Chignolin and Trp-cage, we focus on the backbone energy. As anticipated, models that incur fewer local distortions yield backmapped energy distributions that more closely align with the reference (\cref{fig:energies}). In particular, employing the CNF or the Atomistic model produces backmapped structures whose energy distributions are nearly indistinguishable from the reference. By contrast, the Timewarp model, and especially the Noise models, introduce distortions that shift the energy distributions significantly. Consequently, any backmapping strategy applied to these latter models will produce systematically biased energies, which may be problematic depending on the intended application.

%\Leon{plots or table, probably plots are better}
% \begin{table}
%   \caption{Energies of backmapped samples for the different models and datasets}
%   \label{tab:2AA}
%   \centering
%   \begin{tabular}{lcc}
%     \toprule
%     Model & \multicolumn{2}{c}{Energy $(\downarrow)$}\\
%          & Mean & Range \\
%          \midrule
%     TBG  &$0.48\pm0.59\%$&$(0.0\%,1.47\%)$\\
%     \bottomrule
%   \end{tabular}
% \end{table}

\subsection{Generating forces based on existing generative models}\label{sec:dipeptides}
In this section, we move beyond noise‐based kernels and explore alternative general kernels for force derivation, as introduced in \cref{sec:general-kernels} and \cref{sec:forces-via-transition-kernels}. Specifically, we employ the original Timewarp flow model \citep{klein2023timewarp}, pretrained to predict large transitions in time for dipeptides, i.e., $\tau=5\cdot10^5$fs. This model was trained on $200$ short dipeptide trajectories of $50$ns each using the \texttt{AMBER14} \citep{maier2015ff14sb} force field in implicit solvent \citep{klein2023timewarp}. Without any retraining, we apply it to trajectories of unseen dipeptides to generate CG forces via the kernel derivatives. Specifically, we evaluate on the test‐set peptides from \citet{klein2023timewarp}, each trajectory comprising $2\times10^5$ saved frames and spanning $1\mu$s.
%For further dataset specifications, see \citet{klein2023timewarp}.
We adopt the bead resolution of \citet{charron2023navigating}, namely five heavy atoms per residue. Since the all‐atom trajectories were generated without constraints, the CG force map corresponds directly to the subset of forces on these ten beads. Atomistic and noise‐kernel based forces are obtained as before. For these systems, the CGSchNet model includes prior terms only for bonds and angles, omitting priors for dihedral and repulsion interactions.

\begin{figure}[t]
\centering
\includegraphics[width=\columnwidth]{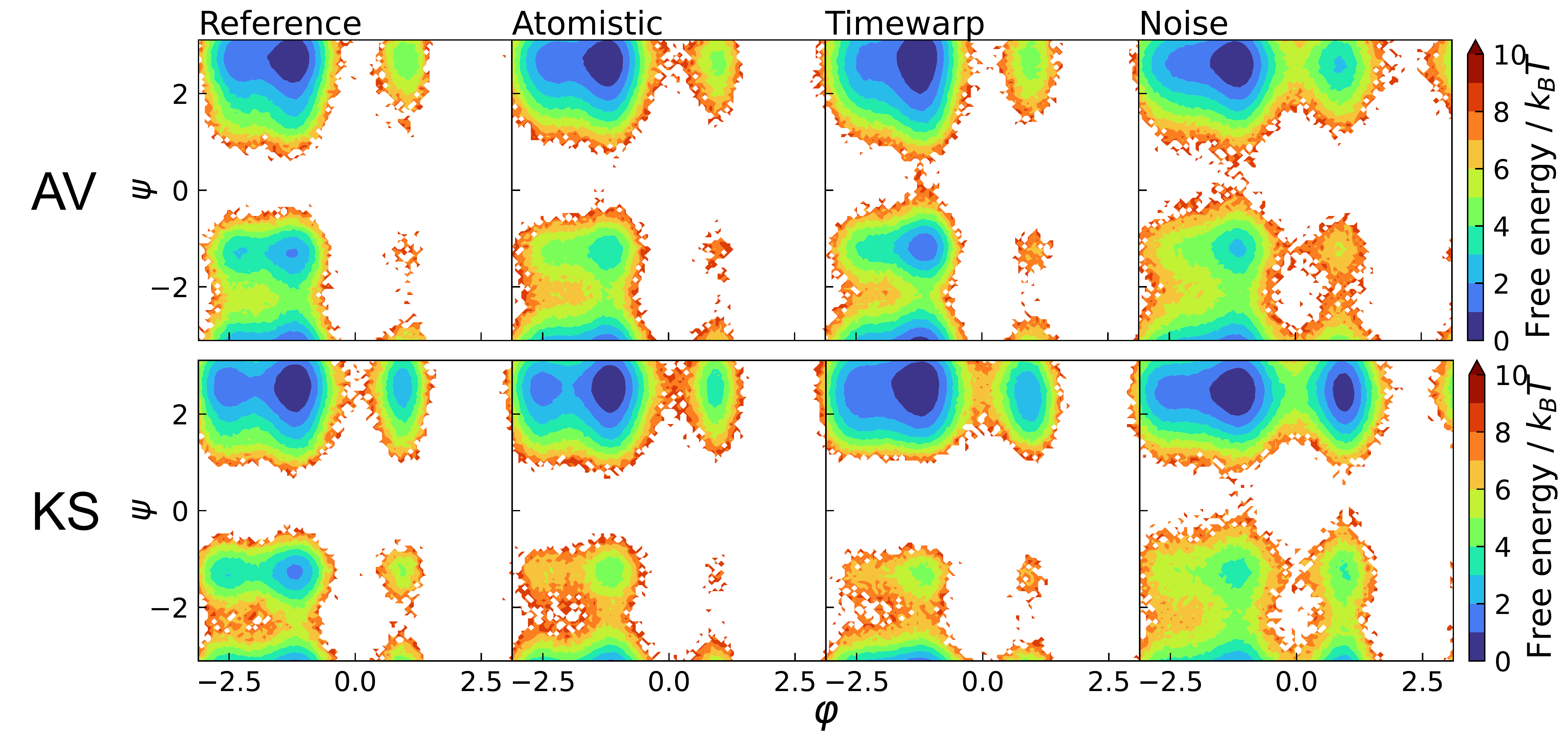}
\caption{Ramachandran plots for the AV and KS dipeptides, samples generated with different models. The pretrained Timewarp model is used to obtain forces via the transition kernel. }
\label{fig:dipeptides}
\end{figure}

Our results show that the models trained with forces from Timewarp‐based transition kernels match or exceed the global‐feature fidelity of the Noise model, evidenced by the difference between Ramachandran distributions, while also better preserving local geometry (\cref{fig:dipeptides} and \cref{tab:dipeptides}). Nevertheless, the reference model trained on direct atomistic forces remains superior across both global and local features. These findings demonstrate that pretrained flow models can be leveraged without retraining to generate CG forces for MLCG applications. For additional results, see \cref{app:additional-dipeptides}.

\begin{table}[ht]
\centering
\begin{tabular}{lcc}
\toprule
\textbf{Model}      
& \textbf{PMF RMS} $(\downarrow)$ & \textbf{Bond distance} $(\downarrow)$ \\
\midrule
Atomistic & $0.286 \pm 0.111$ & $0.0048 \pm 0.0015$ \\
Timewarp  & $0.373 \pm 0.142$ & $0.0112 \pm 0.0009$ \\
Noise     & $0.509 \pm 0.195$ & $0.0305 \pm 0.0009$ \\
\bottomrule
\end{tabular}
\caption{Averaged results for dipeptides. The pretrained Timewarp model is employed to derive coarse‐grained forces by computing the score of its transition kernel. Averaged across four test‐set peptides, each evaluated with three independent CGSchNet models.
}
\label{tab:dipeptides}
\end{table}

\section{Discussion}
In this work, we demonstrate that coarse-grained forces can be obtained from general kernels without requiring atomistic force labels. We introduce two normalizing-flow-based kernels and show that MLCG models trained on these flow-derived forces significantly reduce local distortions while maintaining global conformational accuracy comparable to noise-based kernels. Moreover, we highlight the critical role of local geometry in backmapping: only models trained with flow-derived or true atomistic forces are able to recover low-energy all-atom conformations.
Importantly, all kernel‐based force‐generation methods require only configuration samples, making them ideal for settings without force labels. Moreover, in the small data regime, kernel-based approaches outperform models trained on atomistic force labels. Nevertheless, when local fidelity is less critical, noise-based kernels remain a practical alternative: they often match and sometimes exceed flow-based kernels on global features, especially with sparse training data, and incur no additional overhead beyond tuning the noise variance. 

In addition, we also demonstrate how an existing generative model, which was originally trained to predict large MD time-steps for all atom simulations, can be repurposed within our framework to derive coarse‑grained force kernels without any retraining. As more advanced generative models become available, they too could be integrated directly, leveraging pretrained representations and reducing training overhead. 

There exist numerous alternative normalizing flow architectures tailored for molecular applications, e.g., \citet{midgley2023se, kohler2023rigid, abdin2023pepflow,pmlr-v238-draxler24a, tan2025scalable}, each bringing its own strengths and limitations. We hope future studies explore these and other emerging models to further advance kernel-based CG force generation.

Future work could also investigate hybrid force matching strategies that integrate CG forces generated by our general kernels with atomistic forces, similarly to the approach of \citet{durumeric2024learning} for noise kernels.

%%%%%%%%%%%%%%%%%%%%%%%%%%%%%%%%%%%%%%%%%%%%%%%%%%%%%%%%%%%%%%%%%%%%%%
\begin{acknowledgments}
The authors would like to thank
Nicholas Charron,
Andreas Krämer,
Michael Plainer,
and Maximilian Schebek
for fruitful discussions and their helpful input. 
We gratefully acknowledge support by the Deutsche Forschungsgemeinschaft (SFB1114, Projects No. A04 and No. B08) and the Berlin Mathematics center MATH+ (AA1-10).
Moreover, we gratefully acknowledge the computing time that was made available on the high-performance computer \emph{Lise} at the NHR Center NHR@ZIB. This center is jointly supported by the Federal Ministry of Education and Research and the state governments participating in the NHR (\url{www.nhr-verein.de/unsere-partner}). 

\end{acknowledgments}

\section*{Data Availability Statement}
No new datasets were generated for this study. All datasets used are available as cited in the main text. 
%We here provide direct links to the respective data repositories:  \Leon{Alanine dipeptide?}; for Chignolin and Trp-cage, see the datasets at \url{https://github.com/torchmd/torchmd-protein-thermodynamics/tree/main/Datasets} \citep{majewski2022machine}; for dipeptides, see the Microsoft datasets at \url{https://huggingface.co/datasets/microsoft/timewarp} \citep{klein2023timewarp}.
We additionally release our code and pretrained model weights at \url{https://github.com/noegroup/OperatorForces4CG}.

%%%%%%%%%%%%%%%%%%%%%%%%%%%%%%%%%%%%%%%%%%%%%%%%%%%%%%%%%%%%%%%%%%%%%%%%%%%%%%%%%%
\clearpage
\newpage
\appendix
\section{Theory}
\subsection{Derivation of general kernel forces}\label{app:derivation}
We here provide a simplified derivation for the forces for general kernels based on the theoretical results in \citet{durumeric2023machine, durumeric2024learning}. Note that the following derivation does not encompass atomistic ensembles with constrained bonds and only applies to coarse-graining operations described solely via the removal of variables; however, the more general approach in \citet{ciccotti2005blue} and \citet{durumeric2024learning} implies the same ultimate relationship. We set the thermodynamic beta to 1 and define the density function of the full system as $\joint$ with elements $\ele=(\elem, \elec)\in\mathbb{R}^n$. 
%Our coarse-graining operation removes the last $N<n$ elements from $\ele$, resulting in the marginal distribution $\marginal$ with elements $\elem \in\mathbb{R}^{n-N}$.  
Our coarse-graining operation removes $\elec$ from $\ele$, resulting in the marginal distribution $\marginal$ with elements $\elem \in\mathbb{R}^{N}$; we aim to create a training statement whose minimizer approximates this marginal. 
The corresponding conditional distribution of the eliminated variables $\elec \in\mathbb{R}^{n-N}$ is $\cond=\joint/\marginal$.

%Define our fine-grained (FG) system element to be vector $\ele$ of size $\nfg$. Our coarse-graining (CG) operation removes the last $(\nfg - \ncg)$ elements from $\ele$, resulting in a new vector of length $\ncg$. We assume the FG system is governed by non-zero probability density function $\joint$ and the CG system is governed by distribution $\marginal$. We assume that the CG system is ``ideal'', so it is a suitable marginal distribution of $\joint$ (this assumption goes by the name of ``thermodynamic consistency''). We define $\cond$ to be the conditional distribution of the variables typically integrated out (i.e., $\joint/\marginal$, e.g. what would govern the atomistic system if you were to hold the CG positions fixed). The elements of the gradient of the log of $\marginal$ can be expressed as conditional expectations of the gradient of the log of $\joint$ as follows. $\ind$ denotes the index of an entry of $\ele$ that is \emph{not} integrated out by the CG operation. $\geq\ncg$ refers to the indices that are being integrated out. $\int$ is used to refer to a multidimensional integral (so formally more than one $\int$ symbol). The LHS starts with the gradient of the log of $\marginal$; that is, the CG force in thermal units.

A valid force matching objective to learn forces for the marginal distribution $\marginal$, i.e., \cref{eq:fore-matching}, in this notation is given by
\begin{align}\label{eq:force-matching-app}
    \mathcal{L}(\theta) = \int \left\| \grad_{\elem}\log \joint(\ele) - \hat{F}_{\theta}(\elem) \right\|^2 \joint(x) d\ele.
\end{align}
From standard properties of least-squares regression (see, e.g., \citet{christensen2002plane}, Theorem 6.3.1.), we know that the optimal solution of \cref{eq:force-matching-app} is
\begin{equation}\label{eq:optimal-solution}
    \hat{F}_{\theta}^*(\elem) = \int [\grad_{\elem} \log \joint(\ele) ]\cond(\elec|\elem) \md\elec
\end{equation}

Let us now examine the gradient of $\log \marginal(\elem)$ with respect to the $i$th component of $x_s$:
\begin{align}
	\grad_\ind \log \marginal(\elem) 
	&=  \grad_\ind 
	\log \int \joint(\ele) \md \elec \\ 
	&= 
	\frac{\grad_\ind \int \joint(\ele)  \md \elec}
	{\int \joint(\ele) \md \elec} \\ 
	&= 
	\frac{\int \grad_\ind \joint(\ele) \md \elec}
	{\marginal(\elem)} \\
	&= 
	\frac{\int \left[ \grad_\ind \log \joint(\ele) \right ] \joint(\ele)  \md \elec}
	{\marginal(\elem)} \\
	&= 
	\int \left[ \grad_\ind \log \joint(\ele) \right ] \cond(\elec|\elem) d\elec
    \label{eq:derivation-joint}%\\
\end{align}

% \begin{align}
% 	\grad_\ind \log \marginal(\elm) 
% 	&=  \grad_\ind 
% 	\log \int \joint(\ele) \prod_{x_\secind|\secind\geq\ncg} \md \ele_\secind \\ 
% 	&= 
% 	\frac{\grad_\ind \int \joint(\ele) \prod_{x_\secind|\secind\geq\ncg} \md \ele_\secind}
% 	{\int \joint(\ele) \prod_{x_\secind|\secind\geq\ncg} \md \ele_\secind} \\ 
% 	&= 
% 	\frac{\int \grad_\ind \joint(\ele) \prod_{x_\secind|\secind\geq\ncg} \md \ele_\secind}
% 	{\marginal(\ele_{<\ncg})} \\
% 	&= 
% 	\frac{\int \left[ \grad_\ind \log \joint(\ele) \right ] \joint(\ele) \prod_{x_\secind|\secind\geq\ncg} \md \ele_\secind}
% 	{\marginal(\ele_{<\ncg})} \\
% 	&= 
% 	\int \left[ \grad_\ind \log \joint(\ele) \right ] \frac{\joint(\ele)}{\marginal(\ele_{<\ncg})} \prod_{x_\secind|\secind\geq\ncg} \md \ele_\secind
%     \label{eq:derivation-joint}%\\
% \end{align}

%The RHS is now a conditional expectation over the log gradients of the joint.
%(i.e., the atomistic forces).
Here is an explanation for the above steps:
\begin{itemize}
	\item
		Definition of marginal distribution
	\item
		Chain rule
	\item
		Leibniz integral rule
	\item
		Expand log derivative via chain rule, add factor of 1 
    \item 
        Reorganizing terms and definition of $\cond$
\end{itemize}
Since the RHS of \cref{eq:derivation-joint} coincides with that of \cref{eq:optimal-solution}, we have shown that the point-wise minimizer of \cref{eq:force-matching-app} is the log gradient of $\marginal$. 
Hence, this motivates to target the forces given by $\grad_{\elem} \log \joint(\elem,\elec)$ (again with force matching).

The concrete definition of $\joint$ depends on the CG approach utilized. In the classical CG force matching setting, the joint corresponds to the distribution of the atomistic system. \citet{durumeric2024learning} proposed to instead interpret the joint to be the product of the atomistic system's distribution with a kernel function. For more information on the former setting, which underpins the results in this manuscript focused on atomistic forces, we refer readers to \citet{ciccotti2005blue,noid2008multiscale,kramer2023statistically}. The latter setting is central to kernel forces and is described by
\begin{equation}
    \joint(\ele) = p(\elec)\kappa(\elem|\elec),
\end{equation}
where $p$ is either the atomistic distribution or pre-kernel coarse-grained system. 
As $p(\elec)$ is independent of $\ele_\ind$, we have
\begin{equation}\label{eq:joint-with-kernel}
    \grad_\ind \log \joint(\elem,\elec) = \grad_\ind \log \kappa(\elem|\elec).
\end{equation}
%\begin{equation}
%    	\grad_\ind \log \marginal(\elem) 
%	=\int \left[ \grad_\ind \log \kappa(\elem|\elec) \right ] \cond(\elec|\elem) d\elec. \label{eq:derivation-kernel}
%\end{equation}
In the setting of noise forces, the kernel $\kappa(\elem|\elec)$ is the noise kernel $\kappa_N(R'|R)$ and $p(\elec)$ is the CG data distribution $p(R)$. By substituting \cref{eq:joint-with-kernel} into \cref{eq:force-matching-app} , we directly recover the noise kernel force matching objective (\cref{eq:noise-force-matching}).

Similarly, for our introduced reverse noise kernel, we obtain the force matching objective (\cref{eq:reversenoise-force-matching}), as the kernel $\kappa(\elem|\elec)$ is the reverse noise kernel $\kappa_G(R|R')$ and $p(\elec)$ is the noised CG data distribution $\hat{p}(R')$.

% \Alek{1. integration over the joint is the angle brackets combined with the $E_\kappa$. 2. you have justified the definition of $F_N$ and your other general kernel $F$. You should write this explicitly here}

%\Leon{\citep{noid2008multiscale, ciccotti2005blue}} no bonds constrains.

\section{Additional results}\label{app:additional-experiments}
Here, we present additional results for the experiments described in \cref{sec:results}.

\subsection{Smaller noise levels}\label{app:smaller-noise-levels}
For Trp‐cage some simulations for the Noise (0.05) models diverged when trained on $10\%$ or $2\%$ of the data. All unstable trajectories were excluded from our analysis. For the remaining systems and noise levels reported in \cref{sec:results}, all simulations remained stable. 

We also investigated noise variances below those reported in \cref{sec:results}.
Decreasing the noise level excessively impairs global‐feature fidelity and potentially destabilizes the simulations. Additional results for alanine dipeptide and Chignolin at lower noise variances are presented in \cref{tab:noise001}, with any divergent runs similarly filtered out. 

\begin{table}
\centering
\begin{tabular}{lccc}
\toprule
\textbf{Data size}&  $\mathbf{100\%}$ & $\mathbf{10\%}$& $\mathbf{2\%}$\\
\midrule
\textbf{Model}      
& \multicolumn{3}{c}{\textbf{PMF RMS} $(\downarrow)$}  \\
\midrule
%\cmidrule{2-3}
& \multicolumn{3}{c}{Alanine dipeptide} \\
\cmidrule{2-4}
Noise (0.01)    & $0.23 \pm 0.09$ & $0.30 \pm 0.10$&$11.85 \pm 0.99$ \\
\cmidrule{2-4}
& \multicolumn{3}{c}{Chignolin} \\
\cmidrule{2-4}
Noise (0.01)    & $4.26 \pm 1.27$ & $19.19 \pm 1.06$& - \\
\bottomrule
\end{tabular}
\caption{PMF-RMS errors for the Noise (0.01) model for different systems and their respective training-set fractions. Any simulations that diverged were excluded, which explains the absence of an entry for Chignolin at the $2\%$ split. Errors represent the standard deviation across three CGSchNet models. Each CGSchNet model was trained on a different set of forces generated with the noise kernel.}
\label{tab:noise001}
\end{table}

\subsection{Additional plots}\label{app:additional-plots}
\begin{figure*}[t]
\centering
\includegraphics[width=0.9\textwidth]{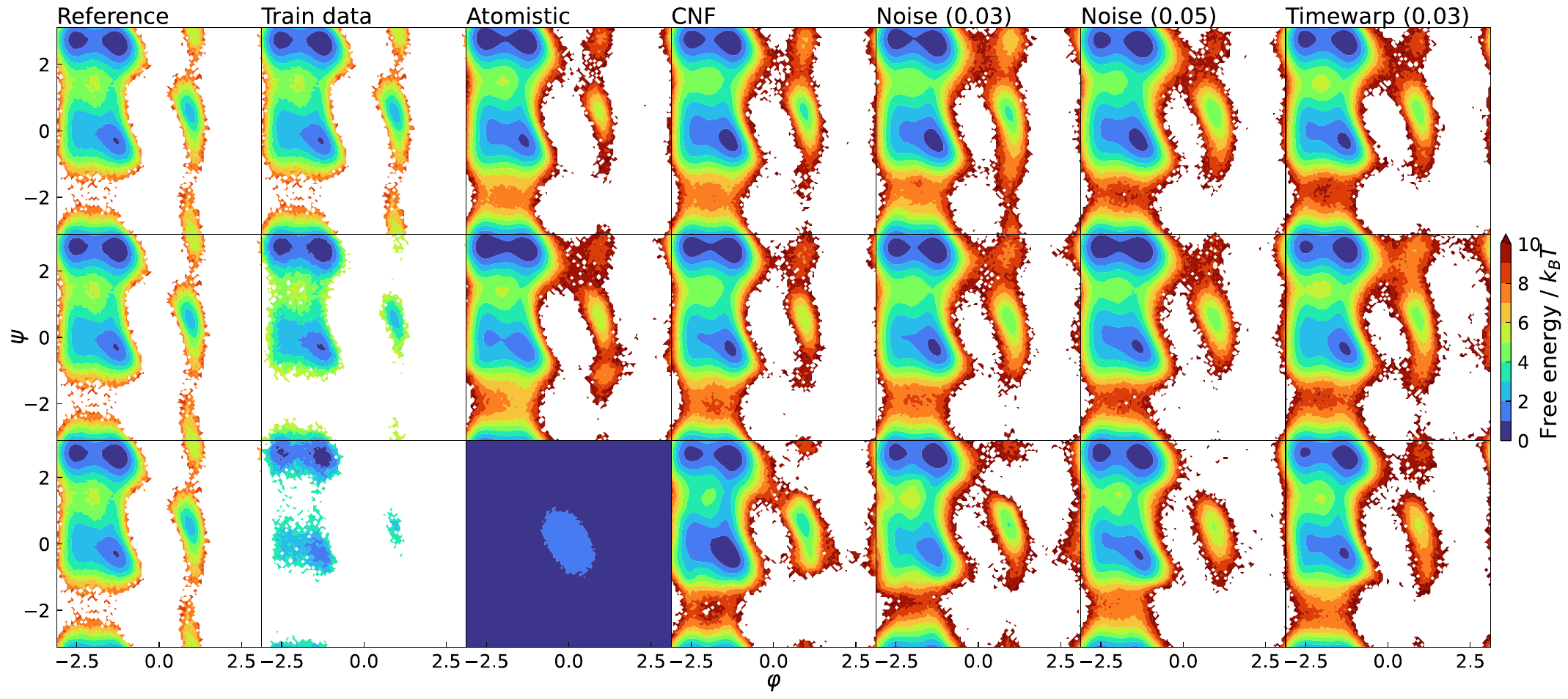}
\caption{Ramachandran plots for alanine dipeptide at varying training‐set sizes: top row $100\%$ of the data; middle row $10\%$; bottom row $2\%$. Note that all Atomistic model simulations diverged at the $2\%$ data level.
}
\label{fig:ala-ramachandran}
\end{figure*}

\begin{figure*}[t]
\centering
\includegraphics[width=0.9\textwidth]{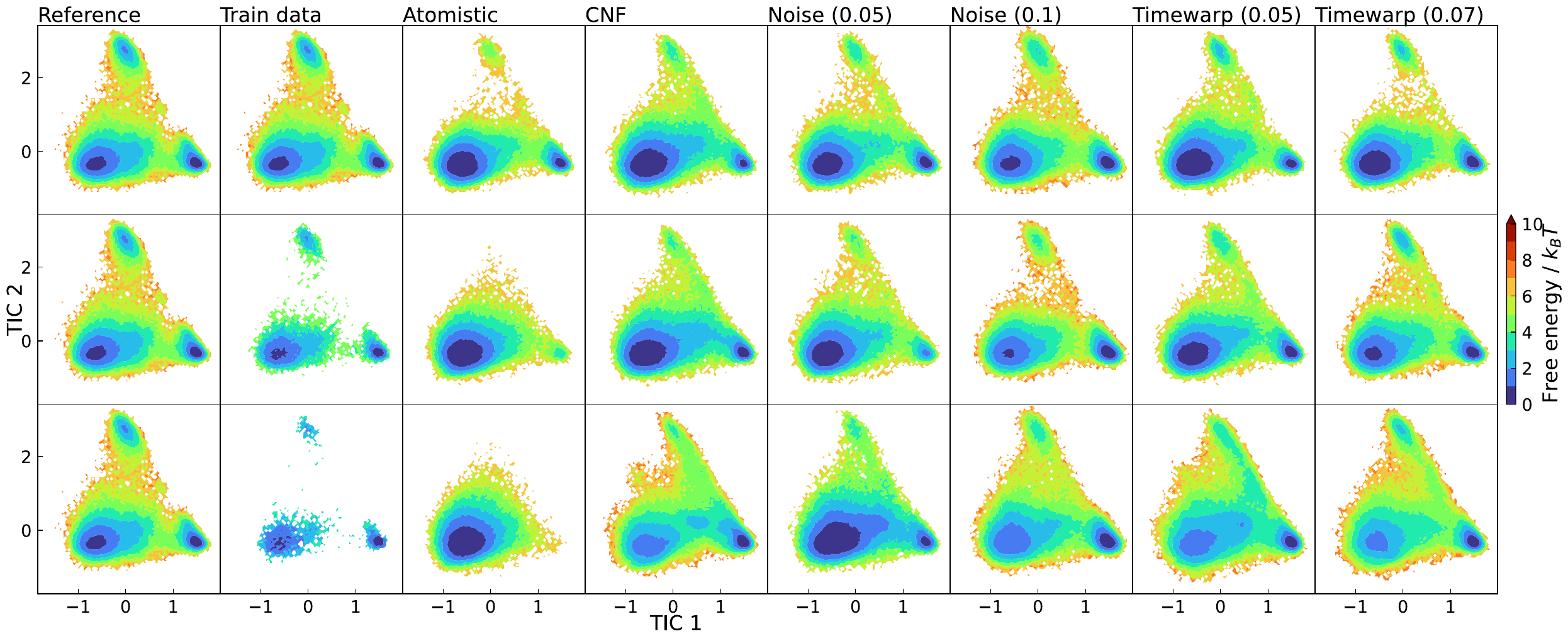}
\caption{TICA plots for Chignolin at varying training‐set sizes: top row $100\%$ of the data; middle row $10\%$; bottom row $2\%$.
}
\label{fig:cln_tica}
\end{figure*}

\begin{figure*}[t]
\centering
\includegraphics[width=0.9\textwidth]{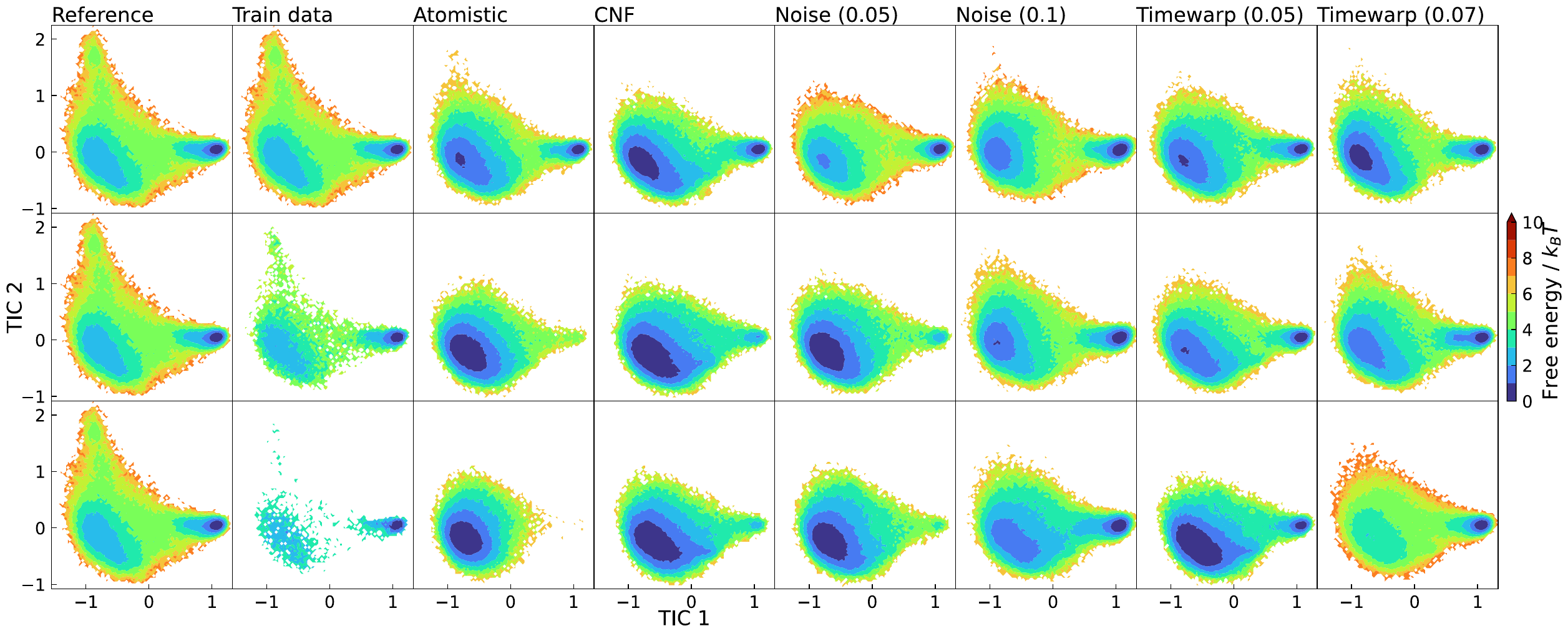}
\caption{TICA plots for Trp-cage at varying training‐set sizes: top row $100\%$ of the data; middle row $10\%$; bottom row $2\%$.
}
\label{fig:trp_tica}
\end{figure*}

We visualize the two‐dimensional free‐energy landscapes of each method for alanine dipeptide, Chignolin, and Trp‐cage in Figures \cref{fig:ala-ramachandran}, \cref{fig:cln_tica}, and \cref{fig:trp_tica}, respectively.

\subsection{Dipeptides}\label{app:additional-dipeptides}
\begin{figure}[t]
\centering
\includegraphics[width=\columnwidth]{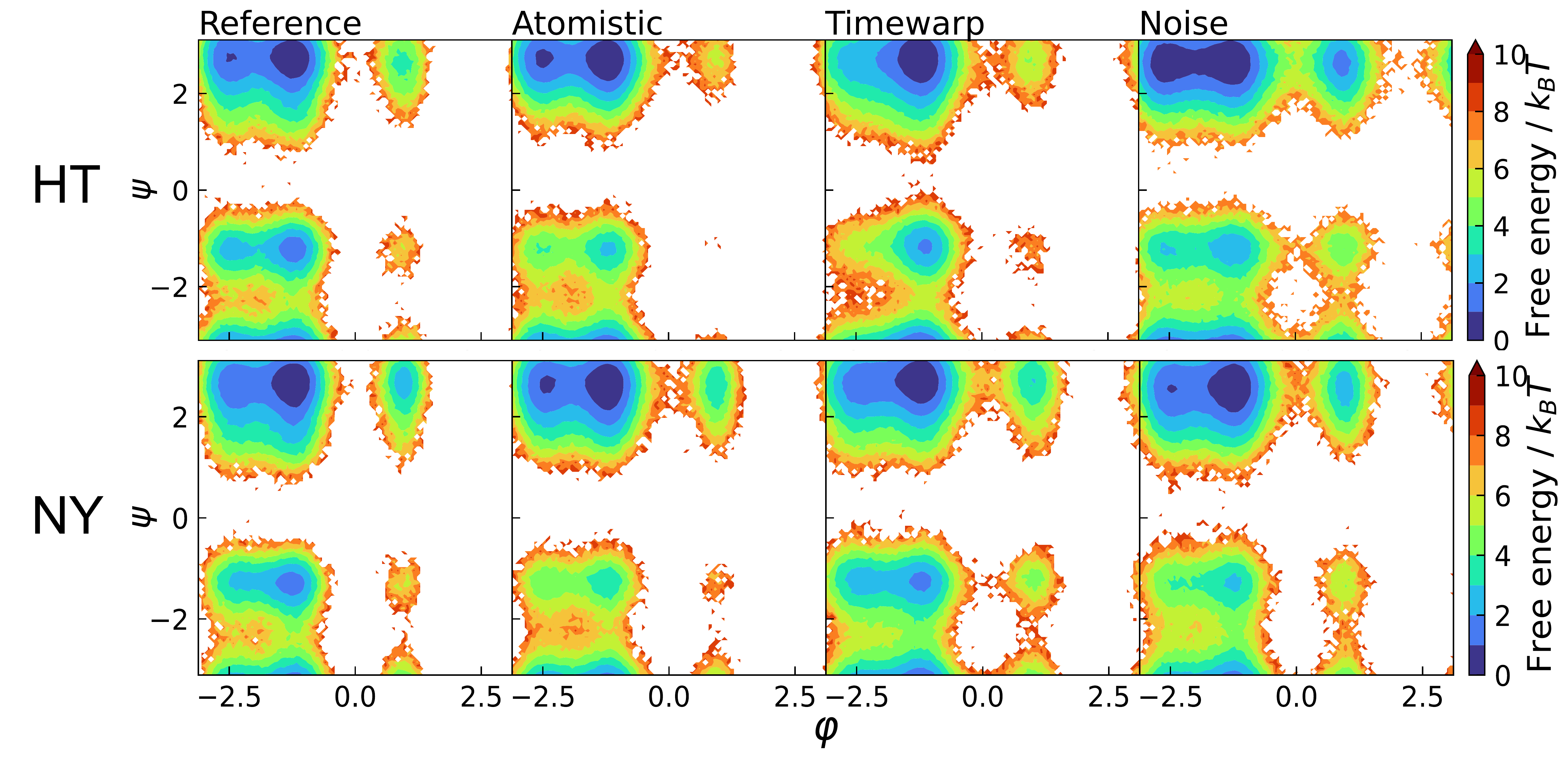}
\caption{Ramachandran plots for the HT and NY dipeptides, samples generated with different models. The pretrained Timewarp model is used to obtain forces via the transition kernel.}
\label{fig:dipeptides_2}
\end{figure}

In \cref{tab:dipeptides-detailed}, we report the individual dipeptide performance metrics for the experiments described in \cref{sec:dipeptides}.  Additional Ramachandran plots for two further dipeptides are shown in \cref{fig:dipeptides_2}. CG forces for the Timewarp model are obtained by evaluating the score of the pretrained Timewarp transition kernel (see \cref{sec:dipeptides}).

\begin{table}[ht]
\centering
\begin{tabular}{lcc}
\toprule
\textbf{Model}      
& \textbf{PMF RMS} $(\downarrow)$ & \textbf{Bond distance} $(\downarrow)$ \\
\midrule
%\cmidrule{2-3}
& \multicolumn{2}{c}{AV dipeptide} \\
\cmidrule{2-3}
Atomistic & $0.198 \pm 0.055$ & $0.0071 \pm 0.0002$ \\
Timewarp  & $0.255 \pm 0.082$ & $0.0122 \pm 0.0004$ \\
Noise     & $0.494 \pm 0.135$ & $0.0307 \pm 0.0010$ \\

\cmidrule{2-3}
& \multicolumn{2}{c}{HT dipeptide} \\
\cmidrule{2-3}
Atomistic & $0.265 \pm 0.014$ & $0.0050 \pm 0.0005$ \\
Timewarp  & $0.402 \pm 0.003$ & $0.0101 \pm 0.0002$ \\
Noise     & $0.446 \pm 0.022$ & $0.0313 \pm 0.0009$ \\

\cmidrule{2-3}
& \multicolumn{2}{c}{KS dipeptide} \\
\cmidrule{2-3}
Atomistic & $0.269 \pm 0.083$ & $0.0036 \pm 0.0003$ \\
Timewarp  & $0.556 \pm 0.030$ & $0.0114 \pm 0.0009$ \\
Noise     & $0.749 \pm 0.164$ & $0.0304 \pm 0.0003$ \\

\cmidrule{2-3}
& \multicolumn{2}{c}{NY dipeptide} \\
\cmidrule{2-3}
Atomistic & $0.410 \pm 0.125$ & $0.0036 \pm 0.0002$ \\
Timewarp  & $0.279 \pm 0.128$ & $0.0109 \pm 0.0005$ \\
Noise     & $0.348 \pm 0.138$ & $0.0297 \pm 0.0001$ \\
\bottomrule
\end{tabular}
\caption{Detailed dipeptide results, averaged over three models each.}
\label{tab:dipeptides-detailed}
\end{table}

%\section{Dataset details}\label{app:dataset-details}
%\Yaoyi{To my understanding, there is no new MD data dedicated to this work, since all datasets have been introduced in previous publications. I already added corresponding citations in the Results section.
% The source data can be accessed by contacting the authors in Frank \& Cecilia's groups.
% Therefore, I don't think there is a need to go to further detail about the MD setup, etc.
% Nonetheless, let me know if you would prefer a summary here about the simulation setup and some other details, like the MSM hyperparameters, CG mapping, etc, or the data accessibility.
% }
%\Leon{perfect}

\section{Implementation details and hyperparameters}\label{app:implementatoin}
\subsection{Continuous normalizing flow model}\label{app:CNF}
In this section, we provide a more detailed description of the continuous normalizing flow (CNF) model. For further details refer to \citet{klein2023equivariant} and \citet{klein2024tbg}.
% \subsubsection{Architecture}
% For our conditional CNF model we use the architecture proposed in \citep{klein2023equivariant, klein2024tbg}, which uses an EGNN \citep{satorras2021graph}as the equivariant vector field. Following the notation of \citet{klein2023equivariant, klein2024tbg}, the graph neural network consists of L
% consecutive layers and the update for the i-th particle is computed as follows
% \begin{align}
%     h_i^0 &= (t, a_i, b_i, c_i), \quad m_{ij}^l=\phi_e\left(h_i^l, h_j^l, d_{ij}^2 \right),\\
%     x_i^{l+1}&=x_i^l+\sum_{j\neq i} \frac{\left(x_i^l-x_j^l \right)}{d_{ij}+1}\phi_d(m_{ij}^l),\\
%     h_i^{l+1}&=\phi_h\left(h_i^l, m_i^l\right), \quad  m_i^l=\sum_{j\neq i} \phi_m(m_{ij}^l) m_{ij}^l,\\
%     v_{\theta}(t,x^0)_i &= x_i^L - x_i^0 - \frac{1}{N}\sum_{j}^N (x_j^L - x_j^0),
% \end{align}
\subsubsection{Training}
To train the conditional CNF model, we employ flow matching \citep{lipman2022flow, albergo2023stochastic, liu2022flow}. We here follow the notation of \citet{klein2023equivariant, klein2024tbg}.
Flow matching enables efficient, simulation-free training of the CNFs. The flow matching training objective allows for the direct training of our conditional vector field $v_{\theta}(x_t, t)$ through
\begin{equation}\label{eq:conditional-fm-loss}
    \mathcal{L}_{\mathrm{CFM}}(\theta)=\mathbb{E}_{t\sim[0,1], x\sim p_t(x|z)}\left|\left|v_{\theta}(x_t,t|R')- u_t(x|z)\right|\right|_2^2. 
\end{equation}
There are many possible parametrizations for the target vector field $u_t(x|z)$ and the target probability path $p_t(x | z)$. We here use the parametrization as introduced in \citet{lipman2022flow} and add our conditioning
\begin{align}
z &= (x_0, x_1) \\p(z) &= q(x_0)p(x_1) \\
u_t(x | z) &= x_1 - x_0 \\ p_t(x | z) &= \mathcal{N}(x | t \cdot x_1 + (1 - t) \cdot x_0, \sigma^2)\\
R' &= R - x_1.
\end{align}
This formulation ensures that the model learns to map base noise $x_0$ to the correction $x_1$ that reverses the corruption of $R$.

\subsubsection{Hyperparameter}
We report the CNF model hyperparameters for the different datasets in \cref{tab:architecture-papameters}. As in \citet{klein2023equivariant, klein2024tbg} all neural networks $\phi_{\alpha}$ have one hidden layer with $n_{\mathrm{hidden}}$ neurons and \textit{SiLU} activation functions. The input size of the embedding $n_{\mathrm{embedding}}$ depends on the system size, as we use a different embedding for each bead. 

\begin{table}
  \caption{Continuous normalizing flow - hyperparameters}
  \label{tab:architecture-papameters}
  \centering
  \begin{tabular}{lccc}
    \toprule
    \textbf{Dataset} & $L$ & $n_{\mathrm{hidden}}$ &  $n_{\mathrm{embedding}}$ \\
    \midrule
    Alanine dipeptide &$5$&$64$&$6$\\
    Chignolin  &$5$&$64$&$10$\\
    Trp-cage  &$7$&$128$&$20$\\
    \bottomrule
  \end{tabular}
\end{table}

\subsection{Coupling flow model}
We adhere closely to the architecture of \citet{klein2023timewarp}, with one key alteration for the denoising task: rather than predicting a time step, the model is trained to predict the denoising kernel. For further details, see \citet{klein2023timewarp}.

\subsubsection{Hyperparameter}
We follow the hyperparameter conventions of \citet{klein2023timewarp} and here report only the settings adjusted for our retrained Timewarp models used in the reverse noise kernel experiments. The original, unmodified Timewarp hyperparameters are retained for the transition kernel experiments and are omitted for brevity.

Depending on the dataset, we employ different Timewarp configurations. For alanine dipeptide the multihead kernel self‐attention layer uses six heads with length‐scales $\ell_i=\{0.5, 1.0, 1.5, 2.0, 3.0, 5.0\}$nm, given in nanometers. For the small proteins Chignolin and Trp‐cage, we instead use $\ell_i=\{0.5, 1, 2, 4, 8, 16\}$nm.
We summarize the number of RealNVP layers $n_{\text{RealNVP}}$, number of Transformer layers $n_{\text{T}}$, and the atom embedding dimension $n_{\text{emb}}$ for the different datasets in \cref{tab:timewarp-hyperparameters}. All other parameters are the same as in \citet{klein2023timewarp}.

\begin{table}
  \caption{Timewarp  hyperparameters}
  \label{tab:timewarp-hyperparameters}
  \centering
  %\begin{ruledtabular}
  \begin{tabular}{lccc}
    \toprule
    \textbf{Dataset}       &    $n_{\text{RealNVP}}$ & $n_{\text{T}}$ & $n_{\text{emb}}$\\
    \midrule
   Alanine dipeptide  &   8 & 3 & 32 \\ 
   Chignolin          &   8 & 3 & 32\\ 
    Trp-cage          &   24 & 4 & 32\\ 
    \bottomrule
  \end{tabular}
  %\end{ruledtabular}
\end{table}

%\Leon{todo}
%The $\phi_{\textrm{in}}$ and $\phi_{\textrm{out}}$ MLPs use SiLUs as activation functions, while the Transformer MLPs use ReLUs. The shapes of these MLPs vary for the different datasets as shown in \cref{tab:mlp-shapes}.
%The first linear layers in the kernel self-attention module always has the shape $[128, 768]$, and the second (after concatenating the output of head head) has the shape $[768, 128]$. The transformer feature dimension $D$ is for all datasets $128$.
% \begin{table}[!htbp]
%   \caption{Timewarp MLP layer sizes \Leon{todo}}
%   \label{tab:mlp-shapes}
%   \centering
%   \begin{tabular}{lccc}
%     \toprule
%     Dataset       & $\phi_{\textrm{in}}$ MLP& $\phi_{\textrm{out}}$ MLP&Transformer MLP\\
%     \midrule
%    Alanine dipeptide   &$[70, 256, 128]$&$[128,256, 3]$&$[128, 256, 128]$\\ 
%    2AA          &$[70, 256, 128]$&$[128,256, 3]$&$[128, 256, 128]$\\ 
%     \bottomrule
%   \end{tabular}
% \end{table}

\subsection{CGSchNet}
As mentioned previously in the main text, we use the architecture as described in \citet{husic2020coarse}. 
%The prior energy term $\hat{U}_{\text{prior}}(R)$ for this work contains the \Yaoyi{List the hyperparameters for each model, and how they were fitted!}
%\Yaoyi{Text and table below directly copied from the optimized force map paper. Paraphrasing required!}
%The CG force field was defined as a sum of a prior model and a modified SchNet GNN\cite{schutt2018schnet}. The prior model was a Hamiltonian that restrained all sequential $\calpha$-$\calpha$ pseudobonds and $\calpha$-$\calpha$-$\calpha$ pseudoangles using harmonic interactions parameterized through Boltzmann inversion of the all-atom data, as well as  sequential $\calpha$ quadruplet pseudodihedrals via a fifth degree sine/cosine expansion and non-bonded power 6 repulsions. The power 6 nonbonded terms were parametrized by residue-type dependent minimum observed distances in the all-atom dataset. These nonbonded interactions were only applied to CG sites which were not involved in bonds or angle terms together. We note that while our choices of hyperparameters were informed by previous work\cite{husic2020coarse}, hyperparameters were not scanned over for this publication; furthermore, the hyperparameters used for Trp Cage were borrowed from those used for CLN025 without modification or experimentation.
The employed hyperparameters for CGSchNet models in this work are shown in table \cref{tab:networkhypers}, following the naming conventions of \citet{durumeric2024learning}. All models use the ExpNormal \citep{unke2019physnet,tholke2022equivariant} as the radial basis function and a cosine cutoff \citep{schutt2018schnetpack,tholke2022equivariant} for the filers. 

\begin{center}
    \begin{table*}
        \centering
        \caption{CGSchNet hyperparameters}
        \label{tab:networkhypers}
        \begin{tabular}{lcc} 
            \toprule
            & \multicolumn{2}{c}{\textbf{Dataset}} \\
            \cmidrule{2-3}
            \textbf{Hyperparameter} & Alanine dipeptide/Dipeptides & Chignolin/Trp-cage\\ 
            \midrule
            Embedding Strategy & Unique per bead & Amino acid type (unique termini)\\
            Activation Function & Tanh&Tanh \\ 
            Distance Cutoff & 0 to 15 \AA{}& 0 to 30 \AA{}\\
            Num Radial Basis Functions & 20  & 128 \\
            Num Filters & 128  & 128\\
            Interaction Blocks & 2  &2\\
            Terminal Network Layer Widths & [128] &[128, 64]\\
            \bottomrule
        \end{tabular}
    \end{table*}
\end{center}

%\subsection{Evaluation metric}\label{app:evaluation-metrics}
%\Leon{discuss the details a bit and show potential alternatives, which penalize outliers more}.
%Can have large error bars. 

\newpage
% Create the reference section using BibTeX:
\bibliography{literature}

%merlin.mbs aipnum4-1.bst 2010-07-25 4.21a (PWD, AO, DPC) hacked
%Control: key (0)
%Control: author (8) initials jnrlst
%Control: editor formatted (1) identically to author
%Control: production of article title (0) allowed
%Control: page (1) range
%Control: year (1) truncated
%Control: production of eprint (0) enabled
\begin{thebibliography}{98}%
\makeatletter
\providecommand \@ifxundefined [1]{%
 \@ifx{#1\undefined}
}%
\providecommand \@ifnum [1]{%
 \ifnum #1\expandafter \@firstoftwo
 \else \expandafter \@secondoftwo
 \fi
}%
\providecommand \@ifx [1]{%
 \ifx #1\expandafter \@firstoftwo
 \else \expandafter \@secondoftwo
 \fi
}%
\providecommand \natexlab [1]{#1}%
\providecommand \enquote  [1]{``#1''}%
\providecommand \bibnamefont  [1]{#1}%
\providecommand \bibfnamefont [1]{#1}%
\providecommand \citenamefont [1]{#1}%
\providecommand \href@noop [0]{\@secondoftwo}%
\providecommand \href [0]{\begingroup \@sanitize@url \@href}%
\providecommand \@href[1]{\@@startlink{#1}\@@href}%
\providecommand \@@href[1]{\endgroup#1\@@endlink}%
\providecommand \@sanitize@url [0]{\catcode `\\12\catcode `\$12\catcode `\&12\catcode `\#12\catcode `\^12\catcode `\_12\catcode `\%12\relax}%
\providecommand \@@startlink[1]{}%
\providecommand \@@endlink[0]{}%
\providecommand \url  [0]{\begingroup\@sanitize@url \@url }%
\providecommand \@url [1]{\endgroup\@href {#1}{\urlprefix }}%
\providecommand \urlprefix  [0]{URL }%
\providecommand \Eprint [0]{\href }%
\providecommand \doibase [0]{http://dx.doi.org/}%
\providecommand \selectlanguage [0]{\@gobble}%
\providecommand \bibinfo  [0]{\@secondoftwo}%
\providecommand \bibfield  [0]{\@secondoftwo}%
\providecommand \translation [1]{[#1]}%
\providecommand \BibitemOpen [0]{}%
\providecommand \bibitemStop [0]{}%
\providecommand \bibitemNoStop [0]{.\EOS\space}%
\providecommand \EOS [0]{\spacefactor3000\relax}%
\providecommand \BibitemShut  [1]{\csname bibitem#1\endcsname}%
\let\auto@bib@innerbib\@empty
%</preamble>
\bibitem [{\citenamefont {Adcock}\ and\ \citenamefont {McCammon}(2006)}]{adcock2006molecular}%
  \BibitemOpen
  \bibfield  {author} {\bibinfo {author} {\bibfnamefont {S.~A.}\ \bibnamefont {Adcock}}\ and\ \bibinfo {author} {\bibfnamefont {J.~A.}\ \bibnamefont {McCammon}},\ }\bibfield  {title} {\enquote {\bibinfo {title} {Molecular dynamics: survey of methods for simulating the activity of proteins},}\ }\href@noop {} {\bibfield  {journal} {\bibinfo  {journal} {Chem. Rev.}\ }\textbf {\bibinfo {volume} {106}},\ \bibinfo {pages} {1589--1615} (\bibinfo {year} {2006})}\BibitemShut {NoStop}%
\bibitem [{\citenamefont {Hospital}\ \emph {et~al.}(2015)\citenamefont {Hospital}, \citenamefont {Go{\~n}i}, \citenamefont {Orozco},\ and\ \citenamefont {Gelp{\'\i}}}]{hospital2015molecular}%
  \BibitemOpen
  \bibfield  {author} {\bibinfo {author} {\bibfnamefont {A.}~\bibnamefont {Hospital}}, \bibinfo {author} {\bibfnamefont {J.~R.}\ \bibnamefont {Go{\~n}i}}, \bibinfo {author} {\bibfnamefont {M.}~\bibnamefont {Orozco}}, \ and\ \bibinfo {author} {\bibfnamefont {J.~L.}\ \bibnamefont {Gelp{\'\i}}},\ }\bibfield  {title} {\enquote {\bibinfo {title} {Molecular dynamics simulations: advances and applications},}\ }\href@noop {} {\bibfield  {journal} {\bibinfo  {journal} {Adv. Appl. Bioinform. Chem.}\ ,\ \bibinfo {pages} {37--47}} (\bibinfo {year} {2015})}\BibitemShut {NoStop}%
\bibitem [{\citenamefont {Hollingsworth}\ and\ \citenamefont {Dror}(2018)}]{hollingsworth2018molecular}%
  \BibitemOpen
  \bibfield  {author} {\bibinfo {author} {\bibfnamefont {S.~A.}\ \bibnamefont {Hollingsworth}}\ and\ \bibinfo {author} {\bibfnamefont {R.~O.}\ \bibnamefont {Dror}},\ }\bibfield  {title} {\enquote {\bibinfo {title} {Molecular dynamics simulation for all},}\ }\href@noop {} {\bibfield  {journal} {\bibinfo  {journal} {Neuron}\ }\textbf {\bibinfo {volume} {99}},\ \bibinfo {pages} {1129--1143} (\bibinfo {year} {2018})}\BibitemShut {NoStop}%
\bibitem [{\citenamefont {Voelz}\ \emph {et~al.}(2010)\citenamefont {Voelz}, \citenamefont {Bowman}, \citenamefont {Beauchamp},\ and\ \citenamefont {Pande}}]{voelz2010molecular}%
  \BibitemOpen
  \bibfield  {author} {\bibinfo {author} {\bibfnamefont {V.~A.}\ \bibnamefont {Voelz}}, \bibinfo {author} {\bibfnamefont {G.~R.}\ \bibnamefont {Bowman}}, \bibinfo {author} {\bibfnamefont {K.}~\bibnamefont {Beauchamp}}, \ and\ \bibinfo {author} {\bibfnamefont {V.~S.}\ \bibnamefont {Pande}},\ }\bibfield  {title} {\enquote {\bibinfo {title} {Molecular simulation of ab initio protein folding for a millisecond folder ntl9 (1- 39)},}\ }\href@noop {} {\bibfield  {journal} {\bibinfo  {journal} {J. Am. Chem. Soc.}\ }\textbf {\bibinfo {volume} {132}},\ \bibinfo {pages} {1526--1528} (\bibinfo {year} {2010})}\BibitemShut {NoStop}%
\bibitem [{\citenamefont {Lindorff-Larsen}\ \emph {et~al.}(2011)\citenamefont {Lindorff-Larsen}, \citenamefont {Piana}, \citenamefont {Dror},\ and\ \citenamefont {Shaw}}]{lindorff2011fast}%
  \BibitemOpen
  \bibfield  {author} {\bibinfo {author} {\bibfnamefont {K.}~\bibnamefont {Lindorff-Larsen}}, \bibinfo {author} {\bibfnamefont {S.}~\bibnamefont {Piana}}, \bibinfo {author} {\bibfnamefont {R.~O.}\ \bibnamefont {Dror}}, \ and\ \bibinfo {author} {\bibfnamefont {D.~E.}\ \bibnamefont {Shaw}},\ }\bibfield  {title} {\enquote {\bibinfo {title} {How fast-folding proteins fold},}\ }\href@noop {} {\bibfield  {journal} {\bibinfo  {journal} {Science}\ }\textbf {\bibinfo {volume} {334}},\ \bibinfo {pages} {517--520} (\bibinfo {year} {2011})}\BibitemShut {NoStop}%
\bibitem [{\citenamefont {Plattner}\ \emph {et~al.}(2017)\citenamefont {Plattner}, \citenamefont {Doerr}, \citenamefont {De~Fabritiis},\ and\ \citenamefont {No{\'e}}}]{plattner2017complete}%
  \BibitemOpen
  \bibfield  {author} {\bibinfo {author} {\bibfnamefont {N.}~\bibnamefont {Plattner}}, \bibinfo {author} {\bibfnamefont {S.}~\bibnamefont {Doerr}}, \bibinfo {author} {\bibfnamefont {G.}~\bibnamefont {De~Fabritiis}}, \ and\ \bibinfo {author} {\bibfnamefont {F.}~\bibnamefont {No{\'e}}},\ }\bibfield  {title} {\enquote {\bibinfo {title} {Complete protein--protein association kinetics in atomic detail revealed by molecular dynamics simulations and {M}arkov modelling},}\ }\href@noop {} {\bibfield  {journal} {\bibinfo  {journal} {Nature chemistry}\ }\textbf {\bibinfo {volume} {9}},\ \bibinfo {pages} {1005--1011} (\bibinfo {year} {2017})}\BibitemShut {NoStop}%
\bibitem [{\citenamefont {Robustelli}\ \emph {et~al.}(2022)\citenamefont {Robustelli}, \citenamefont {Ibanez-de Opakua}, \citenamefont {Campbell-Bezat}, \citenamefont {Giordanetto}, \citenamefont {Becker}, \citenamefont {Zweckstetter}, \citenamefont {Pan},\ and\ \citenamefont {Shaw}}]{robustelli2022molecular}%
  \BibitemOpen
  \bibfield  {author} {\bibinfo {author} {\bibfnamefont {P.}~\bibnamefont {Robustelli}}, \bibinfo {author} {\bibfnamefont {A.}~\bibnamefont {Ibanez-de Opakua}}, \bibinfo {author} {\bibfnamefont {C.}~\bibnamefont {Campbell-Bezat}}, \bibinfo {author} {\bibfnamefont {F.}~\bibnamefont {Giordanetto}}, \bibinfo {author} {\bibfnamefont {S.}~\bibnamefont {Becker}}, \bibinfo {author} {\bibfnamefont {M.}~\bibnamefont {Zweckstetter}}, \bibinfo {author} {\bibfnamefont {A.~C.}\ \bibnamefont {Pan}}, \ and\ \bibinfo {author} {\bibfnamefont {D.~E.}\ \bibnamefont {Shaw}},\ }\bibfield  {title} {\enquote {\bibinfo {title} {Molecular basis of small-molecule binding to $\alpha$-synuclein},}\ }\href@noop {} {\bibfield  {journal} {\bibinfo  {journal} {J. Am. Chem. Soc.}\ }\textbf {\bibinfo {volume} {144}},\ \bibinfo {pages} {2501--2510} (\bibinfo {year} {2022})}\BibitemShut {NoStop}%
\bibitem [{\citenamefont {Clementi}(2008)}]{clementi2008coarse}%
  \BibitemOpen
  \bibfield  {author} {\bibinfo {author} {\bibfnamefont {C.}~\bibnamefont {Clementi}},\ }\bibfield  {title} {\enquote {\bibinfo {title} {Coarse-grained models of protein folding: toy models or predictive tools?}}\ }\href@noop {} {\bibfield  {journal} {\bibinfo  {journal} {Current opinion in structural biology}\ }\textbf {\bibinfo {volume} {18}},\ \bibinfo {pages} {10--15} (\bibinfo {year} {2008})}\BibitemShut {NoStop}%
\bibitem [{\citenamefont {Noid}(2013)}]{noid2013perspective}%
  \BibitemOpen
  \bibfield  {author} {\bibinfo {author} {\bibfnamefont {W.~G.}\ \bibnamefont {Noid}},\ }\bibfield  {title} {\enquote {\bibinfo {title} {Perspective: Coarse-grained models for biomolecular systems},}\ }\href@noop {} {\bibfield  {journal} {\bibinfo  {journal} {The Journal of chemical physics}\ }\textbf {\bibinfo {volume} {139}},\ \bibinfo {pages} {09B201\_1} (\bibinfo {year} {2013})}\BibitemShut {NoStop}%
\bibitem [{\citenamefont {Jin}\ \emph {et~al.}(2022)\citenamefont {Jin}, \citenamefont {Pak}, \citenamefont {Durumeric}, \citenamefont {Loose},\ and\ \citenamefont {Voth}}]{jin2022bottom}%
  \BibitemOpen
  \bibfield  {author} {\bibinfo {author} {\bibfnamefont {J.}~\bibnamefont {Jin}}, \bibinfo {author} {\bibfnamefont {A.~J.}\ \bibnamefont {Pak}}, \bibinfo {author} {\bibfnamefont {A.~E.}\ \bibnamefont {Durumeric}}, \bibinfo {author} {\bibfnamefont {T.~D.}\ \bibnamefont {Loose}}, \ and\ \bibinfo {author} {\bibfnamefont {G.~A.}\ \bibnamefont {Voth}},\ }\bibfield  {title} {\enquote {\bibinfo {title} {Bottom-up coarse-graining: Principles and perspectives},}\ }\href@noop {} {\bibfield  {journal} {\bibinfo  {journal} {J. Chem. Theory Comput.}\ }\textbf {\bibinfo {volume} {18}},\ \bibinfo {pages} {5759--5791} (\bibinfo {year} {2022})}\BibitemShut {NoStop}%
\bibitem [{\citenamefont {Noid}(2023)}]{noid2023perspective}%
  \BibitemOpen
  \bibfield  {author} {\bibinfo {author} {\bibfnamefont {W.~G.}\ \bibnamefont {Noid}},\ }\bibfield  {title} {\enquote {\bibinfo {title} {Perspective: Advances, challenges, and insight for predictive coarse-grained models},}\ }\href@noop {} {\bibfield  {journal} {\bibinfo  {journal} {J. Phys. Chem. B}\ }\textbf {\bibinfo {volume} {127}},\ \bibinfo {pages} {4174--4207} (\bibinfo {year} {2023})}\BibitemShut {NoStop}%
\bibitem [{\citenamefont {Borges-Ara{\'u}jo}\ \emph {et~al.}(2023)\citenamefont {Borges-Ara{\'u}jo}, \citenamefont {Patmanidis}, \citenamefont {Singh}, \citenamefont {Santos}, \citenamefont {Sieradzan}, \citenamefont {Vanni}, \citenamefont {Czaplewski}, \citenamefont {Pantano}, \citenamefont {Shinoda}, \citenamefont {Monticelli} \emph {et~al.}}]{borges2023pragmatic}%
  \BibitemOpen
  \bibfield  {author} {\bibinfo {author} {\bibfnamefont {L.}~\bibnamefont {Borges-Ara{\'u}jo}}, \bibinfo {author} {\bibfnamefont {I.}~\bibnamefont {Patmanidis}}, \bibinfo {author} {\bibfnamefont {A.~P.}\ \bibnamefont {Singh}}, \bibinfo {author} {\bibfnamefont {L.~H.}\ \bibnamefont {Santos}}, \bibinfo {author} {\bibfnamefont {A.~K.}\ \bibnamefont {Sieradzan}}, \bibinfo {author} {\bibfnamefont {S.}~\bibnamefont {Vanni}}, \bibinfo {author} {\bibfnamefont {C.}~\bibnamefont {Czaplewski}}, \bibinfo {author} {\bibfnamefont {S.}~\bibnamefont {Pantano}}, \bibinfo {author} {\bibfnamefont {W.}~\bibnamefont {Shinoda}}, \bibinfo {author} {\bibfnamefont {L.}~\bibnamefont {Monticelli}},  \emph {et~al.},\ }\bibfield  {title} {\enquote {\bibinfo {title} {Pragmatic coarse-graining of proteins: models and applications},}\ }\href@noop {} {\bibfield  {journal} {\bibinfo  {journal} {J. Chem. Theory Comput.}\ }\textbf {\bibinfo {volume} {19}},\ \bibinfo {pages} {7112--7135} (\bibinfo {year} {2023})}\BibitemShut {NoStop}%
\bibitem [{\citenamefont {Marrink}\ \emph {et~al.}(2023)\citenamefont {Marrink}, \citenamefont {Monticelli}, \citenamefont {Melo}, \citenamefont {Alessandri}, \citenamefont {Tieleman},\ and\ \citenamefont {Souza}}]{marrink2023two}%
  \BibitemOpen
  \bibfield  {author} {\bibinfo {author} {\bibfnamefont {S.~J.}\ \bibnamefont {Marrink}}, \bibinfo {author} {\bibfnamefont {L.}~\bibnamefont {Monticelli}}, \bibinfo {author} {\bibfnamefont {M.~N.}\ \bibnamefont {Melo}}, \bibinfo {author} {\bibfnamefont {R.}~\bibnamefont {Alessandri}}, \bibinfo {author} {\bibfnamefont {D.~P.}\ \bibnamefont {Tieleman}}, \ and\ \bibinfo {author} {\bibfnamefont {P.~C.}\ \bibnamefont {Souza}},\ }\bibfield  {title} {\enquote {\bibinfo {title} {Two decades of martini: Better beads, broader scope},}\ }\href@noop {} {\bibfield  {journal} {\bibinfo  {journal} {Wiley Interdisciplinary Reviews: Computational Molecular Science}\ }\textbf {\bibinfo {volume} {13}},\ \bibinfo {pages} {e1620} (\bibinfo {year} {2023})}\BibitemShut {NoStop}%
\bibitem [{\citenamefont {Izvekov}\ and\ \citenamefont {Voth}(2005{\natexlab{a}})}]{izvekov2005multiscale}%
  \BibitemOpen
  \bibfield  {author} {\bibinfo {author} {\bibfnamefont {S.}~\bibnamefont {Izvekov}}\ and\ \bibinfo {author} {\bibfnamefont {G.~A.}\ \bibnamefont {Voth}},\ }\bibfield  {title} {\enquote {\bibinfo {title} {A multiscale coarse-graining method for biomolecular systems},}\ }\href@noop {} {\bibfield  {journal} {\bibinfo  {journal} {The Journal of Physical Chemistry B}\ }\textbf {\bibinfo {volume} {109}},\ \bibinfo {pages} {2469--2473} (\bibinfo {year} {2005}{\natexlab{a}})}\BibitemShut {NoStop}%
\bibitem [{\citenamefont {Izvekov}\ and\ \citenamefont {Voth}(2005{\natexlab{b}})}]{izvekov2005multiscale2}%
  \BibitemOpen
  \bibfield  {author} {\bibinfo {author} {\bibfnamefont {S.}~\bibnamefont {Izvekov}}\ and\ \bibinfo {author} {\bibfnamefont {G.~A.}\ \bibnamefont {Voth}},\ }\bibfield  {title} {\enquote {\bibinfo {title} {Multiscale coarse graining of liquid-state systems},}\ }\href@noop {} {\bibfield  {journal} {\bibinfo  {journal} {The Journal of chemical physics}\ }\textbf {\bibinfo {volume} {123}} (\bibinfo {year} {2005}{\natexlab{b}})}\BibitemShut {NoStop}%
\bibitem [{\citenamefont {Noid}\ \emph {et~al.}(2008)\citenamefont {Noid}, \citenamefont {Chu}, \citenamefont {Ayton}, \citenamefont {Krishna}, \citenamefont {Izvekov}, \citenamefont {Voth}, \citenamefont {Das},\ and\ \citenamefont {Andersen}}]{noid2008multiscale}%
  \BibitemOpen
  \bibfield  {author} {\bibinfo {author} {\bibfnamefont {W.~G.}\ \bibnamefont {Noid}}, \bibinfo {author} {\bibfnamefont {J.~W.}\ \bibnamefont {Chu}}, \bibinfo {author} {\bibfnamefont {G.~S.}\ \bibnamefont {Ayton}}, \bibinfo {author} {\bibfnamefont {V.}~\bibnamefont {Krishna}}, \bibinfo {author} {\bibfnamefont {S.}~\bibnamefont {Izvekov}}, \bibinfo {author} {\bibfnamefont {G.~A.}\ \bibnamefont {Voth}}, \bibinfo {author} {\bibfnamefont {A.}~\bibnamefont {Das}}, \ and\ \bibinfo {author} {\bibfnamefont {H.~C.}\ \bibnamefont {Andersen}},\ }\bibfield  {title} {\enquote {\bibinfo {title} {The multiscale coarse-graining method. i. a rigorous bridge between atomistic and coarse-grained models},}\ }\href {\doibase 10.1063/1.2938860} {\bibfield  {journal} {\bibinfo  {journal} {J. Chem. Phys.}\ }\textbf {\bibinfo {volume} {128}},\ \bibinfo {pages} {244114} (\bibinfo {year} {2008})}\BibitemShut {NoStop}%
\bibitem [{\citenamefont {Lu}\ and\ \citenamefont {Voth}(2011)}]{lu2011multiscale}%
  \BibitemOpen
  \bibfield  {author} {\bibinfo {author} {\bibfnamefont {L.}~\bibnamefont {Lu}}\ and\ \bibinfo {author} {\bibfnamefont {G.~A.}\ \bibnamefont {Voth}},\ }\bibfield  {title} {\enquote {\bibinfo {title} {The multiscale coarse-graining method. vii. free energy decomposition of coarse-grained effective potentials},}\ }\href@noop {} {\bibfield  {journal} {\bibinfo  {journal} {The Journal of chemical physics}\ }\textbf {\bibinfo {volume} {134}} (\bibinfo {year} {2011})}\BibitemShut {NoStop}%
\bibitem [{\citenamefont {Lemke}\ and\ \citenamefont {Peter}(2017)}]{lemke2017neural}%
  \BibitemOpen
  \bibfield  {author} {\bibinfo {author} {\bibfnamefont {T.}~\bibnamefont {Lemke}}\ and\ \bibinfo {author} {\bibfnamefont {C.}~\bibnamefont {Peter}},\ }\bibfield  {title} {\enquote {\bibinfo {title} {Neural network based prediction of conformational free energies-a new route toward coarse-grained simulation models},}\ }\href@noop {} {\bibfield  {journal} {\bibinfo  {journal} {Journal of chemical theory and computation}\ }\textbf {\bibinfo {volume} {13}},\ \bibinfo {pages} {6213--6221} (\bibinfo {year} {2017})}\BibitemShut {NoStop}%
\bibitem [{\citenamefont {Wang}\ \emph {et~al.}(2019)\citenamefont {Wang}, \citenamefont {Olsson}, \citenamefont {Wehmeyer}, \citenamefont {P{\'e}rez}, \citenamefont {Charron}, \citenamefont {Fabritiis}, \citenamefont {No{\'e}},\ and\ \citenamefont {Clementi}}]{wang2019machine}%
  \BibitemOpen
  \bibfield  {author} {\bibinfo {author} {\bibfnamefont {J.}~\bibnamefont {Wang}}, \bibinfo {author} {\bibfnamefont {S.}~\bibnamefont {Olsson}}, \bibinfo {author} {\bibfnamefont {C.}~\bibnamefont {Wehmeyer}}, \bibinfo {author} {\bibfnamefont {A.}~\bibnamefont {P{\'e}rez}}, \bibinfo {author} {\bibfnamefont {N.~E.}\ \bibnamefont {Charron}}, \bibinfo {author} {\bibfnamefont {G.~D.}\ \bibnamefont {Fabritiis}}, \bibinfo {author} {\bibfnamefont {F.}~\bibnamefont {No{\'e}}}, \ and\ \bibinfo {author} {\bibfnamefont {C.}~\bibnamefont {Clementi}},\ }\bibfield  {title} {\enquote {\bibinfo {title} {Machine learning of coarse-grained molecular dynamics force fields},}\ }\href {\doibase 10.1021/acscentsci.8b00913} {\bibfield  {journal} {\bibinfo  {journal} {ACS Cent. Sci.}\ }\textbf {\bibinfo {volume} {5}},\ \bibinfo {pages} {755--767} (\bibinfo {year} {2019})}\BibitemShut {NoStop}%
\bibitem [{\citenamefont {Husic}\ \emph {et~al.}(2020)\citenamefont {Husic}, \citenamefont {Charron}, \citenamefont {Lemm}, \citenamefont {Wang}, \citenamefont {P{\'e}rez}, \citenamefont {Majewski}, \citenamefont {Kr{\"a}mer}, \citenamefont {Chen}, \citenamefont {Olsson}, \citenamefont {De~Fabritiis} \emph {et~al.}}]{husic2020coarse}%
  \BibitemOpen
  \bibfield  {author} {\bibinfo {author} {\bibfnamefont {B.~E.}\ \bibnamefont {Husic}}, \bibinfo {author} {\bibfnamefont {N.~E.}\ \bibnamefont {Charron}}, \bibinfo {author} {\bibfnamefont {D.}~\bibnamefont {Lemm}}, \bibinfo {author} {\bibfnamefont {J.}~\bibnamefont {Wang}}, \bibinfo {author} {\bibfnamefont {A.}~\bibnamefont {P{\'e}rez}}, \bibinfo {author} {\bibfnamefont {M.}~\bibnamefont {Majewski}}, \bibinfo {author} {\bibfnamefont {A.}~\bibnamefont {Kr{\"a}mer}}, \bibinfo {author} {\bibfnamefont {Y.}~\bibnamefont {Chen}}, \bibinfo {author} {\bibfnamefont {S.}~\bibnamefont {Olsson}}, \bibinfo {author} {\bibfnamefont {G.}~\bibnamefont {De~Fabritiis}},  \emph {et~al.},\ }\bibfield  {title} {\enquote {\bibinfo {title} {Coarse graining molecular dynamics with graph neural networks},}\ }\href@noop {} {\bibfield  {journal} {\bibinfo  {journal} {The Journal of chemical physics}\ }\textbf {\bibinfo {volume} {153}} (\bibinfo {year} {2020})}\BibitemShut {NoStop}%
\bibitem [{\citenamefont {Chen}\ \emph {et~al.}(2021)\citenamefont {Chen}, \citenamefont {Kr{\"a}mer}, \citenamefont {Charron}, \citenamefont {Husic}, \citenamefont {Clementi},\ and\ \citenamefont {No{\'e}}}]{chen2021implicit}%
  \BibitemOpen
  \bibfield  {author} {\bibinfo {author} {\bibfnamefont {Y.}~\bibnamefont {Chen}}, \bibinfo {author} {\bibfnamefont {A.}~\bibnamefont {Kr{\"a}mer}}, \bibinfo {author} {\bibfnamefont {N.~E.}\ \bibnamefont {Charron}}, \bibinfo {author} {\bibfnamefont {B.~E.}\ \bibnamefont {Husic}}, \bibinfo {author} {\bibfnamefont {C.}~\bibnamefont {Clementi}}, \ and\ \bibinfo {author} {\bibfnamefont {F.}~\bibnamefont {No{\'e}}},\ }\bibfield  {title} {\enquote {\bibinfo {title} {Machine learning implicit solvation for molecular dynamics},}\ }\href {\doibase 10.1063/5.0059915} {\bibfield  {journal} {\bibinfo  {journal} {J. Chem. Phys.}\ }\textbf {\bibinfo {volume} {155}},\ \bibinfo {pages} {084101} (\bibinfo {year} {2021})}\BibitemShut {NoStop}%
\bibitem [{\citenamefont {Wang}\ \emph {et~al.}(2021)\citenamefont {Wang}, \citenamefont {Charron}, \citenamefont {Husic}, \citenamefont {Olsson}, \citenamefont {No{\'e}},\ and\ \citenamefont {Clementi}}]{wang2021multi}%
  \BibitemOpen
  \bibfield  {author} {\bibinfo {author} {\bibfnamefont {J.}~\bibnamefont {Wang}}, \bibinfo {author} {\bibfnamefont {N.}~\bibnamefont {Charron}}, \bibinfo {author} {\bibfnamefont {B.}~\bibnamefont {Husic}}, \bibinfo {author} {\bibfnamefont {S.}~\bibnamefont {Olsson}}, \bibinfo {author} {\bibfnamefont {F.}~\bibnamefont {No{\'e}}}, \ and\ \bibinfo {author} {\bibfnamefont {C.}~\bibnamefont {Clementi}},\ }\bibfield  {title} {\enquote {\bibinfo {title} {Multi-body effects in a coarse-grained protein force field},}\ }\href@noop {} {\bibfield  {journal} {\bibinfo  {journal} {The Journal of Chemical Physics}\ }\textbf {\bibinfo {volume} {154}} (\bibinfo {year} {2021})}\BibitemShut {NoStop}%
\bibitem [{\citenamefont {Majewski}\ \emph {et~al.}(2023)\citenamefont {Majewski}, \citenamefont {P{\'e}rez}, \citenamefont {Th\"{o}lke}, \citenamefont {Doerr}, \citenamefont {Charron}, \citenamefont {Giorgino}, \citenamefont {Husic}, \citenamefont {Clementi}, \citenamefont {No{\'e}},\ and\ \citenamefont {De~Fabritiis}}]{majewski2022machine}%
  \BibitemOpen
  \bibfield  {author} {\bibinfo {author} {\bibfnamefont {M.}~\bibnamefont {Majewski}}, \bibinfo {author} {\bibfnamefont {A.}~\bibnamefont {P{\'e}rez}}, \bibinfo {author} {\bibfnamefont {P.}~\bibnamefont {Th\"{o}lke}}, \bibinfo {author} {\bibfnamefont {S.}~\bibnamefont {Doerr}}, \bibinfo {author} {\bibfnamefont {N.~E.}\ \bibnamefont {Charron}}, \bibinfo {author} {\bibfnamefont {T.}~\bibnamefont {Giorgino}}, \bibinfo {author} {\bibfnamefont {B.~E.}\ \bibnamefont {Husic}}, \bibinfo {author} {\bibfnamefont {C.}~\bibnamefont {Clementi}}, \bibinfo {author} {\bibfnamefont {F.}~\bibnamefont {No{\'e}}}, \ and\ \bibinfo {author} {\bibfnamefont {G.}~\bibnamefont {De~Fabritiis}},\ }\bibfield  {title} {\enquote {\bibinfo {title} {Machine learning coarse-grained potentials of protein thermodynamics},}\ }\href {http://dx.doi.org/10.1038/s41467-023-41343-1} {\bibfield  {journal} {\bibinfo  {journal} {Nat. Commun.}\ }\textbf {\bibinfo {volume} {14}} (\bibinfo {year} {2023})}\BibitemShut {NoStop}%
\bibitem [{\citenamefont {Arts}\ \emph {et~al.}(2023)\citenamefont {Arts}, \citenamefont {Garcia~Satorras}, \citenamefont {Huang}, \citenamefont {Zugner}, \citenamefont {Federici}, \citenamefont {Clementi}, \citenamefont {No{\'e}}, \citenamefont {Pinsler},\ and\ \citenamefont {van~den Berg}}]{arts2023two}%
  \BibitemOpen
  \bibfield  {author} {\bibinfo {author} {\bibfnamefont {M.}~\bibnamefont {Arts}}, \bibinfo {author} {\bibfnamefont {V.}~\bibnamefont {Garcia~Satorras}}, \bibinfo {author} {\bibfnamefont {C.-W.}\ \bibnamefont {Huang}}, \bibinfo {author} {\bibfnamefont {D.}~\bibnamefont {Zugner}}, \bibinfo {author} {\bibfnamefont {M.}~\bibnamefont {Federici}}, \bibinfo {author} {\bibfnamefont {C.}~\bibnamefont {Clementi}}, \bibinfo {author} {\bibfnamefont {F.}~\bibnamefont {No{\'e}}}, \bibinfo {author} {\bibfnamefont {R.}~\bibnamefont {Pinsler}}, \ and\ \bibinfo {author} {\bibfnamefont {R.}~\bibnamefont {van~den Berg}},\ }\bibfield  {title} {\enquote {\bibinfo {title} {Two for one: Diffusion models and force fields for coarse-grained molecular dynamics},}\ }\href@noop {} {\bibfield  {journal} {\bibinfo  {journal} {Journal of Chemical Theory and Computation}\ }\textbf {\bibinfo {volume} {19}},\ \bibinfo {pages} {6151--6159} (\bibinfo {year} {2023})}\BibitemShut {NoStop}%
\bibitem [{\citenamefont {Durumeric}\ \emph {et~al.}(2023)\citenamefont {Durumeric}, \citenamefont {Charron}, \citenamefont {Templeton}, \citenamefont {Musil}, \citenamefont {Bonneau}, \citenamefont {Pasos-Trejo}, \citenamefont {Chen}, \citenamefont {Kelkar}, \citenamefont {No{\'e}},\ and\ \citenamefont {Clementi}}]{durumeric2023machine}%
  \BibitemOpen
  \bibfield  {author} {\bibinfo {author} {\bibfnamefont {A.~E.}\ \bibnamefont {Durumeric}}, \bibinfo {author} {\bibfnamefont {N.~E.}\ \bibnamefont {Charron}}, \bibinfo {author} {\bibfnamefont {C.}~\bibnamefont {Templeton}}, \bibinfo {author} {\bibfnamefont {F.}~\bibnamefont {Musil}}, \bibinfo {author} {\bibfnamefont {K.}~\bibnamefont {Bonneau}}, \bibinfo {author} {\bibfnamefont {A.~S.}\ \bibnamefont {Pasos-Trejo}}, \bibinfo {author} {\bibfnamefont {Y.}~\bibnamefont {Chen}}, \bibinfo {author} {\bibfnamefont {A.}~\bibnamefont {Kelkar}}, \bibinfo {author} {\bibfnamefont {F.}~\bibnamefont {No{\'e}}}, \ and\ \bibinfo {author} {\bibfnamefont {C.}~\bibnamefont {Clementi}},\ }\bibfield  {title} {\enquote {\bibinfo {title} {Machine learned coarse-grained protein force-fields: Are we there yet?}}\ }\href@noop {} {\bibfield  {journal} {\bibinfo  {journal} {Current opinion in structural biology}\ }\textbf {\bibinfo {volume} {79}},\ \bibinfo {pages} {102533} (\bibinfo {year} {2023})}\BibitemShut {NoStop}%
\bibitem [{\citenamefont {Charron}\ \emph {et~al.}(2023)\citenamefont {Charron}, \citenamefont {Musil}, \citenamefont {Guljas}, \citenamefont {Chen}, \citenamefont {Bonneau}, \citenamefont {Pasos-Trejo}, \citenamefont {Venturin}, \citenamefont {Gusew}, \citenamefont {Zaporozhets}, \citenamefont {Kr{\"a}mer}, \citenamefont {Templeton}, \citenamefont {Kelkar}, \citenamefont {Durumeric}, \citenamefont {Olsson}, \citenamefont {P{\'e}rez}, \citenamefont {Mejewski}, \citenamefont {Husic}, \citenamefont {Patel}, \citenamefont {Gianni}, \citenamefont {No{\'e}},\ and\ \citenamefont {Clementi}}]{charron2023navigating}%
  \BibitemOpen
  \bibfield  {author} {\bibinfo {author} {\bibfnamefont {N.~E.}\ \bibnamefont {Charron}}, \bibinfo {author} {\bibfnamefont {F.}~\bibnamefont {Musil}}, \bibinfo {author} {\bibfnamefont {A.}~\bibnamefont {Guljas}}, \bibinfo {author} {\bibfnamefont {Y.}~\bibnamefont {Chen}}, \bibinfo {author} {\bibfnamefont {K.}~\bibnamefont {Bonneau}}, \bibinfo {author} {\bibfnamefont {A.~S.}\ \bibnamefont {Pasos-Trejo}}, \bibinfo {author} {\bibfnamefont {J.}~\bibnamefont {Venturin}}, \bibinfo {author} {\bibfnamefont {D.}~\bibnamefont {Gusew}}, \bibinfo {author} {\bibfnamefont {I.}~\bibnamefont {Zaporozhets}}, \bibinfo {author} {\bibfnamefont {A.}~\bibnamefont {Kr{\"a}mer}}, \bibinfo {author} {\bibfnamefont {C.}~\bibnamefont {Templeton}}, \bibinfo {author} {\bibfnamefont {A.}~\bibnamefont {Kelkar}}, \bibinfo {author} {\bibfnamefont {A.~E.~P.}\ \bibnamefont {Durumeric}}, \bibinfo {author} {\bibfnamefont {S.}~\bibnamefont {Olsson}}, \bibinfo {author} {\bibfnamefont {A.}~\bibnamefont {P{\'e}rez}}, \bibinfo {author} {\bibfnamefont
  {M.}~\bibnamefont {Mejewski}}, \bibinfo {author} {\bibfnamefont {B.~E.}\ \bibnamefont {Husic}}, \bibinfo {author} {\bibfnamefont {A.}~\bibnamefont {Patel}}, \bibinfo {author} {\bibfnamefont {F.~D.}\ \bibnamefont {Gianni}}, \bibinfo {author} {\bibfnamefont {F.}~\bibnamefont {No{\'e}}}, \ and\ \bibinfo {author} {\bibfnamefont {C.}~\bibnamefont {Clementi}},\ }\bibfield  {title} {\enquote {\bibinfo {title} {Navigating protein landscapes with a machine-learned transferable coarse-grained model},}\ }\href@noop {} {\bibfield  {journal} {\bibinfo  {journal} {arXiv preprint arXiv:2310.18278}\ } (\bibinfo {year} {2023})}\BibitemShut {NoStop}%
\bibitem [{\citenamefont {Wellawatte}, \citenamefont {Hocky},\ and\ \citenamefont {White}(2023)}]{wellawatte2023neural}%
  \BibitemOpen
  \bibfield  {author} {\bibinfo {author} {\bibfnamefont {G.~P.}\ \bibnamefont {Wellawatte}}, \bibinfo {author} {\bibfnamefont {G.~M.}\ \bibnamefont {Hocky}}, \ and\ \bibinfo {author} {\bibfnamefont {A.~D.}\ \bibnamefont {White}},\ }\bibfield  {title} {\enquote {\bibinfo {title} {Neural potentials of proteins extrapolate beyond training data},}\ }\href@noop {} {\bibfield  {journal} {\bibinfo  {journal} {The Journal of Chemical Physics}\ }\textbf {\bibinfo {volume} {159}} (\bibinfo {year} {2023})}\BibitemShut {NoStop}%
\bibitem [{\citenamefont {K{\"o}hler}\ \emph {et~al.}(2023)\citenamefont {K{\"o}hler}, \citenamefont {Chen}, \citenamefont {Kr{\"a}mer}, \citenamefont {Clementi},\ and\ \citenamefont {No{\'e}}}]{kohler2023flow}%
  \BibitemOpen
  \bibfield  {author} {\bibinfo {author} {\bibfnamefont {J.}~\bibnamefont {K{\"o}hler}}, \bibinfo {author} {\bibfnamefont {Y.}~\bibnamefont {Chen}}, \bibinfo {author} {\bibfnamefont {A.}~\bibnamefont {Kr{\"a}mer}}, \bibinfo {author} {\bibfnamefont {C.}~\bibnamefont {Clementi}}, \ and\ \bibinfo {author} {\bibfnamefont {F.}~\bibnamefont {No{\'e}}},\ }\bibfield  {title} {\enquote {\bibinfo {title} {Flow-matching: Efficient coarse-graining of molecular dynamics without forces},}\ }\href@noop {} {\bibfield  {journal} {\bibinfo  {journal} {Journal of Chemical Theory and Computation}\ }\textbf {\bibinfo {volume} {19}},\ \bibinfo {pages} {942--952} (\bibinfo {year} {2023})}\BibitemShut {NoStop}%
\bibitem [{\citenamefont {Kr{\"a}mer}\ \emph {et~al.}(2023)\citenamefont {Kr{\"a}mer}, \citenamefont {Durumeric}, \citenamefont {Charron}, \citenamefont {Chen}, \citenamefont {Clementi},\ and\ \citenamefont {No{\'e}}}]{kramer2023statistically}%
  \BibitemOpen
  \bibfield  {author} {\bibinfo {author} {\bibfnamefont {A.}~\bibnamefont {Kr{\"a}mer}}, \bibinfo {author} {\bibfnamefont {A.~E.}\ \bibnamefont {Durumeric}}, \bibinfo {author} {\bibfnamefont {N.~E.}\ \bibnamefont {Charron}}, \bibinfo {author} {\bibfnamefont {Y.}~\bibnamefont {Chen}}, \bibinfo {author} {\bibfnamefont {C.}~\bibnamefont {Clementi}}, \ and\ \bibinfo {author} {\bibfnamefont {F.}~\bibnamefont {No{\'e}}},\ }\bibfield  {title} {\enquote {\bibinfo {title} {Statistically optimal force aggregation for coarse-graining molecular dynamics},}\ }\href@noop {} {\bibfield  {journal} {\bibinfo  {journal} {The Journal of Physical Chemistry Letters}\ }\textbf {\bibinfo {volume} {14}},\ \bibinfo {pages} {3970--3979} (\bibinfo {year} {2023})}\BibitemShut {NoStop}%
\bibitem [{\citenamefont {Chennakesavalu}, \citenamefont {Toomer},\ and\ \citenamefont {Rotskoff}(2023)}]{chennakesavalu2023ensuring}%
  \BibitemOpen
  \bibfield  {author} {\bibinfo {author} {\bibfnamefont {S.}~\bibnamefont {Chennakesavalu}}, \bibinfo {author} {\bibfnamefont {D.~J.}\ \bibnamefont {Toomer}}, \ and\ \bibinfo {author} {\bibfnamefont {G.~M.}\ \bibnamefont {Rotskoff}},\ }\bibfield  {title} {\enquote {\bibinfo {title} {Ensuring thermodynamic consistency with invertible coarse-graining},}\ }\href@noop {} {\bibfield  {journal} {\bibinfo  {journal} {The Journal of Chemical Physics}\ }\textbf {\bibinfo {volume} {158}} (\bibinfo {year} {2023})}\BibitemShut {NoStop}%
\bibitem [{\citenamefont {Mirarchi}\ \emph {et~al.}(2024)\citenamefont {Mirarchi}, \citenamefont {Pel{\'a}ez}, \citenamefont {Simeon},\ and\ \citenamefont {De~Fabritiis}}]{mirarchi2024amaro}%
  \BibitemOpen
  \bibfield  {author} {\bibinfo {author} {\bibfnamefont {A.}~\bibnamefont {Mirarchi}}, \bibinfo {author} {\bibfnamefont {R.~P.}\ \bibnamefont {Pel{\'a}ez}}, \bibinfo {author} {\bibfnamefont {G.}~\bibnamefont {Simeon}}, \ and\ \bibinfo {author} {\bibfnamefont {G.}~\bibnamefont {De~Fabritiis}},\ }\bibfield  {title} {\enquote {\bibinfo {title} {Amaro: all heavy-atom transferable neural network potentials of protein thermodynamics},}\ }\href@noop {} {\bibfield  {journal} {\bibinfo  {journal} {Journal of Chemical Theory and Computation}\ }\textbf {\bibinfo {volume} {20}},\ \bibinfo {pages} {9871--9878} (\bibinfo {year} {2024})}\BibitemShut {NoStop}%
\bibitem [{\citenamefont {Shell}(2008)}]{shell2008relative}%
  \BibitemOpen
  \bibfield  {author} {\bibinfo {author} {\bibfnamefont {M.~S.}\ \bibnamefont {Shell}},\ }\bibfield  {title} {\enquote {\bibinfo {title} {The relative entropy is fundamental to multiscale and inverse thermodynamic problems},}\ }\href@noop {} {\bibfield  {journal} {\bibinfo  {journal} {The Journal of chemical physics}\ }\textbf {\bibinfo {volume} {129}} (\bibinfo {year} {2008})}\BibitemShut {NoStop}%
\bibitem [{\citenamefont {Carmichael}\ and\ \citenamefont {Shell}(2012)}]{carmichael2012new}%
  \BibitemOpen
  \bibfield  {author} {\bibinfo {author} {\bibfnamefont {S.~P.}\ \bibnamefont {Carmichael}}\ and\ \bibinfo {author} {\bibfnamefont {M.~S.}\ \bibnamefont {Shell}},\ }\bibfield  {title} {\enquote {\bibinfo {title} {A new multiscale algorithm and its application to coarse-grained peptide models for self-assembly},}\ }\href@noop {} {\bibfield  {journal} {\bibinfo  {journal} {The Journal of Physical Chemistry B}\ }\textbf {\bibinfo {volume} {116}},\ \bibinfo {pages} {8383--8393} (\bibinfo {year} {2012})}\BibitemShut {NoStop}%
\bibitem [{\citenamefont {Thaler}, \citenamefont {Stupp},\ and\ \citenamefont {Zavadlav}(2022)}]{thaler2022deep}%
  \BibitemOpen
  \bibfield  {author} {\bibinfo {author} {\bibfnamefont {S.}~\bibnamefont {Thaler}}, \bibinfo {author} {\bibfnamefont {M.}~\bibnamefont {Stupp}}, \ and\ \bibinfo {author} {\bibfnamefont {J.}~\bibnamefont {Zavadlav}},\ }\bibfield  {title} {\enquote {\bibinfo {title} {Deep coarse-grained potentials via relative entropy minimization},}\ }\href@noop {} {\bibfield  {journal} {\bibinfo  {journal} {The Journal of Chemical Physics}\ }\textbf {\bibinfo {volume} {157}} (\bibinfo {year} {2022})}\BibitemShut {NoStop}%
\bibitem [{\citenamefont {Thaler}\ and\ \citenamefont {Zavadlav}(2021)}]{thaler2021learning}%
  \BibitemOpen
  \bibfield  {author} {\bibinfo {author} {\bibfnamefont {S.}~\bibnamefont {Thaler}}\ and\ \bibinfo {author} {\bibfnamefont {J.}~\bibnamefont {Zavadlav}},\ }\bibfield  {title} {\enquote {\bibinfo {title} {Learning neural network potentials from experimental data via differentiable trajectory reweighting},}\ }\href@noop {} {\bibfield  {journal} {\bibinfo  {journal} {Nature communications}\ }\textbf {\bibinfo {volume} {12}},\ \bibinfo {pages} {6884} (\bibinfo {year} {2021})}\BibitemShut {NoStop}%
\bibitem [{\citenamefont {Ding}\ and\ \citenamefont {Zhang}(2022)}]{ding2022contrastive}%
  \BibitemOpen
  \bibfield  {author} {\bibinfo {author} {\bibfnamefont {X.}~\bibnamefont {Ding}}\ and\ \bibinfo {author} {\bibfnamefont {B.}~\bibnamefont {Zhang}},\ }\bibfield  {title} {\enquote {\bibinfo {title} {Contrastive learning of coarse-grained force fields},}\ }\href@noop {} {\bibfield  {journal} {\bibinfo  {journal} {Journal of chemical theory and computation}\ }\textbf {\bibinfo {volume} {18}},\ \bibinfo {pages} {6334--6344} (\bibinfo {year} {2022})}\BibitemShut {NoStop}%
\bibitem [{\citenamefont {Plainer}\ \emph {et~al.}(2025)\citenamefont {Plainer}, \citenamefont {Wu}, \citenamefont {Klein}, \citenamefont {G\"{u}nnemann},\ and\ \citenamefont {No\'{e}}}]{plainer2025consistentsamplingsimulationmolecular}%
  \BibitemOpen
  \bibfield  {author} {\bibinfo {author} {\bibfnamefont {M.}~\bibnamefont {Plainer}}, \bibinfo {author} {\bibfnamefont {H.}~\bibnamefont {Wu}}, \bibinfo {author} {\bibfnamefont {L.}~\bibnamefont {Klein}}, \bibinfo {author} {\bibfnamefont {S.}~\bibnamefont {G\"{u}nnemann}}, \ and\ \bibinfo {author} {\bibfnamefont {F.}~\bibnamefont {No\'{e}}},\ }\bibfield  {title} {\enquote {\bibinfo {title} {Consistent sampling and simulation: Molecular dynamics with energy-based diffusion models},}\ }\href@noop {} {\bibfield  {journal} {\bibinfo  {journal} {arXiv preprint arXiv:2506.17139}\ } (\bibinfo {year} {2025})}\BibitemShut {NoStop}%
\bibitem [{\citenamefont {Hinton}(2002)}]{hinton2002training}%
  \BibitemOpen
  \bibfield  {author} {\bibinfo {author} {\bibfnamefont {G.~E.}\ \bibnamefont {Hinton}},\ }\bibfield  {title} {\enquote {\bibinfo {title} {Training products of experts by minimizing contrastive divergence},}\ }\href@noop {} {\bibfield  {journal} {\bibinfo  {journal} {Neural computation}\ }\textbf {\bibinfo {volume} {14}},\ \bibinfo {pages} {1771--1800} (\bibinfo {year} {2002})}\BibitemShut {NoStop}%
\bibitem [{\citenamefont {Hyv{\"a}rinen}\ and\ \citenamefont {Dayan}(2005)}]{hyvarinen2005estimation}%
  \BibitemOpen
  \bibfield  {author} {\bibinfo {author} {\bibfnamefont {A.}~\bibnamefont {Hyv{\"a}rinen}}\ and\ \bibinfo {author} {\bibfnamefont {P.}~\bibnamefont {Dayan}},\ }\bibfield  {title} {\enquote {\bibinfo {title} {Estimation of non-normalized statistical models by score matching.}}\ }\href@noop {} {\bibfield  {journal} {\bibinfo  {journal} {Journal of Machine Learning Research}\ }\textbf {\bibinfo {volume} {6}} (\bibinfo {year} {2005})}\BibitemShut {NoStop}%
\bibitem [{\citenamefont {Gutmann}\ and\ \citenamefont {Hyv{\"a}rinen}(2010)}]{gutmann2010noise}%
  \BibitemOpen
  \bibfield  {author} {\bibinfo {author} {\bibfnamefont {M.}~\bibnamefont {Gutmann}}\ and\ \bibinfo {author} {\bibfnamefont {A.}~\bibnamefont {Hyv{\"a}rinen}},\ }\bibfield  {title} {\enquote {\bibinfo {title} {Noise-contrastive estimation: A new estimation principle for unnormalized statistical models},}\ }in\ \href@noop {} {\emph {\bibinfo {booktitle} {Proceedings of the thirteenth international conference on artificial intelligence and statistics}}}\ (\bibinfo {organization} {JMLR Workshop and Conference Proceedings},\ \bibinfo {year} {2010})\ pp.\ \bibinfo {pages} {297--304}\BibitemShut {NoStop}%
\bibitem [{\citenamefont {Vincent}(2011)}]{vincent2011connection}%
  \BibitemOpen
  \bibfield  {author} {\bibinfo {author} {\bibfnamefont {P.}~\bibnamefont {Vincent}},\ }\bibfield  {title} {\enquote {\bibinfo {title} {A connection between score matching and denoising autoencoders},}\ }\href@noop {} {\bibfield  {journal} {\bibinfo  {journal} {Neural computation}\ }\textbf {\bibinfo {volume} {23}},\ \bibinfo {pages} {1661--1674} (\bibinfo {year} {2011})}\BibitemShut {NoStop}%
\bibitem [{\citenamefont {Song}\ \emph {et~al.}(2020)\citenamefont {Song}, \citenamefont {Garg}, \citenamefont {Shi},\ and\ \citenamefont {Ermon}}]{song2020sliced}%
  \BibitemOpen
  \bibfield  {author} {\bibinfo {author} {\bibfnamefont {Y.}~\bibnamefont {Song}}, \bibinfo {author} {\bibfnamefont {S.}~\bibnamefont {Garg}}, \bibinfo {author} {\bibfnamefont {J.}~\bibnamefont {Shi}}, \ and\ \bibinfo {author} {\bibfnamefont {S.}~\bibnamefont {Ermon}},\ }\bibfield  {title} {\enquote {\bibinfo {title} {Sliced score matching: A scalable approach to density and score estimation},}\ }in\ \href@noop {} {\emph {\bibinfo {booktitle} {Uncertainty in artificial intelligence}}}\ (\bibinfo {organization} {PMLR},\ \bibinfo {year} {2020})\ pp.\ \bibinfo {pages} {574--584}\BibitemShut {NoStop}%
\bibitem [{\citenamefont {Durumeric}\ \emph {et~al.}(2024)\citenamefont {Durumeric}, \citenamefont {Chen}, \citenamefont {No{\'e}},\ and\ \citenamefont {Clementi}}]{durumeric2024learning}%
  \BibitemOpen
  \bibfield  {author} {\bibinfo {author} {\bibfnamefont {A.~E.}\ \bibnamefont {Durumeric}}, \bibinfo {author} {\bibfnamefont {Y.}~\bibnamefont {Chen}}, \bibinfo {author} {\bibfnamefont {F.}~\bibnamefont {No{\'e}}}, \ and\ \bibinfo {author} {\bibfnamefont {C.}~\bibnamefont {Clementi}},\ }\bibfield  {title} {\enquote {\bibinfo {title} {Learning data efficient coarse-grained molecular dynamics from forces and noise},}\ }\href@noop {} {\bibfield  {journal} {\bibinfo  {journal} {arXiv preprint arXiv:2407.01286}\ } (\bibinfo {year} {2024})}\BibitemShut {NoStop}%
\bibitem [{\citenamefont {Song}\ \emph {et~al.}(2021)\citenamefont {Song}, \citenamefont {Sohl-Dickstein}, \citenamefont {Kingma}, \citenamefont {Kumar}, \citenamefont {Ermon},\ and\ \citenamefont {Poole}}]{song2021}%
  \BibitemOpen
  \bibfield  {author} {\bibinfo {author} {\bibfnamefont {Y.}~\bibnamefont {Song}}, \bibinfo {author} {\bibfnamefont {J.}~\bibnamefont {Sohl-Dickstein}}, \bibinfo {author} {\bibfnamefont {D.~P.}\ \bibnamefont {Kingma}}, \bibinfo {author} {\bibfnamefont {A.}~\bibnamefont {Kumar}}, \bibinfo {author} {\bibfnamefont {S.}~\bibnamefont {Ermon}}, \ and\ \bibinfo {author} {\bibfnamefont {B.}~\bibnamefont {Poole}},\ }\bibfield  {title} {\enquote {\bibinfo {title} {Score-based generative modeling through stochastic differential equations},}\ }in\ \href@noop {} {\emph {\bibinfo {booktitle} {International Conference on Learning Representations}}}\ (\bibinfo {year} {2021})\BibitemShut {NoStop}%
\bibitem [{\citenamefont {Ho}, \citenamefont {Jain},\ and\ \citenamefont {Abbeel}(2020)}]{ho2020denoising}%
  \BibitemOpen
  \bibfield  {author} {\bibinfo {author} {\bibfnamefont {J.}~\bibnamefont {Ho}}, \bibinfo {author} {\bibfnamefont {A.}~\bibnamefont {Jain}}, \ and\ \bibinfo {author} {\bibfnamefont {P.}~\bibnamefont {Abbeel}},\ }\bibfield  {title} {\enquote {\bibinfo {title} {Denoising diffusion probabilistic models},}\ }\href@noop {} {\bibfield  {journal} {\bibinfo  {journal} {Advances in Neural Information Processing Systems}\ }\textbf {\bibinfo {volume} {33}},\ \bibinfo {pages} {6840--6851} (\bibinfo {year} {2020})}\BibitemShut {NoStop}%
\bibitem [{\citenamefont {Badaczewska-Dawid}, \citenamefont {Kolinski},\ and\ \citenamefont {Kmiecik}(2020)}]{badaczewska2020computational}%
  \BibitemOpen
  \bibfield  {author} {\bibinfo {author} {\bibfnamefont {A.~E.}\ \bibnamefont {Badaczewska-Dawid}}, \bibinfo {author} {\bibfnamefont {A.}~\bibnamefont {Kolinski}}, \ and\ \bibinfo {author} {\bibfnamefont {S.}~\bibnamefont {Kmiecik}},\ }\bibfield  {title} {\enquote {\bibinfo {title} {Computational reconstruction of atomistic protein structures from coarse-grained models},}\ }\href@noop {} {\bibfield  {journal} {\bibinfo  {journal} {Computational and structural biotechnology journal}\ }\textbf {\bibinfo {volume} {18}},\ \bibinfo {pages} {162--176} (\bibinfo {year} {2020})}\BibitemShut {NoStop}%
\bibitem [{\citenamefont {Shmilovich}\ \emph {et~al.}(2022)\citenamefont {Shmilovich}, \citenamefont {Stieffenhofer}, \citenamefont {Charron},\ and\ \citenamefont {Hoffmann}}]{shmilovich2022temporally}%
  \BibitemOpen
  \bibfield  {author} {\bibinfo {author} {\bibfnamefont {K.}~\bibnamefont {Shmilovich}}, \bibinfo {author} {\bibfnamefont {M.}~\bibnamefont {Stieffenhofer}}, \bibinfo {author} {\bibfnamefont {N.~E.}\ \bibnamefont {Charron}}, \ and\ \bibinfo {author} {\bibfnamefont {M.}~\bibnamefont {Hoffmann}},\ }\bibfield  {title} {\enquote {\bibinfo {title} {Temporally coherent backmapping of molecular trajectories from coarse-grained to atomistic resolution},}\ }\href@noop {} {\bibfield  {journal} {\bibinfo  {journal} {The Journal of Physical Chemistry A}\ }\textbf {\bibinfo {volume} {126}},\ \bibinfo {pages} {9124--9139} (\bibinfo {year} {2022})}\BibitemShut {NoStop}%
\bibitem [{\citenamefont {Wang}\ \emph {et~al.}(2022)\citenamefont {Wang}, \citenamefont {Xu}, \citenamefont {Cai}, \citenamefont {Miller}, \citenamefont {Smidt}, \citenamefont {Wang}, \citenamefont {Tang},\ and\ \citenamefont {G{\'o}mez-Bombarelli}}]{wang2022generative}%
  \BibitemOpen
  \bibfield  {author} {\bibinfo {author} {\bibfnamefont {W.}~\bibnamefont {Wang}}, \bibinfo {author} {\bibfnamefont {M.}~\bibnamefont {Xu}}, \bibinfo {author} {\bibfnamefont {C.}~\bibnamefont {Cai}}, \bibinfo {author} {\bibfnamefont {B.~K.}\ \bibnamefont {Miller}}, \bibinfo {author} {\bibfnamefont {T.}~\bibnamefont {Smidt}}, \bibinfo {author} {\bibfnamefont {Y.}~\bibnamefont {Wang}}, \bibinfo {author} {\bibfnamefont {J.}~\bibnamefont {Tang}}, \ and\ \bibinfo {author} {\bibfnamefont {R.}~\bibnamefont {G{\'o}mez-Bombarelli}},\ }\bibfield  {title} {\enquote {\bibinfo {title} {Generative coarse-graining of molecular conformations},}\ }\href@noop {} {\bibfield  {journal} {\bibinfo  {journal} {arXiv preprint arXiv:2201.12176}\ } (\bibinfo {year} {2022})}\BibitemShut {NoStop}%
\bibitem [{\citenamefont {Jones}, \citenamefont {Shmilovich},\ and\ \citenamefont {Ferguson}(2023)}]{jones2023diamondback}%
  \BibitemOpen
  \bibfield  {author} {\bibinfo {author} {\bibfnamefont {M.~S.}\ \bibnamefont {Jones}}, \bibinfo {author} {\bibfnamefont {K.}~\bibnamefont {Shmilovich}}, \ and\ \bibinfo {author} {\bibfnamefont {A.~L.}\ \bibnamefont {Ferguson}},\ }\bibfield  {title} {\enquote {\bibinfo {title} {Diamondback: Diffusion-denoising autoregressive model for non-deterministic backmapping of c$\alpha$ protein traces},}\ }\href@noop {} {\bibfield  {journal} {\bibinfo  {journal} {Journal of Chemical Theory and Computation}\ }\textbf {\bibinfo {volume} {19}},\ \bibinfo {pages} {7908--7923} (\bibinfo {year} {2023})}\BibitemShut {NoStop}%
\bibitem [{\citenamefont {Yang}\ and\ \citenamefont {G\'{o}mez-Bombarelli}(2023)}]{yang2023chemically}%
  \BibitemOpen
  \bibfield  {author} {\bibinfo {author} {\bibfnamefont {S.}~\bibnamefont {Yang}}\ and\ \bibinfo {author} {\bibfnamefont {R.}~\bibnamefont {G\'{o}mez-Bombarelli}},\ }\bibfield  {title} {\enquote {\bibinfo {title} {Chemically transferable generative backmapping of coarse-grained proteins},}\ }in\ \href@noop {} {\emph {\bibinfo {booktitle} {Proceedings of the 40th International Conference on Machine Learning}}},\ \bibinfo {series and number} {ICML'23}\ (\bibinfo  {publisher} {JMLR.org},\ \bibinfo {year} {2023})\BibitemShut {NoStop}%
\bibitem [{\citenamefont {Pang}, \citenamefont {Yang},\ and\ \citenamefont {Gumbart}(2024)}]{pang2024simple}%
  \BibitemOpen
  \bibfield  {author} {\bibinfo {author} {\bibfnamefont {Y.~T.}\ \bibnamefont {Pang}}, \bibinfo {author} {\bibfnamefont {L.}~\bibnamefont {Yang}}, \ and\ \bibinfo {author} {\bibfnamefont {J.~C.}\ \bibnamefont {Gumbart}},\ }\bibfield  {title} {\enquote {\bibinfo {title} {From simple to complex: Reconstructing all-atom structures from coarse-grained models using cg2all},}\ }\href@noop {} {\bibfield  {journal} {\bibinfo  {journal} {Structure}\ }\textbf {\bibinfo {volume} {32}},\ \bibinfo {pages} {5--7} (\bibinfo {year} {2024})}\BibitemShut {NoStop}%
\bibitem [{\citenamefont {Sahrmann}\ and\ \citenamefont {Voth}(2025)}]{sahrmann2025emergence}%
  \BibitemOpen
  \bibfield  {author} {\bibinfo {author} {\bibfnamefont {P.~G.}\ \bibnamefont {Sahrmann}}\ and\ \bibinfo {author} {\bibfnamefont {G.~A.}\ \bibnamefont {Voth}},\ }\bibfield  {title} {\enquote {\bibinfo {title} {On the emergence of machine-learning methods in bottom-up coarse-graining},}\ }\href@noop {} {\bibfield  {journal} {\bibinfo  {journal} {Current Opinion in Structural Biology}\ }\textbf {\bibinfo {volume} {90}},\ \bibinfo {pages} {102972} (\bibinfo {year} {2025})}\BibitemShut {NoStop}%
\bibitem [{\citenamefont {Goodfellow}\ \emph {et~al.}(2014)\citenamefont {Goodfellow}, \citenamefont {Pouget-Abadie}, \citenamefont {Mirza}, \citenamefont {Xu}, \citenamefont {Warde-Farley}, \citenamefont {Ozair}, \citenamefont {Courville},\ and\ \citenamefont {Bengio}}]{goodfellow2014generative}%
  \BibitemOpen
  \bibfield  {author} {\bibinfo {author} {\bibfnamefont {I.~J.}\ \bibnamefont {Goodfellow}}, \bibinfo {author} {\bibfnamefont {J.}~\bibnamefont {Pouget-Abadie}}, \bibinfo {author} {\bibfnamefont {M.}~\bibnamefont {Mirza}}, \bibinfo {author} {\bibfnamefont {B.}~\bibnamefont {Xu}}, \bibinfo {author} {\bibfnamefont {D.}~\bibnamefont {Warde-Farley}}, \bibinfo {author} {\bibfnamefont {S.}~\bibnamefont {Ozair}}, \bibinfo {author} {\bibfnamefont {A.}~\bibnamefont {Courville}}, \ and\ \bibinfo {author} {\bibfnamefont {Y.}~\bibnamefont {Bengio}},\ }\bibfield  {title} {\enquote {\bibinfo {title} {Generative adversarial nets},}\ }\href@noop {} {\bibfield  {journal} {\bibinfo  {journal} {Advances in neural information processing systems}\ }\textbf {\bibinfo {volume} {27}} (\bibinfo {year} {2014})}\BibitemShut {NoStop}%
\bibitem [{\citenamefont {Vaswani}\ \emph {et~al.}(2017)\citenamefont {Vaswani}, \citenamefont {Shazeer}, \citenamefont {Parmar}, \citenamefont {Uszkoreit}, \citenamefont {Jones}, \citenamefont {Gomez}, \citenamefont {Kaiser},\ and\ \citenamefont {Polosukhin}}]{vaswani17_atten_is_all_you_need}%
  \BibitemOpen
  \bibfield  {author} {\bibinfo {author} {\bibfnamefont {A.}~\bibnamefont {Vaswani}}, \bibinfo {author} {\bibfnamefont {N.}~\bibnamefont {Shazeer}}, \bibinfo {author} {\bibfnamefont {N.}~\bibnamefont {Parmar}}, \bibinfo {author} {\bibfnamefont {J.}~\bibnamefont {Uszkoreit}}, \bibinfo {author} {\bibfnamefont {L.}~\bibnamefont {Jones}}, \bibinfo {author} {\bibfnamefont {A.~N.}\ \bibnamefont {Gomez}}, \bibinfo {author} {\bibfnamefont {{\L}.}~\bibnamefont {Kaiser}}, \ and\ \bibinfo {author} {\bibfnamefont {I.}~\bibnamefont {Polosukhin}},\ }\bibfield  {title} {\enquote {\bibinfo {title} {Attention is all you need},}\ }\href@noop {} {\bibfield  {journal} {\bibinfo  {journal} {Advances in {N}eural {I}nformation {P}rocessing {S}ystems}\ }\textbf {\bibinfo {volume} {30}} (\bibinfo {year} {2017})}\BibitemShut {NoStop}%
\bibitem [{\citenamefont {Devlin}\ \emph {et~al.}(2019)\citenamefont {Devlin}, \citenamefont {Chang}, \citenamefont {Lee},\ and\ \citenamefont {Toutanova}}]{devlin2019bert}%
  \BibitemOpen
  \bibfield  {author} {\bibinfo {author} {\bibfnamefont {J.}~\bibnamefont {Devlin}}, \bibinfo {author} {\bibfnamefont {M.-W.}\ \bibnamefont {Chang}}, \bibinfo {author} {\bibfnamefont {K.}~\bibnamefont {Lee}}, \ and\ \bibinfo {author} {\bibfnamefont {K.}~\bibnamefont {Toutanova}},\ }\bibfield  {title} {\enquote {\bibinfo {title} {Bert: Pre-training of deep bidirectional transformers for language understanding},}\ }in\ \href@noop {} {\emph {\bibinfo {booktitle} {Proceedings of the 2019 conference of the North American chapter of the association for computational linguistics: human language technologies, volume 1 (long and short papers)}}}\ (\bibinfo {year} {2019})\ pp.\ \bibinfo {pages} {4171--4186}\BibitemShut {NoStop}%
\bibitem [{\citenamefont {Brown}\ \emph {et~al.}(2020)\citenamefont {Brown}, \citenamefont {Mann}, \citenamefont {Ryder}, \citenamefont {Subbiah}, \citenamefont {Kaplan}, \citenamefont {Dhariwal}, \citenamefont {Neelakantan}, \citenamefont {Shyam}, \citenamefont {Sastry}, \citenamefont {Askell} \emph {et~al.}}]{brown2020language}%
  \BibitemOpen
  \bibfield  {author} {\bibinfo {author} {\bibfnamefont {T.}~\bibnamefont {Brown}}, \bibinfo {author} {\bibfnamefont {B.}~\bibnamefont {Mann}}, \bibinfo {author} {\bibfnamefont {N.}~\bibnamefont {Ryder}}, \bibinfo {author} {\bibfnamefont {M.}~\bibnamefont {Subbiah}}, \bibinfo {author} {\bibfnamefont {J.~D.}\ \bibnamefont {Kaplan}}, \bibinfo {author} {\bibfnamefont {P.}~\bibnamefont {Dhariwal}}, \bibinfo {author} {\bibfnamefont {A.}~\bibnamefont {Neelakantan}}, \bibinfo {author} {\bibfnamefont {P.}~\bibnamefont {Shyam}}, \bibinfo {author} {\bibfnamefont {G.}~\bibnamefont {Sastry}}, \bibinfo {author} {\bibfnamefont {A.}~\bibnamefont {Askell}},  \emph {et~al.},\ }\bibfield  {title} {\enquote {\bibinfo {title} {Language models are few-shot learners},}\ }\href@noop {} {\bibfield  {journal} {\bibinfo  {journal} {Advances in neural information processing systems}\ }\textbf {\bibinfo {volume} {33}},\ \bibinfo {pages} {1877--1901} (\bibinfo {year} {2020})}\BibitemShut {NoStop}%
\bibitem [{\citenamefont {Achiam}\ \emph {et~al.}(2023)\citenamefont {Achiam}, \citenamefont {Adler}, \citenamefont {Agarwal}, \citenamefont {Ahmad}, \citenamefont {Akkaya}, \citenamefont {Aleman}, \citenamefont {Almeida}, \citenamefont {Altenschmidt}, \citenamefont {Altman}, \citenamefont {Anadkat} \emph {et~al.}}]{achiam2023gpt}%
  \BibitemOpen
  \bibfield  {author} {\bibinfo {author} {\bibfnamefont {J.}~\bibnamefont {Achiam}}, \bibinfo {author} {\bibfnamefont {S.}~\bibnamefont {Adler}}, \bibinfo {author} {\bibfnamefont {S.}~\bibnamefont {Agarwal}}, \bibinfo {author} {\bibfnamefont {L.}~\bibnamefont {Ahmad}}, \bibinfo {author} {\bibfnamefont {I.}~\bibnamefont {Akkaya}}, \bibinfo {author} {\bibfnamefont {F.~L.}\ \bibnamefont {Aleman}}, \bibinfo {author} {\bibfnamefont {D.}~\bibnamefont {Almeida}}, \bibinfo {author} {\bibfnamefont {J.}~\bibnamefont {Altenschmidt}}, \bibinfo {author} {\bibfnamefont {S.}~\bibnamefont {Altman}}, \bibinfo {author} {\bibfnamefont {S.}~\bibnamefont {Anadkat}},  \emph {et~al.},\ }\bibfield  {title} {\enquote {\bibinfo {title} {Gpt-4 technical report},}\ }\href@noop {} {\bibfield  {journal} {\bibinfo  {journal} {arXiv preprint arXiv:2303.08774}\ } (\bibinfo {year} {2023})}\BibitemShut {NoStop}%
\bibitem [{\citenamefont {Jumper}\ \emph {et~al.}(2021)\citenamefont {Jumper}, \citenamefont {Evans}, \citenamefont {Pritzel}, \citenamefont {Green}, \citenamefont {Figurnov}, \citenamefont {Ronneberger}, \citenamefont {Tunyasuvunakool}, \citenamefont {Bates}, \citenamefont {{\v{Z}}{\'\i}dek}, \citenamefont {Potapenko} \emph {et~al.}}]{jumper2021highly}%
  \BibitemOpen
  \bibfield  {author} {\bibinfo {author} {\bibfnamefont {J.}~\bibnamefont {Jumper}}, \bibinfo {author} {\bibfnamefont {R.}~\bibnamefont {Evans}}, \bibinfo {author} {\bibfnamefont {A.}~\bibnamefont {Pritzel}}, \bibinfo {author} {\bibfnamefont {T.}~\bibnamefont {Green}}, \bibinfo {author} {\bibfnamefont {M.}~\bibnamefont {Figurnov}}, \bibinfo {author} {\bibfnamefont {O.}~\bibnamefont {Ronneberger}}, \bibinfo {author} {\bibfnamefont {K.}~\bibnamefont {Tunyasuvunakool}}, \bibinfo {author} {\bibfnamefont {R.}~\bibnamefont {Bates}}, \bibinfo {author} {\bibfnamefont {A.}~\bibnamefont {{\v{Z}}{\'\i}dek}}, \bibinfo {author} {\bibfnamefont {A.}~\bibnamefont {Potapenko}},  \emph {et~al.},\ }\bibfield  {title} {\enquote {\bibinfo {title} {Highly accurate protein structure prediction with alphafold},}\ }\href@noop {} {\bibfield  {journal} {\bibinfo  {journal} {Nature}\ }\textbf {\bibinfo {volume} {596}},\ \bibinfo {pages} {583--589} (\bibinfo {year} {2021})}\BibitemShut {NoStop}%
\bibitem [{\citenamefont {Abramson}\ \emph {et~al.}(2024)\citenamefont {Abramson}, \citenamefont {Adler}, \citenamefont {Dunger}, \citenamefont {Evans}, \citenamefont {Green}, \citenamefont {Pritzel}, \citenamefont {Ronneberger}, \citenamefont {Willmore}, \citenamefont {Ballard}, \citenamefont {Bambrick} \emph {et~al.}}]{abramson2024accurate}%
  \BibitemOpen
  \bibfield  {author} {\bibinfo {author} {\bibfnamefont {J.}~\bibnamefont {Abramson}}, \bibinfo {author} {\bibfnamefont {J.}~\bibnamefont {Adler}}, \bibinfo {author} {\bibfnamefont {J.}~\bibnamefont {Dunger}}, \bibinfo {author} {\bibfnamefont {R.}~\bibnamefont {Evans}}, \bibinfo {author} {\bibfnamefont {T.}~\bibnamefont {Green}}, \bibinfo {author} {\bibfnamefont {A.}~\bibnamefont {Pritzel}}, \bibinfo {author} {\bibfnamefont {O.}~\bibnamefont {Ronneberger}}, \bibinfo {author} {\bibfnamefont {L.}~\bibnamefont {Willmore}}, \bibinfo {author} {\bibfnamefont {A.~J.}\ \bibnamefont {Ballard}}, \bibinfo {author} {\bibfnamefont {J.}~\bibnamefont {Bambrick}},  \emph {et~al.},\ }\bibfield  {title} {\enquote {\bibinfo {title} {Accurate structure prediction of biomolecular interactions with alphafold 3},}\ }\href@noop {} {\bibfield  {journal} {\bibinfo  {journal} {Nature}\ ,\ \bibinfo {pages} {1--3}} (\bibinfo {year} {2024})}\BibitemShut {NoStop}%
\bibitem [{\citenamefont {No{\'e}}\ \emph {et~al.}(2019)\citenamefont {No{\'e}}, \citenamefont {Olsson}, \citenamefont {K{\"o}hler},\ and\ \citenamefont {Wu}}]{noe2019boltzmann}%
  \BibitemOpen
  \bibfield  {author} {\bibinfo {author} {\bibfnamefont {F.}~\bibnamefont {No{\'e}}}, \bibinfo {author} {\bibfnamefont {S.}~\bibnamefont {Olsson}}, \bibinfo {author} {\bibfnamefont {J.}~\bibnamefont {K{\"o}hler}}, \ and\ \bibinfo {author} {\bibfnamefont {H.}~\bibnamefont {Wu}},\ }\bibfield  {title} {\enquote {\bibinfo {title} {Boltzmann generators --- sampling equilibrium states of many-body systems with deep learning},}\ }\href@noop {} {\bibfield  {journal} {\bibinfo  {journal} {Science}\ }\textbf {\bibinfo {volume} {365}},\ \bibinfo {pages} {eaaw1147} (\bibinfo {year} {2019})}\BibitemShut {NoStop}%
\bibitem [{\citenamefont {Rezende}\ and\ \citenamefont {Mohamed}(2015)}]{RezendeEtAl_NormalizingFlows}%
  \BibitemOpen
  \bibfield  {author} {\bibinfo {author} {\bibfnamefont {D.}~\bibnamefont {Rezende}}\ and\ \bibinfo {author} {\bibfnamefont {S.}~\bibnamefont {Mohamed}},\ }\bibfield  {title} {\enquote {\bibinfo {title} {Variational inference with normalizing flows},}\ }in\ \href@noop {} {\emph {\bibinfo {booktitle} {International conference on machine learning}}}\ (\bibinfo {organization} {PMLR},\ \bibinfo {year} {2015})\ pp.\ \bibinfo {pages} {1530--1538}\BibitemShut {NoStop}%
\bibitem [{\citenamefont {Papamakarios}\ \emph {et~al.}(2021)\citenamefont {Papamakarios}, \citenamefont {Nalisnick}, \citenamefont {Rezende}, \citenamefont {Mohamed},\ and\ \citenamefont {Lakshminarayanan}}]{papamakarios2021normalizing}%
  \BibitemOpen
  \bibfield  {author} {\bibinfo {author} {\bibfnamefont {G.}~\bibnamefont {Papamakarios}}, \bibinfo {author} {\bibfnamefont {E.~T.}\ \bibnamefont {Nalisnick}}, \bibinfo {author} {\bibfnamefont {D.~J.}\ \bibnamefont {Rezende}}, \bibinfo {author} {\bibfnamefont {S.}~\bibnamefont {Mohamed}}, \ and\ \bibinfo {author} {\bibfnamefont {B.}~\bibnamefont {Lakshminarayanan}},\ }\bibfield  {title} {\enquote {\bibinfo {title} {Normalizing flows for probabilistic modeling and inference.}}\ }\href@noop {} {\bibfield  {journal} {\bibinfo  {journal} {J. Mach. Learn. Res.}\ }\textbf {\bibinfo {volume} {22}},\ \bibinfo {pages} {1--64} (\bibinfo {year} {2021})}\BibitemShut {NoStop}%
\bibitem [{\citenamefont {P{\'e}rez-Hern{\'a}ndez}\ \emph {et~al.}(2013)\citenamefont {P{\'e}rez-Hern{\'a}ndez}, \citenamefont {Paul}, \citenamefont {Giorgino}, \citenamefont {De~Fabritiis},\ and\ \citenamefont {No{\'e}}}]{perez2013identification}%
  \BibitemOpen
  \bibfield  {author} {\bibinfo {author} {\bibfnamefont {G.}~\bibnamefont {P{\'e}rez-Hern{\'a}ndez}}, \bibinfo {author} {\bibfnamefont {F.}~\bibnamefont {Paul}}, \bibinfo {author} {\bibfnamefont {T.}~\bibnamefont {Giorgino}}, \bibinfo {author} {\bibfnamefont {G.}~\bibnamefont {De~Fabritiis}}, \ and\ \bibinfo {author} {\bibfnamefont {F.}~\bibnamefont {No{\'e}}},\ }\bibfield  {title} {\enquote {\bibinfo {title} {Identification of slow molecular order parameters for {M}arkov model construction},}\ }\href@noop {} {\bibfield  {journal} {\bibinfo  {journal} {The Journal of chemical physics}\ }\textbf {\bibinfo {volume} {139}},\ \bibinfo {pages} {07B604\_1} (\bibinfo {year} {2013})}\BibitemShut {NoStop}%
\bibitem [{\citenamefont {Ercolessi}\ and\ \citenamefont {Adams}(1994)}]{ercolessi1994interatomic}%
  \BibitemOpen
  \bibfield  {author} {\bibinfo {author} {\bibfnamefont {F.}~\bibnamefont {Ercolessi}}\ and\ \bibinfo {author} {\bibfnamefont {J.~B.}\ \bibnamefont {Adams}},\ }\bibfield  {title} {\enquote {\bibinfo {title} {Interatomic potentials from first-principles calculations: the force-matching method},}\ }\href@noop {} {\bibfield  {journal} {\bibinfo  {journal} {Europhysics Letters}\ }\textbf {\bibinfo {volume} {26}},\ \bibinfo {pages} {583} (\bibinfo {year} {1994})}\BibitemShut {NoStop}%
\bibitem [{\citenamefont {Ciccotti}, \citenamefont {Kapral},\ and\ \citenamefont {Vanden-Eijnden}(2005)}]{ciccotti2005blue}%
  \BibitemOpen
  \bibfield  {author} {\bibinfo {author} {\bibfnamefont {G.}~\bibnamefont {Ciccotti}}, \bibinfo {author} {\bibfnamefont {R.}~\bibnamefont {Kapral}}, \ and\ \bibinfo {author} {\bibfnamefont {E.}~\bibnamefont {Vanden-Eijnden}},\ }\bibfield  {title} {\enquote {\bibinfo {title} {Blue moon sampling, vectorial reaction coordinates, and unbiased constrained dynamics},}\ }\href@noop {} {\bibfield  {journal} {\bibinfo  {journal} {ChemPhysChem}\ }\textbf {\bibinfo {volume} {6}},\ \bibinfo {pages} {1809--1814} (\bibinfo {year} {2005})}\BibitemShut {NoStop}%
\bibitem [{\citenamefont {Prinz}\ \emph {et~al.}(2011)\citenamefont {Prinz}, \citenamefont {Wu}, \citenamefont {Sarich}, \citenamefont {Keller}, \citenamefont {Senne}, \citenamefont {Held}, \citenamefont {Chodera}, \citenamefont {Sch{\"u}tte},\ and\ \citenamefont {No{\'e}}}]{prinz2011markov}%
  \BibitemOpen
  \bibfield  {author} {\bibinfo {author} {\bibfnamefont {J.-H.}\ \bibnamefont {Prinz}}, \bibinfo {author} {\bibfnamefont {H.}~\bibnamefont {Wu}}, \bibinfo {author} {\bibfnamefont {M.}~\bibnamefont {Sarich}}, \bibinfo {author} {\bibfnamefont {B.}~\bibnamefont {Keller}}, \bibinfo {author} {\bibfnamefont {M.}~\bibnamefont {Senne}}, \bibinfo {author} {\bibfnamefont {M.}~\bibnamefont {Held}}, \bibinfo {author} {\bibfnamefont {J.~D.}\ \bibnamefont {Chodera}}, \bibinfo {author} {\bibfnamefont {C.}~\bibnamefont {Sch{\"u}tte}}, \ and\ \bibinfo {author} {\bibfnamefont {F.}~\bibnamefont {No{\'e}}},\ }\bibfield  {title} {\enquote {\bibinfo {title} {Markov models of molecular kinetics: Generation and validation},}\ }\href@noop {} {\bibfield  {journal} {\bibinfo  {journal} {The Journal of chemical physics}\ }\textbf {\bibinfo {volume} {134}},\ \bibinfo {pages} {174105} (\bibinfo {year} {2011})}\BibitemShut {NoStop}%
\bibitem [{\citenamefont {Bowman}, \citenamefont {Pande},\ and\ \citenamefont {No{\'e}}(2013)}]{bowman2013introduction}%
  \BibitemOpen
  \bibfield  {author} {\bibinfo {author} {\bibfnamefont {G.~R.}\ \bibnamefont {Bowman}}, \bibinfo {author} {\bibfnamefont {V.~S.}\ \bibnamefont {Pande}}, \ and\ \bibinfo {author} {\bibfnamefont {F.}~\bibnamefont {No{\'e}}},\ }\href@noop {} {\emph {\bibinfo {title} {An introduction to Markov state models and their application to long timescale molecular simulation}}},\ Vol.\ \bibinfo {volume} {797}\ (\bibinfo  {publisher} {Springer Science \& Business Media},\ \bibinfo {year} {2013})\BibitemShut {NoStop}%
\bibitem [{\citenamefont {Klein}\ \emph {et~al.}(2023)\citenamefont {Klein}, \citenamefont {Foong}, \citenamefont {Fjelde}, \citenamefont {Mlodozeniec}, \citenamefont {Brockschmidt}, \citenamefont {Nowozin}, \citenamefont {Noe},\ and\ \citenamefont {Tomioka}}]{klein2023timewarp}%
  \BibitemOpen
  \bibfield  {author} {\bibinfo {author} {\bibfnamefont {L.}~\bibnamefont {Klein}}, \bibinfo {author} {\bibfnamefont {A.~Y.~K.}\ \bibnamefont {Foong}}, \bibinfo {author} {\bibfnamefont {T.~E.}\ \bibnamefont {Fjelde}}, \bibinfo {author} {\bibfnamefont {B.~K.}\ \bibnamefont {Mlodozeniec}}, \bibinfo {author} {\bibfnamefont {M.}~\bibnamefont {Brockschmidt}}, \bibinfo {author} {\bibfnamefont {S.}~\bibnamefont {Nowozin}}, \bibinfo {author} {\bibfnamefont {F.}~\bibnamefont {Noe}}, \ and\ \bibinfo {author} {\bibfnamefont {R.}~\bibnamefont {Tomioka}},\ }\bibfield  {title} {\enquote {\bibinfo {title} {Timewarp: Transferable acceleration of molecular dynamics by learning time-coarsened dynamics},}\ }in\ \href {https://openreview.net/forum?id=EjMLpTgvKH} {\emph {\bibinfo {booktitle} {Thirty-seventh Conference on Neural Information Processing Systems}}}\ (\bibinfo {year} {2023})\BibitemShut {NoStop}%
\bibitem [{\citenamefont {Sch{\"u}tt}\ \emph {et~al.}(2018{\natexlab{a}})\citenamefont {Sch{\"u}tt}, \citenamefont {Sauceda}, \citenamefont {Kindermans}, \citenamefont {Tkatchenko},\ and\ \citenamefont {M{\"u}ller}}]{schutt2018schnet}%
  \BibitemOpen
  \bibfield  {author} {\bibinfo {author} {\bibfnamefont {K.~T.}\ \bibnamefont {Sch{\"u}tt}}, \bibinfo {author} {\bibfnamefont {H.~E.}\ \bibnamefont {Sauceda}}, \bibinfo {author} {\bibfnamefont {P.-J.}\ \bibnamefont {Kindermans}}, \bibinfo {author} {\bibfnamefont {A.}~\bibnamefont {Tkatchenko}}, \ and\ \bibinfo {author} {\bibfnamefont {K.-R.}\ \bibnamefont {M{\"u}ller}},\ }\bibfield  {title} {\enquote {\bibinfo {title} {Schnet--a deep learning architecture for molecules and materials},}\ }\href@noop {} {\bibfield  {journal} {\bibinfo  {journal} {J. Chem. Phys.}\ }\textbf {\bibinfo {volume} {148}},\ \bibinfo {pages} {241722} (\bibinfo {year} {2018}{\natexlab{a}})}\BibitemShut {NoStop}%
\bibitem [{\citenamefont {Dinh}, \citenamefont {Krueger},\ and\ \citenamefont {Bengio}(2014)}]{dinh14_nice}%
  \BibitemOpen
  \bibfield  {author} {\bibinfo {author} {\bibfnamefont {L.}~\bibnamefont {Dinh}}, \bibinfo {author} {\bibfnamefont {D.}~\bibnamefont {Krueger}}, \ and\ \bibinfo {author} {\bibfnamefont {Y.}~\bibnamefont {Bengio}},\ }\bibfield  {title} {\enquote {\bibinfo {title} {Nice: Non-linear independent components estimation},}\ }\href {http://arxiv.org/abs/1410.8516v6} {\bibfield  {journal} {\bibinfo  {journal} {CoRR}\ } (\bibinfo {year} {2014})},\ \Eprint {http://arxiv.org/abs/1410.8516} {arXiv:1410.8516 [cs.LG]} \BibitemShut {NoStop}%
\bibitem [{\citenamefont {Dinh}, \citenamefont {Sohl-Dickstein},\ and\ \citenamefont {Bengio}(2017)}]{dinh16_densit_estim_using_real_nvp}%
  \BibitemOpen
  \bibfield  {author} {\bibinfo {author} {\bibfnamefont {L.}~\bibnamefont {Dinh}}, \bibinfo {author} {\bibfnamefont {J.}~\bibnamefont {Sohl-Dickstein}}, \ and\ \bibinfo {author} {\bibfnamefont {S.}~\bibnamefont {Bengio}},\ }\bibfield  {title} {\enquote {\bibinfo {title} {Density estimation using real {NVP}},}\ }in\ \href {https://openreview.net/forum?id=HkpbnH9lx} {\emph {\bibinfo {booktitle} {International Conference on Learning Representations}}}\ (\bibinfo {year} {2017})\BibitemShut {NoStop}%
\bibitem [{\citenamefont {Chen}\ \emph {et~al.}(2018)\citenamefont {Chen}, \citenamefont {Rubanova}, \citenamefont {Bettencourt},\ and\ \citenamefont {Duvenaud}}]{chen2018neural}%
  \BibitemOpen
  \bibfield  {author} {\bibinfo {author} {\bibfnamefont {T.~Q.}\ \bibnamefont {Chen}}, \bibinfo {author} {\bibfnamefont {Y.}~\bibnamefont {Rubanova}}, \bibinfo {author} {\bibfnamefont {J.}~\bibnamefont {Bettencourt}}, \ and\ \bibinfo {author} {\bibfnamefont {D.~K.}\ \bibnamefont {Duvenaud}},\ }\bibfield  {title} {\enquote {\bibinfo {title} {Neural ordinary differential equations},}\ }in\ \href@noop {} {\emph {\bibinfo {booktitle} {Advances in neural information processing systems}}}\ (\bibinfo {year} {2018})\ pp.\ \bibinfo {pages} {6571--6583}\BibitemShut {NoStop}%
\bibitem [{\citenamefont {Grathwohl}\ \emph {et~al.}(2019)\citenamefont {Grathwohl}, \citenamefont {Chen}, \citenamefont {Bettencourt},\ and\ \citenamefont {Duvenaud}}]{grathwohl2018ffjord}%
  \BibitemOpen
  \bibfield  {author} {\bibinfo {author} {\bibfnamefont {W.}~\bibnamefont {Grathwohl}}, \bibinfo {author} {\bibfnamefont {R.~T.~Q.}\ \bibnamefont {Chen}}, \bibinfo {author} {\bibfnamefont {J.}~\bibnamefont {Bettencourt}}, \ and\ \bibinfo {author} {\bibfnamefont {D.}~\bibnamefont {Duvenaud}},\ }\bibfield  {title} {\enquote {\bibinfo {title} {Scalable reversible generative models with free-form continuous dynamics},}\ }in\ \href {https://openreview.net/forum?id=rJxgknCcK7} {\emph {\bibinfo {booktitle} {International Conference on Learning Representations}}}\ (\bibinfo {year} {2019})\BibitemShut {NoStop}%
\bibitem [{\citenamefont {Klein}, \citenamefont {Kr{\"a}mer},\ and\ \citenamefont {Noe}(2023)}]{klein2023equivariant}%
  \BibitemOpen
  \bibfield  {author} {\bibinfo {author} {\bibfnamefont {L.}~\bibnamefont {Klein}}, \bibinfo {author} {\bibfnamefont {A.}~\bibnamefont {Kr{\"a}mer}}, \ and\ \bibinfo {author} {\bibfnamefont {F.}~\bibnamefont {Noe}},\ }\bibfield  {title} {\enquote {\bibinfo {title} {Equivariant flow matching},}\ }in\ \href {https://openreview.net/forum?id=eLH2NFOO1B} {\emph {\bibinfo {booktitle} {Thirty-seventh Conference on Neural Information Processing Systems}}}\ (\bibinfo {year} {2023})\BibitemShut {NoStop}%
\bibitem [{\citenamefont {Klein}\ and\ \citenamefont {No\'{e}}(2024)}]{klein2024tbg}%
  \BibitemOpen
  \bibfield  {author} {\bibinfo {author} {\bibfnamefont {L.}~\bibnamefont {Klein}}\ and\ \bibinfo {author} {\bibfnamefont {F.}~\bibnamefont {No\'{e}}},\ }\bibfield  {title} {\enquote {\bibinfo {title} {Transferable {Boltzmann} generators},}\ }in\ \href@noop {} {\emph {\bibinfo {booktitle} {Advances in Neural Information Processing Systems}}},\ Vol.~\bibinfo {volume} {37},\ \bibinfo {editor} {edited by\ \bibinfo {editor} {\bibfnamefont {A.}~\bibnamefont {Globerson}}, \bibinfo {editor} {\bibfnamefont {L.}~\bibnamefont {Mackey}}, \bibinfo {editor} {\bibfnamefont {D.}~\bibnamefont {Belgrave}}, \bibinfo {editor} {\bibfnamefont {A.}~\bibnamefont {Fan}}, \bibinfo {editor} {\bibfnamefont {U.}~\bibnamefont {Paquet}}, \bibinfo {editor} {\bibfnamefont {J.}~\bibnamefont {Tomczak}}, \ and\ \bibinfo {editor} {\bibfnamefont {C.}~\bibnamefont {Zhang}}}\ (\bibinfo  {publisher} {Curran Associates, Inc.},\ \bibinfo {year} {2024})\ pp.\ \bibinfo {pages} {45281--45314}\BibitemShut {NoStop}%
\bibitem [{\citenamefont {Satorras}, \citenamefont {Hoogeboom},\ and\ \citenamefont {Welling}(2021)}]{satorras2021graph}%
  \BibitemOpen
  \bibfield  {author} {\bibinfo {author} {\bibfnamefont {V.~G.}\ \bibnamefont {Satorras}}, \bibinfo {author} {\bibfnamefont {E.}~\bibnamefont {Hoogeboom}}, \ and\ \bibinfo {author} {\bibfnamefont {M.}~\bibnamefont {Welling}},\ }\bibfield  {title} {\enquote {\bibinfo {title} {E (n) equivariant graph neural networks},}\ }in\ \href@noop {} {\emph {\bibinfo {booktitle} {International conference on machine learning}}}\ (\bibinfo {organization} {PMLR},\ \bibinfo {year} {2021})\ pp.\ \bibinfo {pages} {9323--9332}\BibitemShut {NoStop}%
\bibitem [{\citenamefont {Garcia~Satorras}\ \emph {et~al.}(2021)\citenamefont {Garcia~Satorras}, \citenamefont {Hoogeboom}, \citenamefont {Fuchs}, \citenamefont {Posner},\ and\ \citenamefont {Welling}}]{satorras2021n}%
  \BibitemOpen
  \bibfield  {author} {\bibinfo {author} {\bibfnamefont {V.}~\bibnamefont {Garcia~Satorras}}, \bibinfo {author} {\bibfnamefont {E.}~\bibnamefont {Hoogeboom}}, \bibinfo {author} {\bibfnamefont {F.}~\bibnamefont {Fuchs}}, \bibinfo {author} {\bibfnamefont {I.}~\bibnamefont {Posner}}, \ and\ \bibinfo {author} {\bibfnamefont {M.}~\bibnamefont {Welling}},\ }\bibfield  {title} {\enquote {\bibinfo {title} {E(n) equivariant normalizing flows},}\ }in\ \href {https://proceedings.neurips.cc/paper/2021/file/21b5680d80f75a616096f2e791affac6-Paper.pdf} {\emph {\bibinfo {booktitle} {Advances in Neural Information Processing Systems}}},\ Vol.~\bibinfo {volume} {34},\ \bibinfo {editor} {edited by\ \bibinfo {editor} {\bibfnamefont {M.}~\bibnamefont {Ranzato}}, \bibinfo {editor} {\bibfnamefont {A.}~\bibnamefont {Beygelzimer}}, \bibinfo {editor} {\bibfnamefont {Y.}~\bibnamefont {Dauphin}}, \bibinfo {editor} {\bibfnamefont {P.}~\bibnamefont {Liang}}, \ and\ \bibinfo {editor} {\bibfnamefont {J.~W.}\ \bibnamefont {Vaughan}}}\
  (\bibinfo  {publisher} {Curran Associates, Inc.},\ \bibinfo {year} {2021})\ pp.\ \bibinfo {pages} {4181--4192}\BibitemShut {NoStop}%
\bibitem [{\citenamefont {K{\"o}hler}, \citenamefont {Klein},\ and\ \citenamefont {No{\'e}}(2020)}]{kohler2020equivariant}%
  \BibitemOpen
  \bibfield  {author} {\bibinfo {author} {\bibfnamefont {J.}~\bibnamefont {K{\"o}hler}}, \bibinfo {author} {\bibfnamefont {L.}~\bibnamefont {Klein}}, \ and\ \bibinfo {author} {\bibfnamefont {F.}~\bibnamefont {No{\'e}}},\ }\bibfield  {title} {\enquote {\bibinfo {title} {Equivariant flows: exact likelihood generative learning for symmetric densities},}\ }in\ \href@noop {} {\emph {\bibinfo {booktitle} {International conference on machine learning}}}\ (\bibinfo {organization} {PMLR},\ \bibinfo {year} {2020})\ pp.\ \bibinfo {pages} {5361--5370}\BibitemShut {NoStop}%
\bibitem [{\citenamefont {Lipman}\ \emph {et~al.}(2023)\citenamefont {Lipman}, \citenamefont {Chen}, \citenamefont {Ben-Hamu}, \citenamefont {Nickel},\ and\ \citenamefont {Le}}]{lipman2022flow}%
  \BibitemOpen
  \bibfield  {author} {\bibinfo {author} {\bibfnamefont {Y.}~\bibnamefont {Lipman}}, \bibinfo {author} {\bibfnamefont {R.~T.~Q.}\ \bibnamefont {Chen}}, \bibinfo {author} {\bibfnamefont {H.}~\bibnamefont {Ben-Hamu}}, \bibinfo {author} {\bibfnamefont {M.}~\bibnamefont {Nickel}}, \ and\ \bibinfo {author} {\bibfnamefont {M.}~\bibnamefont {Le}},\ }\bibfield  {title} {\enquote {\bibinfo {title} {Flow matching for generative modeling},}\ }in\ \href {https://openreview.net/forum?id=PqvMRDCJT9t} {\emph {\bibinfo {booktitle} {The Eleventh International Conference on Learning Representations}}}\ (\bibinfo {year} {2023})\BibitemShut {NoStop}%
\bibitem [{\citenamefont {Albergo}, \citenamefont {Boffi},\ and\ \citenamefont {Vanden-Eijnden}(2023)}]{albergo2023stochastic}%
  \BibitemOpen
  \bibfield  {author} {\bibinfo {author} {\bibfnamefont {M.~S.}\ \bibnamefont {Albergo}}, \bibinfo {author} {\bibfnamefont {N.~M.}\ \bibnamefont {Boffi}}, \ and\ \bibinfo {author} {\bibfnamefont {E.}~\bibnamefont {Vanden-Eijnden}},\ }\bibfield  {title} {\enquote {\bibinfo {title} {{S}tochastic interpolants: {A} unifying framework for flows and diffusions},}\ }\href {https://arxiv.org/abs/2303.08797} {\bibfield  {journal} {\bibinfo  {journal} {ArXiv preprint}\ }\textbf {\bibinfo {volume} {abs/2303.08797}} (\bibinfo {year} {2023})}\BibitemShut {NoStop}%
\bibitem [{\citenamefont {Liu}, \citenamefont {Gong},\ and\ \citenamefont {Liu}(2022)}]{liu2022flow}%
  \BibitemOpen
  \bibfield  {author} {\bibinfo {author} {\bibfnamefont {X.}~\bibnamefont {Liu}}, \bibinfo {author} {\bibfnamefont {C.}~\bibnamefont {Gong}}, \ and\ \bibinfo {author} {\bibfnamefont {Q.}~\bibnamefont {Liu}},\ }\bibfield  {title} {\enquote {\bibinfo {title} {Flow straight and fast: Learning to generate and transfer data with rectified flow},}\ }\href@noop {} {\bibfield  {journal} {\bibinfo  {journal} {arXiv preprint arXiv:2209.03003}\ } (\bibinfo {year} {2022})}\BibitemShut {NoStop}%
\bibitem [{\citenamefont {Ramachandran}, \citenamefont {Ramakrishnan},\ and\ \citenamefont {Sasisekharan}(1963)}]{ramachandran1963stereochemistry}%
  \BibitemOpen
  \bibfield  {author} {\bibinfo {author} {\bibfnamefont {G.~N.}\ \bibnamefont {Ramachandran}}, \bibinfo {author} {\bibfnamefont {C.}~\bibnamefont {Ramakrishnan}}, \ and\ \bibinfo {author} {\bibfnamefont {V.}~\bibnamefont {Sasisekharan}},\ }\bibfield  {title} {\enquote {\bibinfo {title} {Stereochemistry of polypeptide chain configurations},}\ }\href@noop {} {\bibfield  {journal} {\bibinfo  {journal} {Journal of Molecular Biology}\ ,\ \bibinfo {pages} {95--99}} (\bibinfo {year} {1963})}\BibitemShut {NoStop}%
\bibitem [{\citenamefont {Deb}(2011)}]{deb2011multi}%
  \BibitemOpen
  \bibfield  {author} {\bibinfo {author} {\bibfnamefont {K.}~\bibnamefont {Deb}},\ }\bibfield  {title} {\enquote {\bibinfo {title} {Multi-objective optimisation using evolutionary algorithms: an introduction},}\ }in\ \href@noop {} {\emph {\bibinfo {booktitle} {Multi-objective evolutionary optimisation for product design and manufacturing}}}\ (\bibinfo  {publisher} {Springer},\ \bibinfo {year} {2011})\ pp.\ \bibinfo {pages} {3--34}\BibitemShut {NoStop}%
\bibitem [{\citenamefont {Piana}, \citenamefont {Lindorff-Larsen},\ and\ \citenamefont {Shaw}(2011)}]{piana2011robust}%
  \BibitemOpen
  \bibfield  {author} {\bibinfo {author} {\bibfnamefont {S.}~\bibnamefont {Piana}}, \bibinfo {author} {\bibfnamefont {K.}~\bibnamefont {Lindorff-Larsen}}, \ and\ \bibinfo {author} {\bibfnamefont {D.~E.}\ \bibnamefont {Shaw}},\ }\bibfield  {title} {\enquote {\bibinfo {title} {How robust are protein folding simulations with respect to force field parameterization?}}\ }\href@noop {} {\bibfield  {journal} {\bibinfo  {journal} {Biophys. J.}\ }\textbf {\bibinfo {volume} {100}},\ \bibinfo {pages} {L47--L49} (\bibinfo {year} {2011})}\BibitemShut {NoStop}%
\bibitem [{\citenamefont {Dibak}\ \emph {et~al.}(2022)\citenamefont {Dibak}, \citenamefont {Klein}, \citenamefont {Kr\"amer},\ and\ \citenamefont {No\'e}}]{dibak2021temperature}%
  \BibitemOpen
  \bibfield  {author} {\bibinfo {author} {\bibfnamefont {M.}~\bibnamefont {Dibak}}, \bibinfo {author} {\bibfnamefont {L.}~\bibnamefont {Klein}}, \bibinfo {author} {\bibfnamefont {A.}~\bibnamefont {Kr\"amer}}, \ and\ \bibinfo {author} {\bibfnamefont {F.}~\bibnamefont {No\'e}},\ }\bibfield  {title} {\enquote {\bibinfo {title} {Temperature steerable flows and {Boltzmann} generators},}\ }\href {\doibase 10.1103/PhysRevResearch.4.L042005} {\bibfield  {journal} {\bibinfo  {journal} {Phys. Rev. Res.}\ }\textbf {\bibinfo {volume} {4}},\ \bibinfo {pages} {L042005} (\bibinfo {year} {2022})}\BibitemShut {NoStop}%
\bibitem [{\citenamefont {K\"{o}hler}, \citenamefont {Kr\"{a}mer},\ and\ \citenamefont {Noé}(2021)}]{kohler2021smooth}%
  \BibitemOpen
  \bibfield  {author} {\bibinfo {author} {\bibfnamefont {J.}~\bibnamefont {K\"{o}hler}}, \bibinfo {author} {\bibfnamefont {A.}~\bibnamefont {Kr\"{a}mer}}, \ and\ \bibinfo {author} {\bibfnamefont {F.}~\bibnamefont {Noé}},\ }\bibfield  {title} {\enquote {\bibinfo {title} {Smooth normalizing flows},}\ }in\ \href {https://proceedings.neurips.cc/paper/2021/file/167434fa6219316417cd4160c0c5e7d2-Paper.pdf} {\emph {\bibinfo {booktitle} {Advances in Neural Information Processing Systems}}},\ Vol.~\bibinfo {volume} {34},\ \bibinfo {editor} {edited by\ \bibinfo {editor} {\bibfnamefont {M.}~\bibnamefont {Ranzato}}, \bibinfo {editor} {\bibfnamefont {A.}~\bibnamefont {Beygelzimer}}, \bibinfo {editor} {\bibfnamefont {Y.}~\bibnamefont {Dauphin}}, \bibinfo {editor} {\bibfnamefont {P.}~\bibnamefont {Liang}}, \ and\ \bibinfo {editor} {\bibfnamefont {J.~W.}\ \bibnamefont {Vaughan}}}\ (\bibinfo  {publisher} {Curran Associates, Inc.},\ \bibinfo {year} {2021})\ pp.\ \bibinfo {pages} {2796--2809}\BibitemShut {NoStop}%
\bibitem [{\citenamefont {Invernizzi}\ \emph {et~al.}(2022)\citenamefont {Invernizzi}, \citenamefont {Kr\"amer}, \citenamefont {Clementi},\ and\ \citenamefont {No{\'e}}}]{invernizzi2022skipping}%
  \BibitemOpen
  \bibfield  {author} {\bibinfo {author} {\bibfnamefont {M.}~\bibnamefont {Invernizzi}}, \bibinfo {author} {\bibfnamefont {A.}~\bibnamefont {Kr\"amer}}, \bibinfo {author} {\bibfnamefont {C.}~\bibnamefont {Clementi}}, \ and\ \bibinfo {author} {\bibfnamefont {F.}~\bibnamefont {No{\'e}}},\ }\bibfield  {title} {\enquote {\bibinfo {title} {Skipping the replica exchange ladder with normalizing flows},}\ }\href@noop {} {\bibfield  {journal} {\bibinfo  {journal} {The Journal of Physical Chemistry Letters}\ }\textbf {\bibinfo {volume} {13}},\ \bibinfo {pages} {11643--11649} (\bibinfo {year} {2022})}\BibitemShut {NoStop}%
\bibitem [{\citenamefont {Honda}\ \emph {et~al.}(2008)\citenamefont {Honda}, \citenamefont {Akiba}, \citenamefont {Kato}, \citenamefont {Sawada}, \citenamefont {Sekijima}, \citenamefont {Ishimura}, \citenamefont {Ooishi}, \citenamefont {Watanabe}, \citenamefont {Odahara},\ and\ \citenamefont {Harata}}]{honda2008crystal}%
  \BibitemOpen
  \bibfield  {author} {\bibinfo {author} {\bibfnamefont {S.}~\bibnamefont {Honda}}, \bibinfo {author} {\bibfnamefont {T.}~\bibnamefont {Akiba}}, \bibinfo {author} {\bibfnamefont {Y.~S.}\ \bibnamefont {Kato}}, \bibinfo {author} {\bibfnamefont {Y.}~\bibnamefont {Sawada}}, \bibinfo {author} {\bibfnamefont {M.}~\bibnamefont {Sekijima}}, \bibinfo {author} {\bibfnamefont {M.}~\bibnamefont {Ishimura}}, \bibinfo {author} {\bibfnamefont {A.}~\bibnamefont {Ooishi}}, \bibinfo {author} {\bibfnamefont {H.}~\bibnamefont {Watanabe}}, \bibinfo {author} {\bibfnamefont {T.}~\bibnamefont {Odahara}}, \ and\ \bibinfo {author} {\bibfnamefont {K.}~\bibnamefont {Harata}},\ }\bibfield  {title} {\enquote {\bibinfo {title} {Crystal structure of a ten-amino acid protein},}\ }\href@noop {} {\bibfield  {journal} {\bibinfo  {journal} {J. Am. Chem. Soc.}\ }\textbf {\bibinfo {volume} {130}},\ \bibinfo {pages} {15327--15331} (\bibinfo {year} {2008})}\BibitemShut {NoStop}%
\bibitem [{\citenamefont {Maier}\ \emph {et~al.}(2015)\citenamefont {Maier}, \citenamefont {Martinez}, \citenamefont {Kasavajhala}, \citenamefont {Wickstrom}, \citenamefont {Hauser},\ and\ \citenamefont {Simmerling}}]{maier2015ff14sb}%
  \BibitemOpen
  \bibfield  {author} {\bibinfo {author} {\bibfnamefont {J.~A.}\ \bibnamefont {Maier}}, \bibinfo {author} {\bibfnamefont {C.}~\bibnamefont {Martinez}}, \bibinfo {author} {\bibfnamefont {K.}~\bibnamefont {Kasavajhala}}, \bibinfo {author} {\bibfnamefont {L.}~\bibnamefont {Wickstrom}}, \bibinfo {author} {\bibfnamefont {K.~E.}\ \bibnamefont {Hauser}}, \ and\ \bibinfo {author} {\bibfnamefont {C.}~\bibnamefont {Simmerling}},\ }\bibfield  {title} {\enquote {\bibinfo {title} {ff14sb: improving the accuracy of protein side chain and backbone parameters from ff99sb},}\ }\href@noop {} {\bibfield  {journal} {\bibinfo  {journal} {Journal of chemical theory and computation}\ }\textbf {\bibinfo {volume} {11}},\ \bibinfo {pages} {3696--3713} (\bibinfo {year} {2015})}\BibitemShut {NoStop}%
\bibitem [{\citenamefont {Midgley}\ \emph {et~al.}(2023)\citenamefont {Midgley}, \citenamefont {Stimper}, \citenamefont {Antoran}, \citenamefont {Mathieu}, \citenamefont {Sch{\"o}lkopf},\ and\ \citenamefont {Hern{\'a}ndez-Lobato}}]{midgley2023se}%
  \BibitemOpen
  \bibfield  {author} {\bibinfo {author} {\bibfnamefont {L.~I.}\ \bibnamefont {Midgley}}, \bibinfo {author} {\bibfnamefont {V.}~\bibnamefont {Stimper}}, \bibinfo {author} {\bibfnamefont {J.}~\bibnamefont {Antoran}}, \bibinfo {author} {\bibfnamefont {E.}~\bibnamefont {Mathieu}}, \bibinfo {author} {\bibfnamefont {B.}~\bibnamefont {Sch{\"o}lkopf}}, \ and\ \bibinfo {author} {\bibfnamefont {J.~M.}\ \bibnamefont {Hern{\'a}ndez-Lobato}},\ }\bibfield  {title} {\enquote {\bibinfo {title} {{SE}(3) equivariant augmented coupling flows},}\ }in\ \href {https://openreview.net/forum?id=KKxO6wwx8p} {\emph {\bibinfo {booktitle} {Thirty-seventh Conference on Neural Information Processing Systems}}}\ (\bibinfo {year} {2023})\BibitemShut {NoStop}%
\bibitem [{\citenamefont {K{\"{o}}hler}\ \emph {et~al.}(2023)\citenamefont {K{\"{o}}hler}, \citenamefont {Invernizzi}, \citenamefont {de~Haan},\ and\ \citenamefont {No{\'{e}}}}]{kohler2023rigid}%
  \BibitemOpen
  \bibfield  {author} {\bibinfo {author} {\bibfnamefont {J.}~\bibnamefont {K{\"{o}}hler}}, \bibinfo {author} {\bibfnamefont {M.}~\bibnamefont {Invernizzi}}, \bibinfo {author} {\bibfnamefont {P.}~\bibnamefont {de~Haan}}, \ and\ \bibinfo {author} {\bibfnamefont {F.}~\bibnamefont {No{\'{e}}}},\ }\bibfield  {title} {\enquote {\bibinfo {title} {Rigid body flows for sampling molecular crystal structures},}\ }in\ \href {https://proceedings.mlr.press/v202/kohler23a.html} {\emph {\bibinfo {booktitle} {International Conference on Machine Learning, {ICML} 2023}}},\ \bibinfo {series} {Proceedings of Machine Learning Research}, Vol.\ \bibinfo {volume} {202}\ (\bibinfo  {publisher} {{PMLR}},\ \bibinfo {year} {2023})\ pp.\ \bibinfo {pages} {17301--17326}\BibitemShut {NoStop}%
\bibitem [{\citenamefont {Abdin}\ and\ \citenamefont {Kim}(2023)}]{abdin2023pepflow}%
  \BibitemOpen
  \bibfield  {author} {\bibinfo {author} {\bibfnamefont {O.}~\bibnamefont {Abdin}}\ and\ \bibinfo {author} {\bibfnamefont {P.~M.}\ \bibnamefont {Kim}},\ }\bibfield  {title} {\enquote {\bibinfo {title} {Pepflow: direct conformational sampling from peptide energy landscapes through hypernetwork-conditioned diffusion},}\ }\href@noop {} {\bibfield  {journal} {\bibinfo  {journal} {bioRxiv}\ ,\ \bibinfo {pages} {2023--06}} (\bibinfo {year} {2023})}\BibitemShut {NoStop}%
\bibitem [{\citenamefont {Draxler}\ \emph {et~al.}(2024)\citenamefont {Draxler}, \citenamefont {Sorrenson}, \citenamefont {Zimmermann}, \citenamefont {Rousselot},\ and\ \citenamefont {K\"{o}the}}]{pmlr-v238-draxler24a}%
  \BibitemOpen
  \bibfield  {author} {\bibinfo {author} {\bibfnamefont {F.}~\bibnamefont {Draxler}}, \bibinfo {author} {\bibfnamefont {P.}~\bibnamefont {Sorrenson}}, \bibinfo {author} {\bibfnamefont {L.}~\bibnamefont {Zimmermann}}, \bibinfo {author} {\bibfnamefont {A.}~\bibnamefont {Rousselot}}, \ and\ \bibinfo {author} {\bibfnamefont {U.}~\bibnamefont {K\"{o}the}},\ }\bibfield  {title} {\enquote {\bibinfo {title} {Free-form flows: Make any architecture a normalizing flow},}\ }in\ \href {https://proceedings.mlr.press/v238/draxler24a.html} {\emph {\bibinfo {booktitle} {Proceedings of The 27th International Conference on Artificial Intelligence and Statistics}}},\ \bibinfo {series} {Proceedings of Machine Learning Research}, Vol.\ \bibinfo {volume} {238},\ \bibinfo {editor} {edited by\ \bibinfo {editor} {\bibfnamefont {S.}~\bibnamefont {Dasgupta}}, \bibinfo {editor} {\bibfnamefont {S.}~\bibnamefont {Mandt}}, \ and\ \bibinfo {editor} {\bibfnamefont {Y.}~\bibnamefont {Li}}}\ (\bibinfo  {publisher} {PMLR},\ \bibinfo {year}
  {2024})\ pp.\ \bibinfo {pages} {2197--2205}\BibitemShut {NoStop}%
\bibitem [{\citenamefont {Tan}\ \emph {et~al.}(2025)\citenamefont {Tan}, \citenamefont {Bose}, \citenamefont {Lin}, \citenamefont {Klein}, \citenamefont {Bronstein},\ and\ \citenamefont {Tong}}]{tan2025scalable}%
  \BibitemOpen
  \bibfield  {author} {\bibinfo {author} {\bibfnamefont {C.~B.}\ \bibnamefont {Tan}}, \bibinfo {author} {\bibfnamefont {A.~J.}\ \bibnamefont {Bose}}, \bibinfo {author} {\bibfnamefont {C.}~\bibnamefont {Lin}}, \bibinfo {author} {\bibfnamefont {L.}~\bibnamefont {Klein}}, \bibinfo {author} {\bibfnamefont {M.~M.}\ \bibnamefont {Bronstein}}, \ and\ \bibinfo {author} {\bibfnamefont {A.}~\bibnamefont {Tong}},\ }\bibfield  {title} {\enquote {\bibinfo {title} {Scalable equilibrium sampling with sequential boltzmann generators},}\ }\href@noop {} {\bibfield  {journal} {\bibinfo  {journal} {arXiv preprint arXiv:2502.18462}\ } (\bibinfo {year} {2025})}\BibitemShut {NoStop}%
\bibitem [{\citenamefont {Christensen}\ \emph {et~al.}(2002)\citenamefont {Christensen} \emph {et~al.}}]{christensen2002plane}%
  \BibitemOpen
  \bibfield  {author} {\bibinfo {author} {\bibfnamefont {R.}~\bibnamefont {Christensen}} \emph {et~al.},\ }\href@noop {} {\emph {\bibinfo {title} {Plane answers to complex questions}}},\ Vol.~\bibinfo {volume} {35}\ (\bibinfo  {publisher} {Springer},\ \bibinfo {year} {2002})\BibitemShut {NoStop}%
\bibitem [{\citenamefont {Unke}\ and\ \citenamefont {Meuwly}(2019)}]{unke2019physnet}%
  \BibitemOpen
  \bibfield  {author} {\bibinfo {author} {\bibfnamefont {O.~T.}\ \bibnamefont {Unke}}\ and\ \bibinfo {author} {\bibfnamefont {M.}~\bibnamefont {Meuwly}},\ }\bibfield  {title} {\enquote {\bibinfo {title} {Physnet: A neural network for predicting energies, forces, dipole moments, and partial charges},}\ }\href {\doibase 10.1021/acs.jctc.9b00181} {\bibfield  {journal} {\bibinfo  {journal} {J. Chem. Theory Comput.}\ }\textbf {\bibinfo {volume} {15}},\ \bibinfo {pages} {3678–3693} (\bibinfo {year} {2019})}\BibitemShut {NoStop}%
\bibitem [{\citenamefont {Th{\"o}lke}\ and\ \citenamefont {De~Fabritiis}(2022)}]{tholke2022equivariant}%
  \BibitemOpen
  \bibfield  {author} {\bibinfo {author} {\bibfnamefont {P.}~\bibnamefont {Th{\"o}lke}}\ and\ \bibinfo {author} {\bibfnamefont {G.}~\bibnamefont {De~Fabritiis}},\ }\bibfield  {title} {\enquote {\bibinfo {title} {Equivariant transformers for neural network based molecular potentials},}\ }in\ \href {https://openreview.net/forum?id=zNHzqZ9wrRB} {\emph {\bibinfo {booktitle} {International Conference on Learning Representations}}}\ (\bibinfo {year} {2022})\BibitemShut {NoStop}%
\bibitem [{\citenamefont {Sch{\"u}tt}\ \emph {et~al.}(2018{\natexlab{b}})\citenamefont {Sch{\"u}tt}, \citenamefont {Kessel}, \citenamefont {Gastegger}, \citenamefont {Nicoli}, \citenamefont {Tkatchenko},\ and\ \citenamefont {M{\"u}ller}}]{schutt2018schnetpack}%
  \BibitemOpen
  \bibfield  {author} {\bibinfo {author} {\bibfnamefont {K.}~\bibnamefont {Sch{\"u}tt}}, \bibinfo {author} {\bibfnamefont {P.}~\bibnamefont {Kessel}}, \bibinfo {author} {\bibfnamefont {M.}~\bibnamefont {Gastegger}}, \bibinfo {author} {\bibfnamefont {K.~A.}\ \bibnamefont {Nicoli}}, \bibinfo {author} {\bibfnamefont {A.}~\bibnamefont {Tkatchenko}}, \ and\ \bibinfo {author} {\bibfnamefont {K.-R.}\ \bibnamefont {M{\"u}ller}},\ }\bibfield  {title} {\enquote {\bibinfo {title} {Schnetpack: A deep learning toolbox for atomistic systems},}\ }\href@noop {} {\bibfield  {journal} {\bibinfo  {journal} {Journal of chemical theory and computation}\ }\textbf {\bibinfo {volume} {15}},\ \bibinfo {pages} {448--455} (\bibinfo {year} {2018}{\natexlab{b}})}\BibitemShut {NoStop}%
\end{thebibliography}%

\end{document}